\documentclass[onecolumn,secnumarabic,amssymb,amsmath,amsfonts nobibnotes,nofootinbib, aps, prd,preprintnumbers]{revtex4-1}

\usepackage{epsfig,graphicx}
\usepackage{chngcntr}
\usepackage{rotating}
\usepackage{wrapfig}
\usepackage{parskip} % Package to tweak paragraph skipping
\usepackage{graphicx}
\usepackage{xcolor}
\usepackage{caption}
\usepackage{enumitem}
\usepackage{subcaption}
\usepackage{lipsum}  
\usepackage[linkcolor=blue, citecolor=red, urlcolor=blue]{hyperref}
\begin{document}

\title{Stability of the Standard Model Vacuum with Vector-Like Leptons: A Critical Examination }%

\author{\sf Kivanc Y. Cingiloglu\footnote{kivanc.cingiloglu@concordia.ca},\; Mariana Frank\footnote{mariana.frank@concordia.ca}}%

\affiliation{ Department of Physics,  
			Concordia University, 7141 Sherbrooke St.West,\\
			Montreal, Quebec, Canada H4B 1R6.\\}
\date{\today}%			
			
			\begin{abstract}

The stability of the Standard Model (SM) Higgs vacuum is a long-standing issue in particle physics. The SM Higgs quartic coupling parameter is expected to become negative at high scales, potentially generating an unstable vacuum well before reaching the Planck scale. We investigate whether the introduction of vector-like leptons (VLLs) in six distinct gauge anomaly-independent representations can stabilize the SM vacuum without introducing additional scalar fields. By analyzing the mass spectrum and mixing angles of these VLLs with the SM leptons, we identify conditions under which the Higgs potential remains stable. We demonstrate that there exists a set of allowed but narrowed VLL spectra that can indeed stabilize the SM vacuum. We also study the effect of these VLLs on electroweak precision observables, particularly the oblique parameters $\mathbb{S}$ and $\mathbb{T}$ to ensure  fit within global experimental constraints. Finally, we perform a comprehensive analysis of the parameter space allowed by electroweak precision data and the parameter space required for stability, highlighting cases that accommodate both. This study opens up new avenues for understanding the role of vector-like leptons in extending the validity of the Standard Model up to the Planck scale.
\end{abstract}
\maketitle
\bigskip

%%%%%%%%%%%%%%%%%%%%%%%%%%%%%%%%%%%%%%%%%%%%%%%%%%%%%%%%%%%%%%

%\tableofcontents
\section{Introduction}
 \label{sec:intro}

While the discovery of the Higgs boson with mass $125.66 \pm 0.30$ GeV and analysis of its properties \cite{CMS:2022dwd,ATLAS:2022vkf} completes the search for the particle content of the Standard Model (SM), confirming Higgs mechanism to be responsible for electroweak symmetry breaking, it also raises questions about naturalness.  In particular, there are hints that the SM is incomplete, or perhaps just an effective theory at large scales, where the model becomes unstable. The metastability of the SM vacuum is driven by the behavior of the couplings in the model at high energy scales. In general, the SM couplings run slowly but at $\mu \sim 10^{10}$ GeV the Higgs quartic coupling $\lambda$ flips sign, as evidenced by a downward spike,   which indicates the onset of vacuum instability. Extending the validity of the SM
to $M_{\rm {Planck}}$, a second, deeper minimum develops, located near the Planck scale,
such that the electroweak vacuum is metastable, i.e., implying that the theoretical transition lifetime of the electroweak vacuum to the deeper minimum is finite with lifetime $\sim 10^{300}$ years \cite{Sher:1988mj,Sher:1993mf,Elias-Miro:2011sqh,Degrassi:2012ry,Anchordoqui:2012fq,Lebedev:2012zw,Bednyakov:2015sca}.
The issue is that the Higgs quartic coupling is renormalized not only by itself ($\lambda$ increasing as the energy scale increases), but also by the Higgs (Yukawa) coupling to the top quark \cite{Alekhin_2012}, which tends to drive it to smaller, even negative values at high scales $\mu$. 

Vacuum stability can be achieved through beyond the Standard Model (BSM) effects, as long as these enhance the Higgs
quartic coupling sufficiently strongly. This is achieved by introducing new particles, which can couple to the gauge or Higgs fields. While coupling to gauge fields modify the SM beta functions and generally result in small changes, couplings to the Higgs can affect the running of the SM couplings more significantly.

Perhaps the simplest remedy to the stability problem is to augment the  SM by an extra (singlet, as it is simplest) scalar boson which interacts solely with the SM Higgs boson \cite{Gonderinger:2012rd,Falkowski:2015iwa,Khan:2014kba,Han:2015hda,Garg:2017iva}. The addition of a boson provides a positive boost to the coupling parameter, counteracting the effect of the top quark and contributing towards repairing the Higgs vacuum stability. In this scenario, the scalar couplings increase with energy scale, compensating for the SM Higgs coupling.  Therefore, the addition of an extra scalar boson to the SM rescues the theory from vacuum instability, as long as its mixing with the SM Higgs boson is non-zero \cite{PhysRevD.107.036018}. It has been shown that such a singlet scalar, if light, can also serve as dark matter (DM) candidate, obeying all constraints from relic abundance and direct detection \cite{Borah:2020nsz}.  The study of the scalar singlet DM has been extended,  with additional portal couplings of the scalar  added on top of the usual Higgs portal interaction. Such possibility is to include DM portal couplings with new vector like leptons \cite{Toma:2013bka,Giacchino:2013bta,Giacchino:2014moa,Ibarra:2014qma,Barman:2019tuo,Barman:2019aku,Barman:2019oda} and/or quarks \cite{Giacchino:2015hvk,Baek:2016lnv,Baek:2017ykw,Colucci:2018vxz,Biondini:2019int}.

Vector-like leptons (VLLs) are color-singlet fermions and vector-like quarks (VLQs) are color-triplet fermions, i.e., fermions with left- and right-handed components transforming the same way under the electroweak gauge symmetry group. Such new states arise in a wide variety of BSM scenarios, including, but not limited
to, supersymmetric models \cite{Martin:2009bg,Zheng:2019kqu,Graham:2009gy,Endo:2011mc,Araz:2018uyi}
 models with extra spatial dimensions \cite{Kong:2010qd,Huang:2012kz}
and grand unified theories. Expansions of the SM with one or more vector-like fermion
families may provide a dark matter candidate \cite{Schwaller:2013hqa,Halverson:2014nwa,Bahrami:2016has,Bhattacharya:2018fus}, and account for the mass hierarchy
between the different generations of particles in the SM via their mixings with the SM fermions \cite{Agashe:2008fe,Redi:2013pga,Falkowski:2013jya,Frank:2014aca}. 
 
Vector-like particles have been considered before in the context of stabilizing the vacuum of the SM in \cite{Xiao:2014kba,Hiller:2022rla}, in the context of baryogenesis \cite{Egana-Ugrinovic:2017jib}, to account for the anomalous magnetic moment of the muon and discrepancies in the $W$ boson mass \cite{He:2022zjz,Hiller:2019mou,PhysRevD.106.055042},  and to help explain the observed excess at 750 GeV \cite{Dhuria:2015ufo,Zhang:2015uuo}. Analyses of vacuum stability have served as guides for beyond the SM model building \cite{Gabrielli:2013hma,Bond:2017wut,Hiller:2023bdb}.

In a previous work \cite{PhysRevD.107.036018} we analyzed the stability of the SM with additional vector-like quarks. We included all the possible non-anomalous representations, and analyzed a complete interplay of all possible vector-like quark representations. We investigated the restrictions on the masses and mixing angles for the all anomaly-free representations of vector-like quarks, as well as the additional  boson field which is needed to be added for vacuum stabilization, including effects and restrictions induced by the vector-like fermions on the electroweak precision observables (EWPOs), $\mathbb{S}$ and $\mathbb{T}$.
Previous works have also performed complementary analyses \cite{Hiller:2024zjp}, some combining vacuum stability constraints with the possibility of allowing the new scalar to be the dark matter candidate \cite{Khoze:2014xha,Borah:2020nsz}.

Here we complement our previous work here by performing an analysis of the stability of the SM with vector-like leptons. The study of the vacuum stability is different here as vector-like leptons do not have QCD couplings. Additionally, unlike the vector-like quarks, the leptons can still be light, the limit from LEP allowing vector-like charged fermions with masses above 100 GeV \cite{L3:2001xsz,DELPHI:2003uqw,10.1093/ptep/ptac097}. There have been previous studies of the vacuum stability of the SM in the presence of vector-like fermions \cite{Ellis:2014dza,Mann:2017wzh,Borah:2020nsz}. Our analysis differs from previous ones in that we once again examine the effects of {\it all} possible non-anomalous vector-like lepton representations. Moreover, we shall show that stability of the SM with vector-like leptons does not require the additional scalar boson mixed with the Higgs bosons, and this result is consistent with parameter space restrictions on the electroweak precision observables, $\mathbb{S, T}$ and $\mathbb{U}$.

Our work is organized as follows. Sec. \ref{sec:instabilityemerg} summarizes the emergence of the stability issue within the current SM framework. In Sec. \ref{sec:VLLmodel}, we introduce six different vector-like lepton representations and describe their connections to the SM leptons through relevant Lagrangian. In Sec. \ref{sec:electroweakprecision}  we present the vector-like lepton contributions to the electroweak precision observables and discuss the regions of parameter space that are consistent with the spectrum that ensures stability.
 Sec. \ref{sec:rgef} is dedicated to the examination of the running of the SM couplings and renormalization group equation (RGE) solutions in the presence of vector-like leptons along with allowed parameter space that satisfy the vacuum stability constraints. Further, we add two loop corrections to RGEs to check how the next-to-next-to-leading order accuracy affects the model couplings up to the Planck scale. Furthermore, we draw our conclusions in Sec. \ref{sec:conclusion} and leave the complete set of one-loop RGEs{\footnote {While we perform a complete analysis of the parameter space to two loops, for brevity we  list only the one loop contributions to the RGE.}} and VLL modified EW couplings for the Appendix.

%%%%%%%%%%%%%%%%%%%%%%%%%%%%%%%%%%%%%%%%%%%%%%%%
\section{The Emergence of the SM Vacuum Instability}
%%%%%%%%%%%%%%%%%%%%%%%%%%%%%%%%%%%%%%%%%%%%%%%%
\label{sec:instabilityemerg}
The effective Higgs potential, central to the mechanism of electroweak symmetry breaking, is described, at  one loop  level as
\begin{equation}
V(\Phi)_{\text{eff}}=V_{0}(\Phi)+\Delta V_{1-loop}(\Phi)\, ,
\end{equation}
where
\begin{equation}
\label{eq:zeropot}
V_{0}(\Phi)=-\frac{1}{2}\kappa^2\Phi^2+\frac{1}{4}\lambda\Phi^4+\mathcal{O}(\Phi^6)\, ,
\end{equation}
and the one loop corrections to the effective Higgs potential are \cite{PhysRevD.7.1888}
\begin{eqnarray}
\label{eq:1loopcorrections}
16\pi^2\Delta V_{1-loop}(\Phi) &=& \frac{1}{4} \left( m^2 + \frac{\lambda}{2} \Phi^2 \right)^2 \left( \ln \frac{m^2 + \frac{\lambda}{2} \Phi^2}{\mu^2} + \gamma - \ln 4\pi - \frac{3}{2} \right) \nonumber \\
&+&\frac{3}{4} \left( m^2 + \frac{\lambda}{6} \Phi^2 \right)^2 \left( \ln \frac{m^2 + \frac{\lambda}{6} \Phi^2}{\mu^2} + \gamma - \ln 4\pi - \frac{3}{2} \right) \nonumber \\
&-&\frac{3y_t^4\Phi^4}{4} \left( \ln \frac{y_t^2\Phi^2}{2\mu^2} + \gamma - \ln 4\pi - \frac{3}{2} \right) \nonumber \\
&+& \frac{3g_2^4\Phi^4}{32} \left( \ln \frac{g_2^2\Phi^2}{4\mu^2} + \gamma - \ln 4\pi - \frac{5}{6} \right) \nonumber \\
&+& \frac{3(g'^2 + g_2^2)^2\Phi^4}{64} \left( \ln \frac{(g'^2 + g_2^2)\Phi^2}{4\mu^2} + \gamma - \ln 4\pi - \frac{5}{6} \right)\, .
\end{eqnarray}
Here $g$ and $g'$ are $SU(2)_L\otimes U(1)_Y$ gauge couplings. The effective potential $V(\Phi)_{\text{eff}}$ satisfies Callan–Symanzik (CS) equation \cite{PhysRevD.2.1541}
\begin{equation}
\left[-\frac{\partial}{\partial\mu}+\sum_{i}\beta_i(\lambda)\frac{\partial}{\partial\lambda_i}+\sum_{j}n_j\gamma_j(\lambda) \right] V(\Phi)_{\text{eff}}(\lambda_i,\mu)=0\, ,
\end{equation}
where all the couplings in the effective potential are assumed $\lambda_i=(\lambda,m^2,g',g_2,g_3,y_t)$. The correction term $\Delta V_{1-loop}(\Phi)$ introduces logarithmic dependencies on the renormalization scale $\mu$, leading to a potential that evolves with the energy scale according to the RGEs:
\begin{equation}
\frac{d\lambda(\mu)}{d\mu}=\beta_{\lambda}(\lambda,g_i) \qquad,\qquad \frac{dg_i(\mu)}{d\mu}=\beta_{i}(\lambda,g_i)\,.
\end{equation}
Eq. \ref{eq:zeropot} indicates that vacuum stability is assured by the positivity of $\lambda(\mu)$. However, as the energy scale increases, the running of $\lambda$ is influenced by various factors,  including the top quark Yukawa coupling, gauge couplings, and the Higgs self-interaction as included in Eq. \ref{eq:1loopcorrections}. The interplay between these contributions can drive $\lambda(\mu)$ to negative values at high scales, signalling the emergence of a deeper minimum in the Higgs potential at large field values. This scenario suggests that the vacuum we inhabit may not be the true ground state of the universe but rather a metastable state that could eventually decay into a more stable vacuum.  This outcome can be inferred from the SM RGEs that govern the Higgs quartic coupling and the top quark Yukawa coupling and
\begin{eqnarray}
\label{eq:SMRGE}
\frac{d \lambda(\mu)}{d \ln \mu^2}&=&\frac{1}{16 \pi^2}\left[4\lambda^2+12 \lambda y_t^2-36 y_t^4-9 \lambda g_1^2-3 \lambda g_2^2 +\frac94g_2^4 +\frac92g_1^2 g_2^2 +\frac{27}{4}g_1^4 \right] \, , \nonumber \\
\frac{dy_t^2(\mu)}{d \ln \mu^2}&=&\frac{y_t^2}{16 \pi^2}\left[\frac92 y_t^2-\frac94 g_2^2-\frac{17}{12}g_1^2-8g_3^2\right]\, ,
\end{eqnarray}
where  $g_1=\sqrt{5/3}g'$. Given the initial conditions for the coupled RGEs at $\mu=m_t$, the solution to these equations shows that $\lambda(\mu)$ becomes negative far before the GUT scale. 
Indeed, solutions to the SM RGEs have been used to constrain absolute stability at the one loop level before the discovery of the Higgs boson. However, an error $\Delta m_H\gtrsim16\%$ is not expected from experimental results, making it inaccurate to condition absolute stability on different values of the Higgs mass. Additionally, fixing $\lambda\gtrsim0$ and scanning for the minimum bound on $m_H$ shows the same energy scale where the instability occurs for $m_t\sim173$ GeV  \cite{PhysRevD.107.036018} .
%as seen in the right panel of Fig. \ref{fig:Higgsmasses}. 
When these two results are compiled within the current particle content of the Standard Model, either the chiral top quark must be lighter or the Higgs boson must be heavier in order to ensure SM vacuum stability.

Given these constraints, it becomes clear that within the SM alone, achieving absolute vacuum stability is problematic. The observed Higgs mass and top quark mass push the vacuum towards metastability, leaving no room for adjusting these parameters within experimental bounds to achieve a stable vacuum. This predicament underscores the need for new physics beyond the Standard Model to modify the behaviour of the Higgs potential at high energy scales.

While the most conventional approach to remedy this issue involves the introduction of new scalar fields, which can alter the running of the Higgs quartic coupling, this work takes a different path. Instead of relying on additional scalars, we explore the introduction of new fermionic degrees of freedom—specifically, non-chiral fermions—as a mechanism to stabilize the vacuum.

The motivation for considering new fermions stems from their ability to contribute to the RGE of $\lambda(\mu)$ in a manner that can counterbalance the destabilizing effects induced by the top quark. By carefully scanning the properties of these fermions, it is possible to achieve a scenario where the Higgs quartic coupling remains positive up to the Planck scale, thereby ensuring the stability of the Higgs potential without necessitating heavy modifications to the Standard Model’s scalar sector.

In the subsequent sections, we undertake a thorough exploration of the newly introduced fermions, focusing on their interaction dynamics, mass spectrum, and their influence on the renormalization group equations. We will demonstrate how these fermions can effectively stabilize the Higgs potential, providing a viable solution to the vacuum instability problem while preserving the SM scalar sector. 
%\begin{figure}[htbp]
%	\centering
%	\begin{subfigure}{.5\textwidth}\hspace{-1.5cm}
%		\includegraphics[height=2.2in]{Higgslambda.png}
%		\caption{}
%	\end{subfigure}\hspace{-0.3cm}
%	\begin{subfigure}{.5\textwidth}
%		\includegraphics[height=2.2in]{Higgstop.png}
%		\caption{}
%	\end{subfigure}
%  \caption{The quartic Higgs coupling as a function of $\mu$ for various values of $m_H$ (a), The minimum limits on $m_H$ as a function of $\mu$ for various values of $m_t$ (b).}
%  \label{fig:Higgsmasses}
%\end{figure}

\section{ A Model with Vector-like Leptons}
\label{sec:VLLmodel}
%%%%%%%%%%%%%%%%%%%%%%%%%%%%%%%%%%%%%%%%%%%%%%%%%%%%%%
%%%%%%%%%%%%%%%%%%%%%%%%%%%%%%%%%%%%%%%%%%%%%%%%%%%%%%
\subsection{Theoretical Framework}
\label{subsec:VLLtheory}
%%%%%%%%%%%%%%%%%%%%%%%%%%%%%%%%%%%%%%%%%%%%%%%%%%%%%%

Here we explore a simple extension of the SM, incorporation only vector-like leptons.
The selection of leptonic mixing via Yukawa interactions with the Higgs field constrains us to a finite set of anomaly-free and renormalizable SU(2) gauge representations and hypercharge assignments for the new fermionic states. We give the list of the VLL representations under $SU(2)_L \times U(1)_Y$ symmetry in Table \ref{tab:VLrepresentations}.
\begin{table}[htbp]
\caption{\label{tab:VLrepresentations} \small Representations of Vector-Like Leptons, with quantum numbers under $SU(2)_L \times U(1)_Y$.}
\centering
\renewcommand{\arraystretch}{1.3}
\setlength{\tabcolsep}{10pt}
\begin{tabular*}{\textwidth}{@{\extracolsep{\fill}}|c|c|c|c|c|c|c|}
\hline
\textbf{Name} & \textbf{${\cal S}_1$} & \textbf{${\cal S}_2$} & \textbf{${\cal D}_1$} & \textbf{${\cal D}_2$} & \textbf{${\cal T}_1$} & \textbf{${\cal T}_2$} \\ 
\hline\hline
\textbf{Type} & Singlet & Singlet & Doublet & Doublet & Triplet & Triplet \\
\hline
 & $L^0$ & $L^-$ & $\begin{pmatrix} L^0 \\ L^- \end{pmatrix}$ & $\begin{pmatrix} L^- \\ L^{--} \end{pmatrix}$ & $\begin{pmatrix} L^+ \\ L^0 \\ L^- \end{pmatrix}$ & $\begin{pmatrix} L^0 \\ L^- \\ L^{--} \end{pmatrix}$ \\
\hline
\textbf{SU(2)$_L$} & 1 & 1 & 2 & 2 & 3 & 3 \\
\hline
\textbf{Y} & 0 & -1 & -1/2 & -3/2 & 0 & -1 \\
\hline
\end{tabular*}
\end{table}

The renormalizable Lagrangian for these model, including the Yukawa interactions and Dirac mass terms of the weak multiplets contains the SM part, and additional interactions corresponding to the different interactions in Table \ref{tab:VLrepresentations}:
\begin{eqnarray}
\label{eq:lagtype1}
{\cal L}_{SM}&=& -y_{\nu} {\bar l}_L \Phi^c \nu_R -y_{\tau}{\bar l}_L \Phi \tau_R \nonumber \\
{\cal L}_{{\cal S}_1, {\cal S}_2}&=& -y_{L^{0}}{\bar l}_L \Phi^c S_{1_R} -y_{L^{-}}{\bar l}_L \Phi S_{2_R}-y_M ({\bar S}_{1_L} \Phi S_{1_R} +{\bar S}_{2_L} \Phi S_{2_R})-M_{L^0} {\bar S}_{1_L} S_{1_R}-M_{L^-} {\bar S}_{2_L} S_{2_R}, \nonumber \\
{\cal L}_{{\cal D}_1, {\cal D}_2}&=& -y_{L^-} {\bar D}_{1_L}\Phi \tau_{R} -y_{L^0}{\bar D}_{1_L} \Phi^c \nu_{R}-y_{L^-}{\bar D}_{2_L}\Phi^c \tau_R -y_M( {\bar D}_{1_L}\Phi D_{1_R} +y_{L^-}{\bar D}_{2_L} \Phi^c D_{2_R}) \nonumber \\
&-&M_{D_1} {\bar D}_{1_L} D_{1_R}-M_{D_2} {\bar D}_{2_L} D_{2_R}, \nonumber \\
{\cal L}_{{\cal T}_1, {\cal T}_2}&=& -y_{{\cal T}_1} {\bar l}_{L}\tau^a \Phi^c  {\cal T}^a_{1_R} -y_{{\cal T}_2}{\bar l}_{L} \tau^a \Phi {\cal T}^a_{2_R}-y_M ({\bar {\cal T} }_{1_L}\tau^a \Phi^c  {\cal T}^a_{1_R} +y_{L^-}{\bar {\cal T}}_{2_L} \tau^a \Phi {\cal T}^a_{2_R})-M_{L^+} {\bar  {\cal T}}_{1_L}  {\cal T}_{1_R}-M_{L^{--}} {\bar  {\cal T}}_{2_L}  {\cal T}_{2_R},\nonumber \\ 
\end{eqnarray}
where the triplet models can equivalently be written as irreducible $SU(2)$ representations
\begin{equation}
\tau^{a} {\cal T}^a_{1_R}= \begin{pmatrix}
\frac{L^{0}}{\sqrt{2}} & L^{+} \\
L^{-} & -\frac{L^{0}}{\sqrt{2}} 
\end{pmatrix}_{R}       \qquad   \tau^{a} {\cal T}_{2_R}=  \begin{pmatrix}
\frac{L^{-}}{\sqrt{2}}  & L^{0} \\
L^{--} & -\frac{L^{-}}{\sqrt{2}} 
\end{pmatrix}_{R}
\end{equation}

Here, $\Phi^c=i \sigma^2 \Phi^\star$,  $y_\tau$, $y_\nu$, $y_{L^0}$, $y_{L^-}$ and $y_{{\cal T}_{1,2}}$ are the Yukawa couplings of the Higgs field $\Phi$ to vector-like leptons and SM leptons\footnote{Although the SM does not inherently include a right-handed neutrino ($\nu_R$), our analysis considers such state as added to the SM.   This right-handed neutrino is required to preserve mixing between the VLL and the SM leptons through the Yukawa interactions present in our study. Although this state is included only in the SM and ${\cal D}_1$ part of Lagrangian in Eq. \ref{eq:lagtype1}, its contribution appears in mass matrices throughout all six representations whenever the neutral sector is considered.}, while $y_M$ is the Yukawa coupling of the Higgs scalar field to vector-like leptons only. We assume that only the third generation SM leptons mix with VLLs in order to avoid unwanted complications from flavour-changing neutral current (FCNC) and lepton flavour violating (LFV) decays. If the VLLs mix with leptons of all generations, such mixing induces  flavor transitions between the SM generations, which would lead to dangerous LFV processes. These  are tightly constrained by experimental data, especially from processes like $\mu\to e \gamma$ decay \cite{Chiappini_2023} and $\mu \to e$ conversion \cite{Pezzullo_2017}, which are highly sensitive to new flavor-changing interactions. By restricting the VLL mixing to only the third-generation leptons, the model avoids these stringent flavor constraints, as the third-generation leptons are less sensitive to flavor-changing processes  \cite{Blechman2010TheFP}. This is additionally motivated by the analysis of EWPOs and of renormalization group equations. Large mass splitting between the members of weak eigenstates in Eq. \ref{eq:gauge_eigenst_fields} can lead to adverse effects on initial conditions for the Yukawa couplings %in Eq. \ref{eq:IC}
in the low mass regime of the first and the second generation SM leptons. Similarly, the EWPOs are sensitively dependent on logarithmic mass splitting between the SM and the vector-like leptons \cite{Bizot:2015zaa} as $\Delta\mathbb{T}\sim\frac{M_W^2}{\alpha_e}(\frac{m_l^2}{M_{\text{VLL}}^2})$, inducing large discrepancies to the global fit of the oblique $\mathbb{T}$-parameter \cite{10.1093/ptep/ptac097} if the  SM leptons of the first two generations couple to VLL.
\\\\
The weak eigenstate lepton fields mix, for both chiralities in the neutral and charged sectors, and are respectively given as
\begin{eqnarray}
{\cal N}_{L,R}=&\left(\begin{matrix}
\nu \\L^0\end{matrix}\right)_{L,R}\ \qquad 
{\cal L}_{L,R}=&\left(\begin{matrix} \tau\\L^-\end{matrix}\right)_{L,R}\ %\nonumber \\
\,
\label{eq:gauge_eigenst_fields}
\end{eqnarray}
The mass eigenstate fields are denoted as $(n_1, n_2)$ and $(l_1, l_2)$ and they correspond to bi-unitary transformation of weak eigenstates,
\begin{eqnarray}
{\mathbf N}_{L,R}&=&\left(\begin{matrix} n_1 \\n_2\end{matrix}\right)_{L,R}=V_{L,R}^0 \left(\begin{matrix} \nu\\L^0\end{matrix}\right)_{L,R}\nonumber \\
{\mathbf L}_{L,R}&=&\left(\begin{matrix} l_1\\l_2\end{matrix}\right)_{L,R}=V_{L,R}^l \left(\begin{matrix} \tau\\L^-\end{matrix}\right)_{L,R}
\, ,
\label{eq:mass_eigenstates}
\end{eqnarray}
where the mixing matrices in neutral $(0)$ and charged sector $(l)$ follow as
\begin{equation}
V_{L,R}^{0}=\left(
\begin{matrix}
\cos\theta^u& -\sin\theta^u\\
\sin\theta^u &\cos\theta^u\end{matrix} 
\right)_{L,R}\, , \qquad
V_{L,R}^{l}=\left(
\begin{matrix}
\cos\theta^d & -\sin\theta^d\\
\sin\theta^d &\cos\theta^d \end{matrix}
\right)_{L,R}\, ,
\label{eq:rotation_matrices}
\end{equation}
By using these rotation operators we construct the diagonal mass matrices 
\begin{equation}
\label{eq:massmatrices}
M^u_{diag}=V_L^0 M^u (V_R^0)^\dagger=\left(\begin{matrix}m_{n_1} & 0 \\0 &m_{n_2}\end{matrix}\right)
\quad , \quad
M^d_{diag}=V_L^l M^d (V_R^l)^\dagger=\left(\begin{matrix}m_{l_1} & 0 \\0 &m_{l_2}\end{matrix}\right)\, .
\end{equation}
Utilizing the gauge eigenstate fields, the mass matrices for both the neutral and charged sectors are obtained following spontaneous symmetry breaking
\begin{eqnarray}
\label{eq:gaugetransform}
-{\cal L}^{u}_{Yuk}&=&\left(\begin{matrix} \nu_L & L^0_L \end{matrix}\right) \left(
\begin{matrix}
y_\nu\frac{v}{\sqrt{2}}& y_{L^0}\frac{v}{\sqrt{2}}\\
y_{L^0}\frac{v}{\sqrt{2}}&y_M\frac{v}{\sqrt{2}}+M_{L^0}\end{matrix} 
\right)\left(\begin{matrix} \nu_R \\ L^0_R\end{matrix}\right)\,, \nonumber \\
-{\cal L}^{d}_{Yuk}&=&\left(\begin{matrix} \tau_L & L^-_L \end{matrix}\right) \left(
\begin{matrix}
y_\tau\frac{v}{\sqrt{2}}& y_{L^-}\frac{v}{\sqrt{2}}\\
y_{L^-}\frac{v}{\sqrt{2}}&y_M\frac{v}{\sqrt{2}}+M_{L^-}\end{matrix} 
\right)\left(\begin{matrix} \tau_R \\ L^-_R\end{matrix}\right)
\end{eqnarray}
The Dirac mass term introduces another free parameter into VLL models and it appears as an uncoupled degree of parametric freedom in RGE level, so to this end, we absorb $M_{L^{0,-}}$ in Eq. \ref{eq:gaugetransform} into $y_M$ for all VLL representations. This is equivalent to assuming that the mass of the vector-like fermion are purely generated from the spontanous symmetry breaking, such that $m_{\text{Dirac}}=0$. The mass eigenvalues for charged partners in VLL model are
\begin{eqnarray}
 m_{l_1,l_2}^2=\frac{1}{4}\left [( y_\tau^2+y_{L^-}^2+y_M^2)v^2 \right] \left[ 1\pm \sqrt{1-\left( \frac{2y_\tau y_M }{ (y_\tau^2+y_{L^-}^2+y_M^2)}\right )^2} \right]
 \label{eq:eigenvecTt}
 \end{eqnarray}

The diagonalization of the mass matrices, as presented in Eq. \ref{eq:massmatrices}, facilitates the expression of the mixing angles for both the up and down sectors in terms of the model's free parameters\footnote{The mixing angle for the neutral sector can be derived by substituting $\theta^d$ with $\theta^u$ and $\tau,L^-$ with $\nu,L^0$.}.
\begin{eqnarray}
 \label{eq:mixing}
 \tan (2\theta^d_L)&=&\frac{2y_My_{L^-}}{ y_M^2-y_\tau^2-y_{L^-}^2}\nonumber \\
\tan (2\theta^d_R)&=& \frac{2y_\tau y_{L^-}}{y_M^2+y_\tau^2-y_{L^-}^2}\, ,
\end{eqnarray}
\\

The charge assignments of the exotic leptons prevent $L^{+}$ and $L^{--}$ fields from mixing with other fermions via Yukawa interactions. Consequently, these vector-like leptons are also mass eigenstates $m_{L^+}=M_{L^+}$ and $m_{L^{--}}=M_{L^{--}}$.
\\

Furthermore, the relationship between the mass eigenstates and the mixing angles of the SM leptons with VLLs in anomaly-free states generates distinct mass splitting relations:
\begin{eqnarray}
 \label{eq:mixingrelations}
{\rm For~ doublets:} 
&(L^0L^-): &m_{L^0}^2 (\cos \theta^u_R)^{2}+m_\nu^2 (\sin \theta^u_R)^{2}=m_{L^-}^2 (\cos \theta^d_R)^{2}+m_\tau^2 (\sin \theta^d_R)^{2}
\nonumber \\
&(L^-L^{--}):&m_{L^{--}}^2=m_{L^-}^2 (\cos \theta^d_R)^{2}+m_\tau^2 (\sin \theta^d_R)^{2}\nonumber \\
\nonumber \\
{\rm For ~triplets:}& (L^+L^0L^-): &m_{L^+}^2=m_{L^0}^2 (\cos \theta^u_L)^{2}+m_\nu^2 (\sin \theta^u_L)^{2}\nonumber \\
 && \phantom{m_{L^+}^2} =m_{L^-}^2 (\cos \theta^d_L)^{2} + m_\tau^2(\sin \theta^d_L)^{2} \, ,\nonumber \\
  &&{\rm where}~\sin(2\theta^d_L)= \sqrt{2}{m_{L^0}^2-m_\nu^2\over  (m_{L^0}^2-m_\tau^2)}\sin(2\theta^u_L)\, .\nonumber \\
  &(L^0L^-L^{--}):& m_{L^{--}}^2=m_{L^-} ^2 (\cos \theta^d_L)^2+m_\tau^2 (\sin \theta^d_L)^{2}\nonumber \\
  &&\phantom{m_{L^{--}}^2}=m_{L^0}^2 (\cos \theta^u_L)^{2}+m_\nu^2 (\sin \theta^u_L)^{2} \, , \nonumber\\
 &&{\rm where}~\sin(2\theta^d_L)= {m_{L^0}^2-m_\nu^2\over \sqrt{2} (m_{L^0}^2-m_\tau^2)}\sin(2\theta^u_L)\, , 
 \label{relations}
\end{eqnarray}
and where 
\begin{eqnarray}
\end{eqnarray}

\begin{eqnarray}
m_{L^0,L^-} (\tan \theta^{u,d}_R)&=m_{\nu,\tau}(\tan \theta^{u,d}_L)\qquad &{\hbox{for~singlets,~triplets}}\nonumber \\
m_{L^0,L^-} (\tan \theta^{u,d}_L)&=m_{\nu,\tau} (\tan \theta^{u,d}_R)\qquad &{\hbox{for~doublets}} \, .
\label{angles}
\end{eqnarray}
Furthermore, initial conditions for all Yukawa couplings are modified with mixing relations.
\begin{eqnarray}
\label{eq:IC}
y_\tau(\mu_0)&=&\frac{\sqrt{2} m_\tau}{v}\frac{1}{\sqrt{\cos^2 \theta_L+x_\tau^2 \sin^2 \theta_L}}\, , \nonumber \\
y_{L^0}(\mu_0)&=&\frac{\sqrt{2} m_{L^0}}{v}\frac{\sin \theta_L \cos \theta_L (1-x_\nu^2)}{\sqrt{\cos^2 \theta_L+x_\nu^2 \sin^2 \theta_L}}\, , \nonumber \\
y_{L^-}(\mu_0)&=&\frac{\sqrt{2} m_{L^-}}{v}\frac{\sin \theta_L \cos \theta_L (1-x_\tau^2)}{\sqrt{\cos^2 \theta_L+x_\tau^2 \sin^2 \theta_L}}\, , \nonumber \\
y_M(\mu_0)&=&\sum_{i=L^+,L^0,L^-,L^{--}}\frac{C_R m_i}{v}\sqrt{\cos^2 \theta_L+x_\nu^2 \sin^2 \theta_L}\, , \nonumber \\
\end{eqnarray}
where $C_R=(\sqrt{2},\frac{1}{\sqrt{2}},\frac{\sqrt{2}}{3}$) are the representation dependent weight factors, with $x_\tau=m_\tau/m_{L^-}$, and $x_\nu=m_\nu/m_{L^0}$. Given that the $L^{+}$ and $L^{--}$ fields do not mix with other fermions in the model, their low-energy Yukawa couplings remain unaffected by mixing relations. Nevertheless, their Yukawa coupling indirectly influences the coupled RGEs, as shown in the Appendix. \ref{sec:apprge}.
\\
%%%%%%%%%%%%%%%%%%%%%%%%%%%%%%%%%%%%%%%%%%%%%%%%%%%%%%%%
\subsection{Restrictions on Vector-like Lepton Masses}
\label{subsec:VLLlimits}
%%%%%%%%%%%%%%%%%%%%%%%%%%%%%%%%%%%%%%%
The CMS Collaboration has carried out three direct searches targeting extensions of the SM with VLLs in $pp$ collisions at $\sqrt{s}=13$ TeV collision data set. In the first of these searches, multilepton final states with electrons and muons were probed using a data set collected during 2016-2017, yielding the first direct constraints on doublet models with vector-like tau leptons in the mass range of 120–790 GeV \cite{CMS:2019hsm}. A second search, targeting both doublet and singlet vector-like tau lepton models and conducted with the larger full Run-2 data set, included additional multilepton final states, including hadronically decaying tau leptons, and superseded the first result \cite{CMS:2022nty}. Lastly, a third search performed by the CMS Collaboration probed non-minimal SM extensions involving VLLs and other BSM states in the context of the 4321 model in all-hadronic final states involving multiple jets and hadronically decaying tau leptons \cite{CMS:2022cpe}. Searches for vector-like leptons have also been performed at ATLAS, most recently for third-generation leptons in \cite{ATLAS:2023sbu}. A description of the expectation at all colliders is summarized in \cite{Sultansoy:2019xiw}, while  for an up-to-date review of searches for vector-like fermions at LHC see \cite{CMS:2024bni}. 

In what follows, in order to allow exploration of the largest parameter space and to remove any model-dependency, we will impose the weakest constraints, as restricted by Particle Data, requiring masses above 100 GeV \cite{L3:2001xsz,DELPHI:2003uqw,10.1093/ptep/ptac097}.

%%%%%%%%%%%%%%%%%%%%%%%%%%%%%%%%%%%%%%%%%%%%%%%%%%%%%%%%%%%%%%
\section{Electroweak Precision Observables}
\label{sec:electroweakprecision}

In addition to the constraints defined in the previous subsections, electroweak precision observables (EWPOs) are essential tools for probing the SM and constraining possible extensions. These observables arise from precise measurements of electroweak processes, such as the properties of the $W$ and $Z$ bosons, and provide stringent tests for any new physics scenarios. The addition of new particles, particularly scalars and leptons, impacts these observables through loop corrections to gauge boson masses, leading to potential signals of new physics. A significant aspect of these constraints is encapsulated in the oblique parameters, also known as the Peskin-Takeuchi parameters, namely $\mathbb{S}$, $\mathbb{T}$, and $\mathbb{U}$ \cite{Peskin:1991sw}. The latest fits for the oblique parameters give $\mathbb{S} = -0.04\pm0.10$ and $\mathbb{T} = 0.01 \pm 0.12$ at $90$\% CL \cite{10.1093/ptep/ptac097}. These parameters quantify the effects of new physics on the vacuum polarization corrections to the gauge bosons. They provide a model-independent way to parameterize deviations from the SM predictions, thus offering a systematic approach to compare different theoretical models. Previous analyses reproducing the oblique corrections for some vector-like representations have appeared in \cite{Cynolter:2008ea,Lavoura:1992np}. In what follows, we investigate the contributions to these parameters by the additional vector-like states in our scenarios.

\subsection{VLL contributions to the $\mathbb{S}$ and $\mathbb{T}$ parameters}
\label{subsec:VLLSTU}
\begin{figure}[htbp]
	\begin{center}
	\includegraphics[width=5.0in]{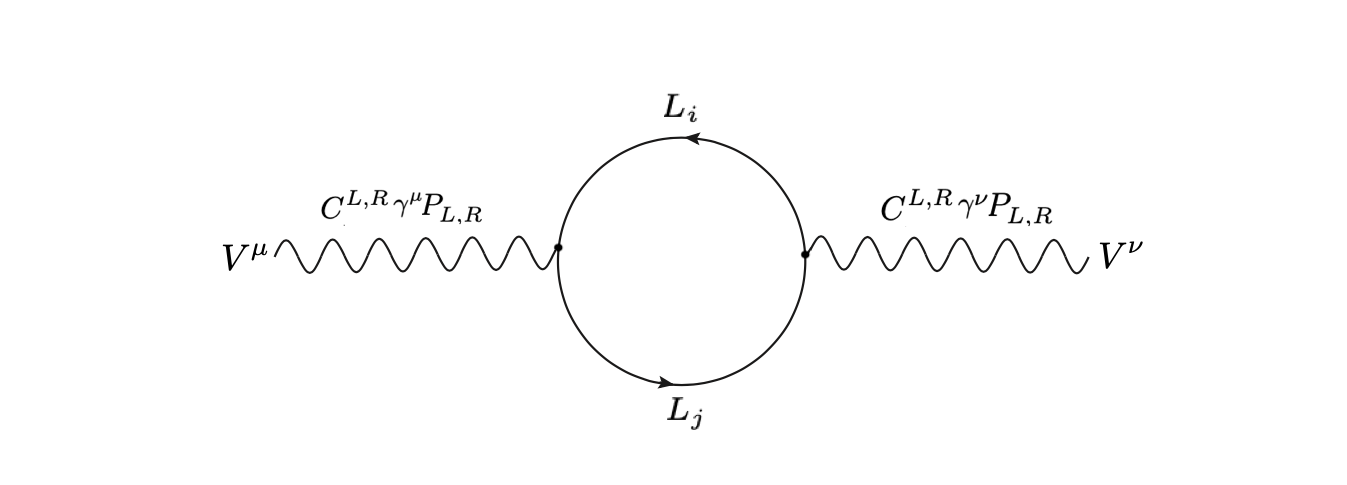}
	\end{center}
	\caption{Vector-like lepton contribution  to vacuum polarization amplitude of the SM gauge bosons. Here  $L_i$ and $L_j$ are the mass eigenstates.}
	\label{fig:vacuumpol}
\end{figure}

In scenarios with additional fermions, such as VLLs, the oblique parameters require special consideration. These fermions can alter the gauge boson self-energies, leading to unique patterns in the oblique corrections. VLLs might transform under certain symmetries, such as  $Z_2$, and may not interact with the SM Higgs boson, allowing their contributions to the oblique parameters to be isolated and studied separately. In VLL extensions, the physical states $L$ and $N$ contribute to the transverse component of the vacuum polarization for the gauge bosons in the SM through Feynman loop  diagrams. To estimate their contributions to $\mathbb{S}$ and $\mathbb{T}$, one needs to compute the one loop diagrams that contribute to the electroweak gauge boson vacuum polarization amplitudes, as shown in Fig. \ref{fig:vacuumpol}.

Adopting the general expression for the $\mathbb{T}$ and $\mathbb{S}$ parameter contributions from additional fermions \cite{garg2013vectorlikeleptonsextended}
\begin{eqnarray}
\mathbb{T}_{VLL}&=&\frac{1}{\alpha_{e}}\left[ \frac{\Pi_{WW}^{VLL}(0)}{M_W^2}-\frac{\Pi_{ZZ}^{VLL}(0)}{M_Z^2}\right], \nonumber \\
\mathbb{S}_{VLL}&=&\frac{4s_W^2c_W^2}{\alpha_eM_Z^2}\left[\Pi_{ZZ}^{VLL}(M_Z^2)-\Pi_{ZZ}^{VLL}(0)-\Pi_{\gamma\gamma}^{VLL}(M_Z^2)-\frac{c_W^2-s_W^2}{c_Ws_W}\Pi_{\gamma Z}^{VLL}(M_Z^2)\right] ,
\end{eqnarray}
 we initially calculate their gauge couplings with the vector bosons. The VLL mass matrices are given in Eq. \ref{eq:mass_eigenstates}, and the components of the diagonalizing matrices  in Eq. \ref{eq:rotation_matrices}. The couplings to $W$-boson and $Z$-boson are been modified by the VLLs through their mixing with SM leptons in the relevant Lagrangian
\begin{eqnarray}
\label{eq:ewlag}
{\cal L}_{W}&=&\frac{g}{\sqrt{2}}\bar{L_0}\gamma^{\mu}(C^L_{L_{0}L_{i}}P_L+C^R_{L_{0}L_{i}}P_R)L_iW_{\mu}^{+}+h.c.\,,\nonumber \\
{\cal L}_{Z}&=&\frac{g}{2c_W}\bar{L_i}\gamma^{\mu}(Y_ZN^L_{L_{i}L_{j}}P_L+Y_ZN^R_{L_{i}L_{j}}P_R)L_jZ_{\mu}\, ,
\end{eqnarray}
where $L_{i,j}$ are any type of leptons in our  electroweak Lagrangian and $Y_Z=T^3-\mathbb{Q}_is_W^2$. The condition $|\mathbb{Q}_i-\mathbb{Q}_j|=1$ holds for all forms of  $W-L_{i}-L_{j}$ interactions. 
\\\\

We further define VLL modified electroweak couplings to $Z$ and $W$ bosons in terms of the weak hypercharge operator and mixing identities
\begin{eqnarray}
\Omega^{L,R}_{WL_{0}L_{i}}= \frac{g}{\sqrt{2}}C^{L,R}_{L_{0}L_{i}} \qquad,\qquad   \Omega^{L,R}_{ZL_{i}L_{j}}= \frac{g}{2c_W}(T^3-\mathbb{Q}s_W^2)N^{L,R}_{L_{i}L_{j}}\,,
\end{eqnarray}
yields the final form of modified electroweak interactions
\footnote{For neutral leptons, operator $Y_Z$ generates a term proportional to $(-\frac{1}{2}\bar{L}_0\gamma^{\mu}L^0)Z_{\mu}$, where the coefficients of these neutral currents are absorbed in $\Omega^{L,R}_{ZL_{i}L_{j}}$ throughout all VLL representations.} 
\begin{eqnarray}
{\cal L}_{W}&\supset&\gamma^{\mu}(\Omega^L_{WL_{0}L_{i}}\mathbb{L}+\Omega^R_{WL_{0}L_{i}}\mathbb{R})W_{\mu}^{+}\,,\nonumber \\
{\cal L}_{Z}&\supset&\gamma^{\mu}(\Omega^L_{ZL_{i}L_{j}}\mathbb{L}+\Omega^R_{ZL_{i}L_{j}}\mathbb{R})Z_{\mu}\,.
\end{eqnarray}
We used {\tt LoopTools} \cite{Hahn:1999mt} to extract gauge boson self-energies. Additionally, we implemented the analytical expressions of Passarino-Veltman (PV) functions in {\tt FeynCalc} \cite{Shtabovenko:2020gxv} to generate to self energy amplitudes in the final expressions
\begin{eqnarray}
\label{eq:obliqueTvll}
\mathbb{T}_{VLL}&=&\frac{1}{\alpha_e}\left[\frac{2s_W}{c_WM_Z^2}\sum_{i}\mathcal{F}_{Z\gamma}(\Omega^L_{ZL_{i}L_{i}},\Omega^R_{ZL_{i}L_{i}},\mathbb{Q}_i,m_i^2,p^2=0)\right. \nonumber \\
&+&\left.\frac{-2}{M_Z^2}\sum_{i\neq j}\delta(\mathbb{Q}_i-\mathbb{Q}_j)\mathcal{F}_{ZZ}(\Omega^L_{ZL_{i}L_{j}},\Omega^R_{ZL_{i}L_{j}},m_i^2,m_j^2,p^2=0)\right. \nonumber \\
&+&\left.\frac{1}{M_Z^2}\sum_{i}\mathcal{F}_{ZZ}(\Omega^L_{ZL_{i}L_{i}},\Omega^R_{ZL_{i}L_{i}},m_i^2,m_i^2,p^2=0)\right. \nonumber \\
&+&\left.\frac{1}{M_W^2}\sum_{i\neq j}\delta(\mathbb{Q}_i-\mathbb{Q}_j)\mathcal{F}_{WW}(\Omega^L_{WL_{i}L_{j}},\Omega^R_{WL_{i}L_{i}},m_i^2,m_j^2,p^2=0) \right]\, ,
\end{eqnarray}
\begin{eqnarray}
\label{eq:obliqueSvll}
\mathbb{S}_{VLL}&=&\frac{4s_W^2c_W^2}{\alpha_e}\left[(\frac{c_W^2-s_W^2}{s_Wc_WM_Z^2})\left(\sum_{i}\mathcal{F}_{Z\gamma}(\Omega^L_{ZL_{i}L_{i}},\Omega^R_{ZL_{i}L_{i}},\mathbb{Q}_i,m_i^2,M_Z^2)+\sum_{i}\mathcal{F}_{Z\gamma}(\Omega^L_{ZL_{i}L_{i}},\Omega^R_{ZL_{i}L_{i}},\mathbb{Q}_i,m_i^2,0)\right)\right.\nonumber \\
&-&\left.\frac{1}{M_Z^2}\sum_{i}\mathcal{F}_{\gamma\gamma}(\mathbb{Q}_i,\mathbb{Q}_i,m_i^2,m_i^2,M_Z^2)+\frac{2}{M_Z^2}\sum_{i\neq j}\delta(\mathbb{Q}_i-\mathbb{Q}_j)\mathcal{F}_{ZZ}(\Omega^L_{ZL_{i}L_{j}},\Omega^R_{ZL_{i}L_{j}},m_i^2,m_j^2,M_Z^2)\right. \nonumber \\
&-&\left.\frac{2}{M_Z^2}\sum_{i\neq j}\delta(\mathbb{Q}_i-\mathbb{Q}_j)\mathcal{F}_{ZZ}(\Omega^L_{ZL_{i}L_{j}},\Omega^R_{ZL_{i}L_{j}},m_i^2,m_j^2,0)+\frac{1}{M_Z^2}\sum_{i}\mathcal{F}_{ZZ}(\Omega^L_{ZL_{i}L_{i}},\Omega^R_{ZL_{i}L_{i}},m_i^2,m_i^2,M_Z^2)\right.\nonumber \\
&-&\left.\frac{1}{M_Z^2}\sum_{i}\mathcal{F}_{ZZ}(\Omega^L_{ZL_{i}L_{i}},\Omega^R_{ZL_{i}L_{i}},m_i^2,m_i^2,0) \right]\, ,
\end{eqnarray}
where the fermion functions $\mathcal{F}_{VV,Z\gamma}$ contributing to the gauge boson two-point functions are calculated as
\begin{eqnarray}
\mathcal{F}_{Z\gamma}(\Omega_1,\Omega_2,\mathbb{Q},m^2,p^2)&=&\frac{1}{8\pi^2}[\mathbb{Q}(\Omega_1+\Omega_2)\left(2B_{00}(p^2,m^2,m^2)-p^2B_1(p^2,m^2,m^2)-A_0(m^2) \right)]\, , \nonumber \\
\mathcal{F}_{VV}(\Omega_1,\Omega_2,m_1^2,m_2^2,p^2)&=&\frac{1}{8\pi^2}\left[\left((\Omega_1^2+\Omega_2^2)m_1^2-2\Omega_1\Omega_2m_1m_2 \right)B_0(p^2,m_1^2,m_2^2)\right. \nonumber \\
&+&\left.(\Omega_1^2+\Omega_2^2)\left(p^2B_1(p^2,m_1^2,m_2^2)-2B_{00}(p^2,m_1^2,m_2^2)+A_0(m_2^2) \right) \right]\, .
\end{eqnarray}
The relevant electroweak couplings $\Omega^{L,R}_{VL_{i}L_{j}}$ are unique to each representation and  given in the Appendix \ref{sec:EWcouplings}. And finally, the largest contribution to $\mathbb{T}$ and $\mathbb{S}$ parameters in the SM  due $t$ and $b$ quarks are
\begin{equation}
\mathbb{T}^{SM}_{f}=\frac{3m_t^2}{4\pi e^2v^2}\qquad, \qquad \mathbb{S}^{SM}_{f}=\frac{1}{2\pi}\left(1-\frac{1}{3}\ln(\frac{m_t^2}{m_b^2}) \right).
\end{equation}
Subtracting these values from Eq. \ref{eq:obliqueTvll} and \ref{eq:obliqueSvll}, we scan the oblique parameters with respect to neutral and charged vector-like leptons.

\begin{figure}[htbp]
	\centering
	\begin{subfigure}{.5\textwidth}\hspace{-1.5cm}
		\includegraphics[height=2.0in]{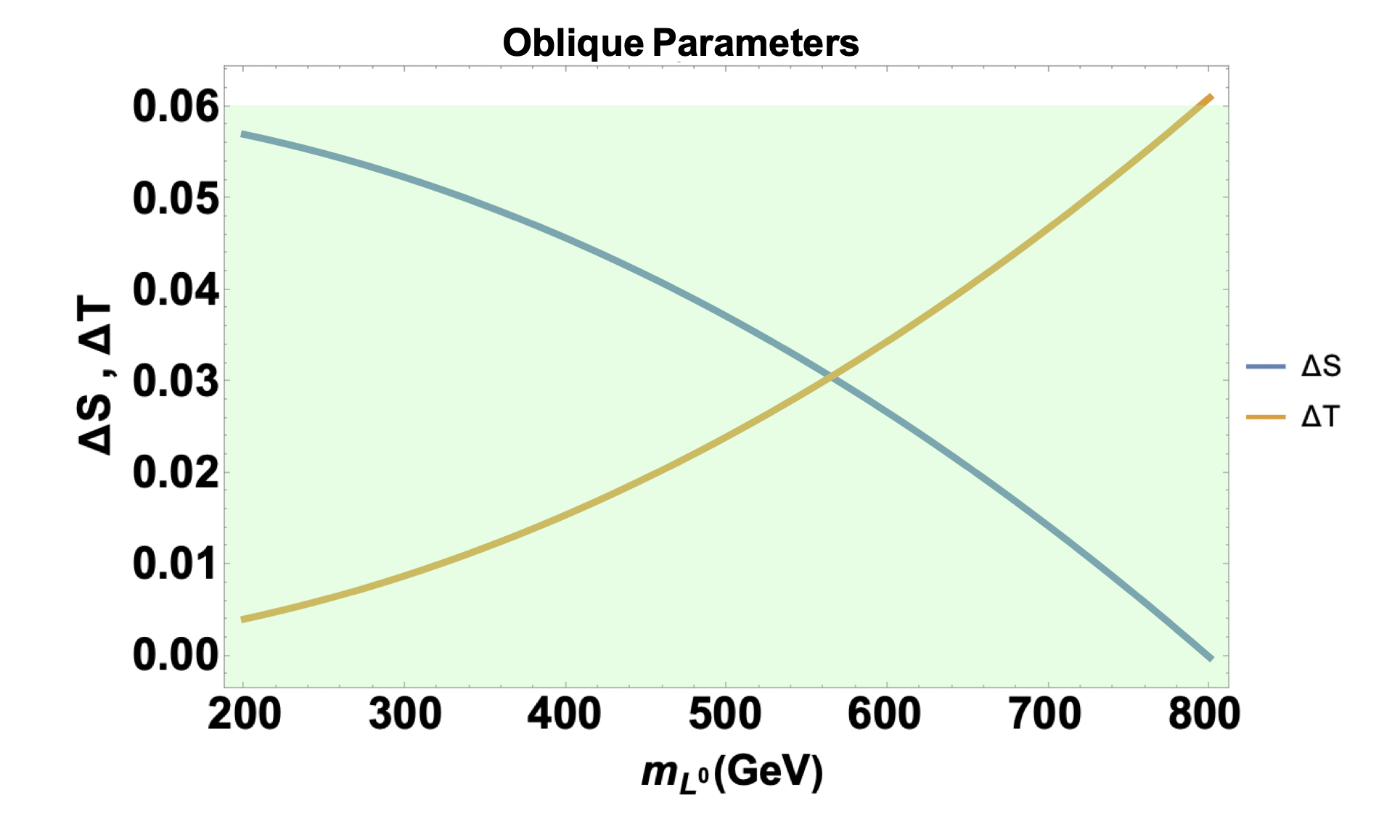}
		\caption{$\mathcal{S}_{1}$}
	\end{subfigure}\hspace{-1.3cm}
	\begin{subfigure}{.5\textwidth}
		\includegraphics[height=2.0in]{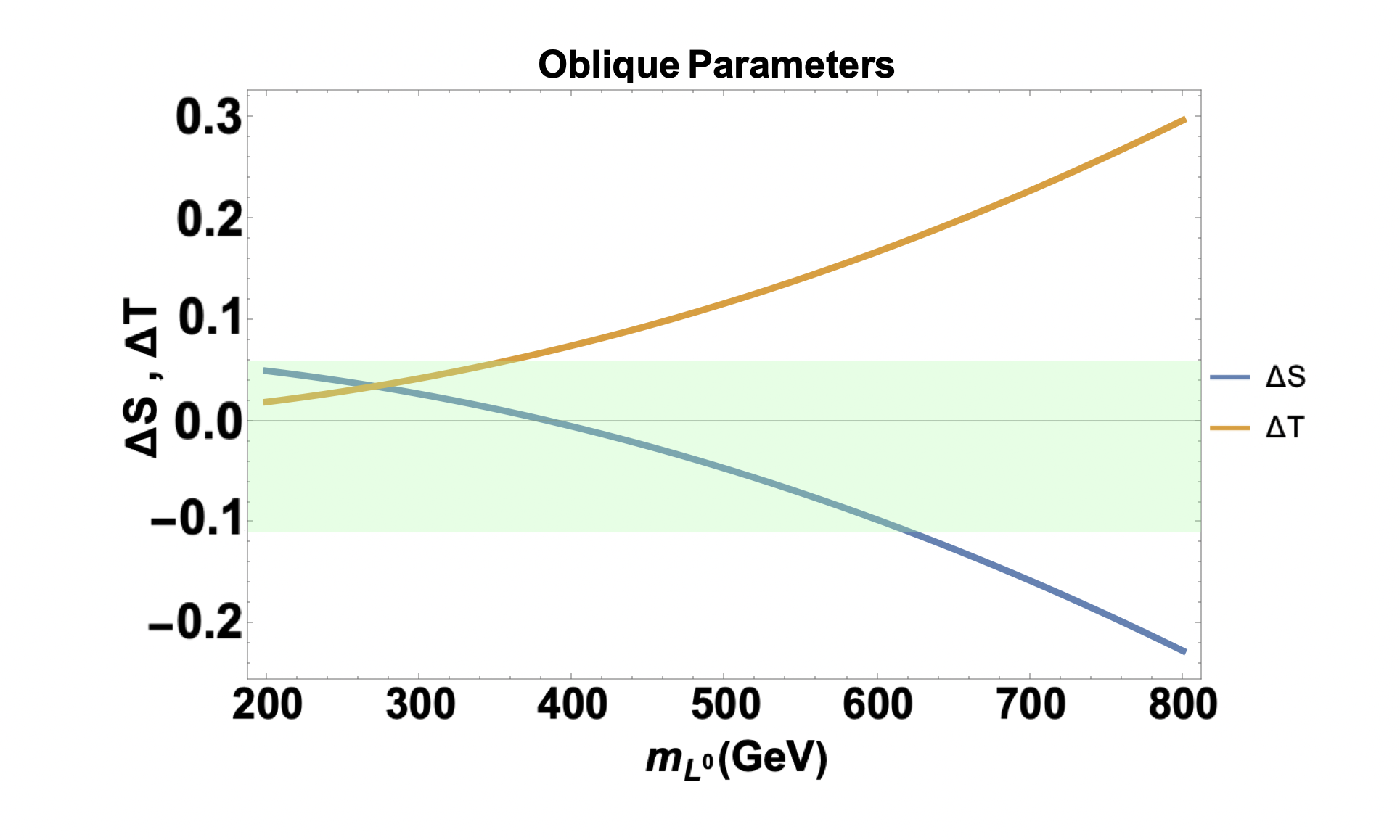}
		\caption{$\mathcal{S}_{1}$}
	\end{subfigure}\\
\begin{subfigure}{.5\textwidth}\hspace{-1.5cm}
		\includegraphics[height=2.0in]{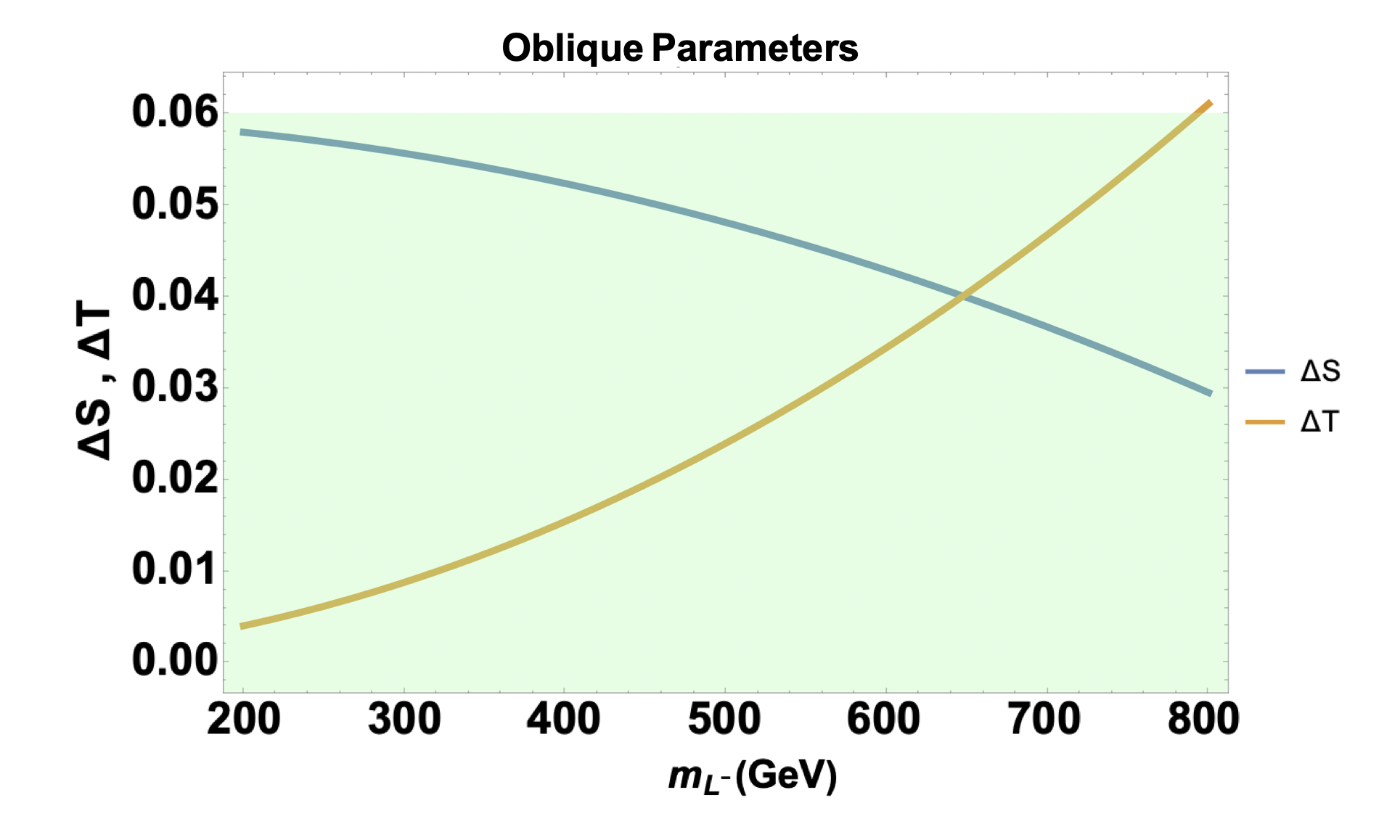}
		\caption{$\mathcal{S}_{2}$}
	\end{subfigure}\hspace{-1.1cm}
	\begin{subfigure}{.5\textwidth}
		\includegraphics[height=2.0in]{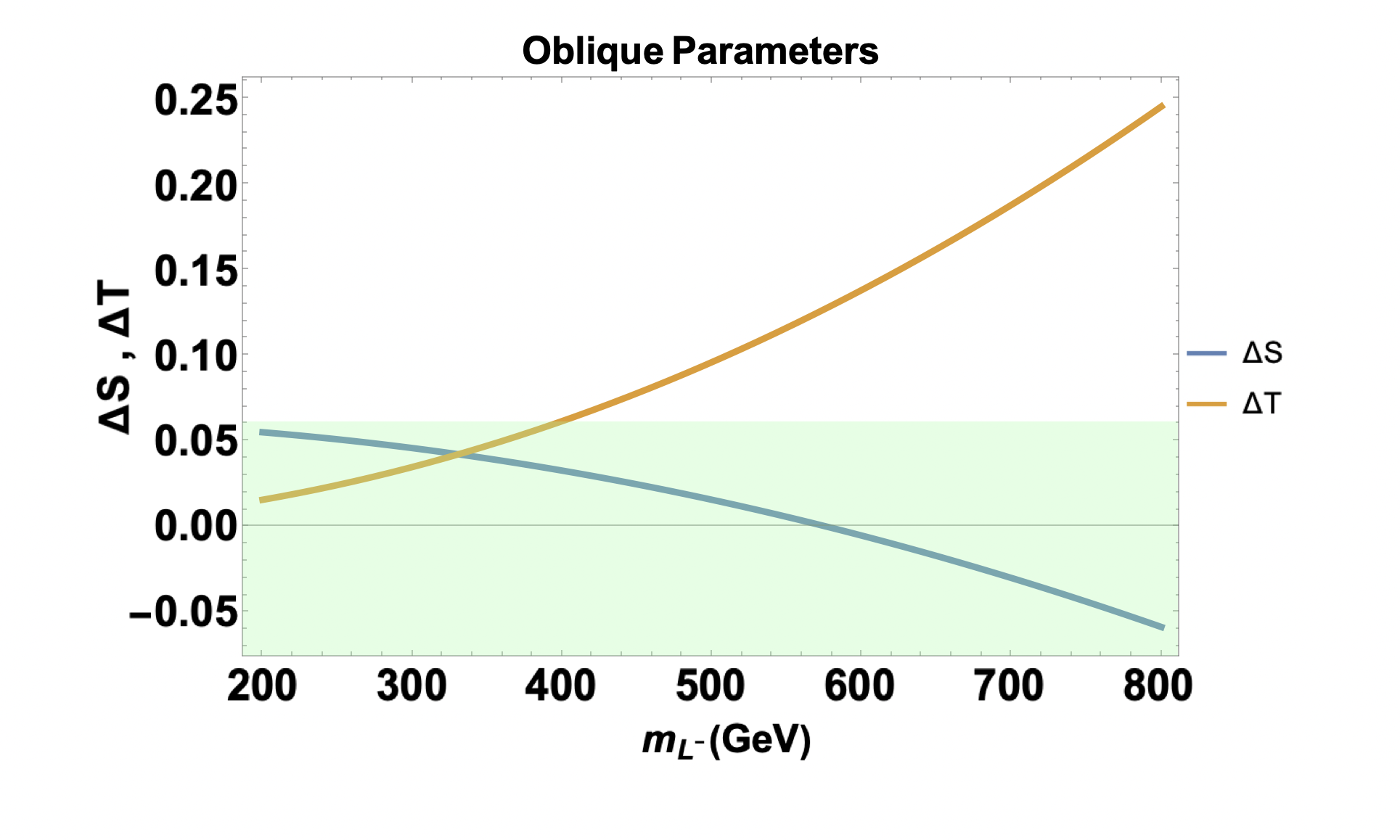}
		\caption{$\mathcal{S}_{2}$}
	\end{subfigure}	
  \caption{New physics contributions to the oblique parameters: $\mathbb{T}$ (orange) and $\mathbb{S}$ (blue) from singlet vector-like lepton representations for different VLL-SM lepton mixing $\sin \theta_L=0.05$ (left) and $\sin \theta_L=0.1$ (right). The green shaded region is the allowed space from the $\mathbb{S}$ and the $\mathbb{T}$ parameters in $2\sigma$ level.}
  \label{fig:VLLsingletoblique}
\end{figure}

In Fig. \ref{fig:VLLsingletoblique}, we illustrate the dependence of the oblique parameters on the mass of VLLs for two specific mixing angles. For a near-decoupling limit, $\sin\theta=0.05$, neither of the oblique parameters imposes a constraint on VLL masses. However, increasing the mixing value to $\sin\theta\rightarrow 0.1$, the $\mathbb{S}$ parameter starts to disfavor $m_{L^0}>520$ GeV in the neutral singlet model $\mathcal{S}_{1}$, whereas the entire spectrum of $m_{L^-}$ is allowed by the $\mathbb{S}$ parameter in the $\mathcal{S}_{2}$ model. This is well-motivated, as the $\mathbb{S}$ parameter is influenced by the hypercharge of the new leptons, fundamentally measuring the difference in the running of the electroweak gauge couplings. In contrast, the $\mathbb{T}$ parameter is more restrictive for both singlet VLLs at larger mixing scales. While $m_{L^0}>540$ GeV falls outside $2\sigma$ region for the $\mathbb{T}$ parameter, the upper bound for charged VLL extends to $m_{L^-}\sim590$ GeV. The $\mathbb{T}$ parameter is more sensitive to weak isospin breaking and to the mass splitting between components of weak isospin multiplets. Due to the large mass splitting between members of neutral mass eigenstates as compared to charged mass eigenstates of $\mathcal{S}_{2}$, the $\mathbb{T}$ parameter imposes more constraint on $\mathcal{S}_{1}$. As expected, having the least number of possible electroweak couplings, singlet VLLs recover the SM limit for $\mathbb{T}\rightarrow 0$ as $\sin\theta \rightarrow 0$ because most terms in the $\mathbb{T}$ parameter are modified by weak isospin breaking. 

\begin{figure}[htbp]
	\centering
	\begin{subfigure}{.5\textwidth}\hspace{-1.5cm}
		\includegraphics[height=2.05in]{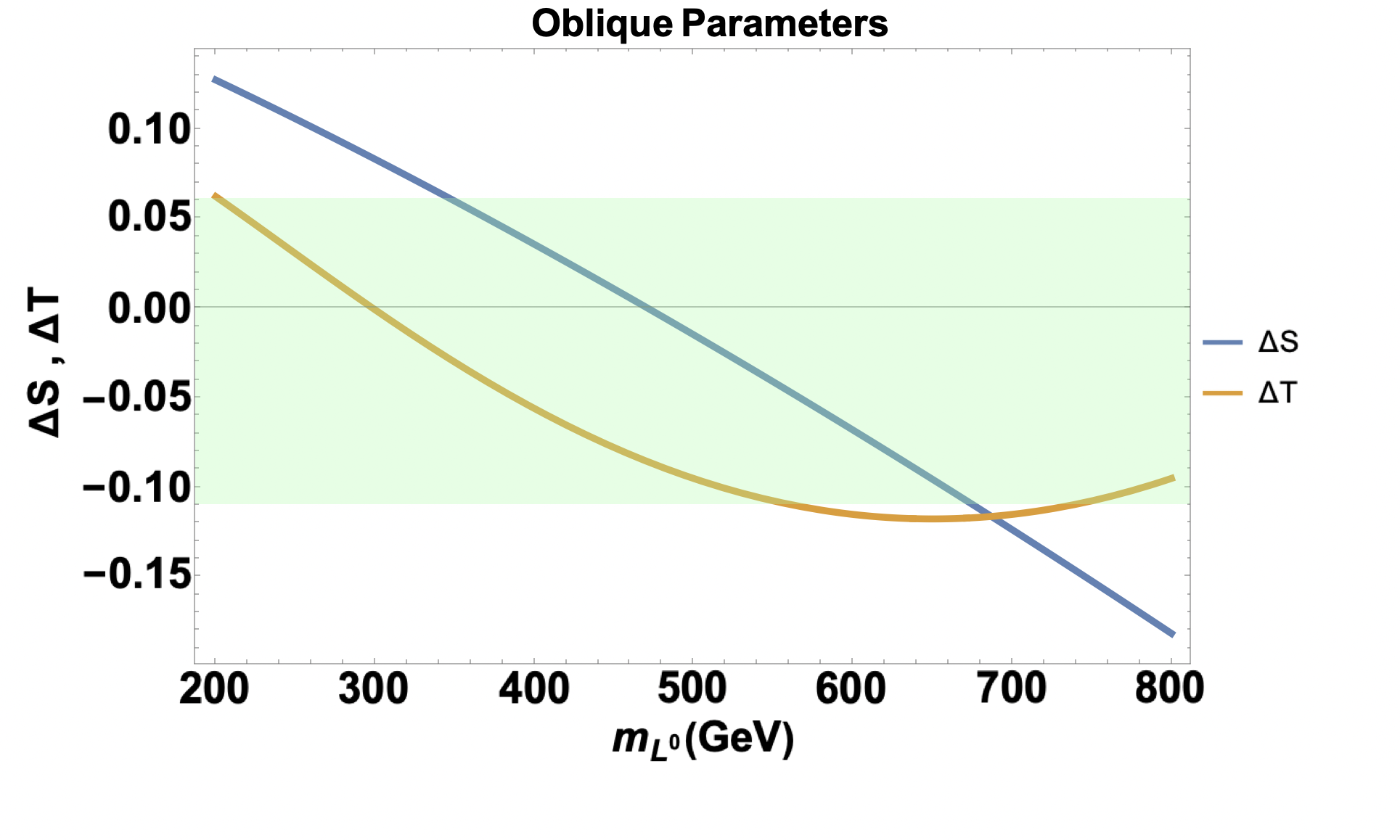}
		\caption{$\mathcal{D}_{1}$}
	\end{subfigure}\hspace{-1.3cm}
	\begin{subfigure}{.5\textwidth}
		\includegraphics[height=2.05in]{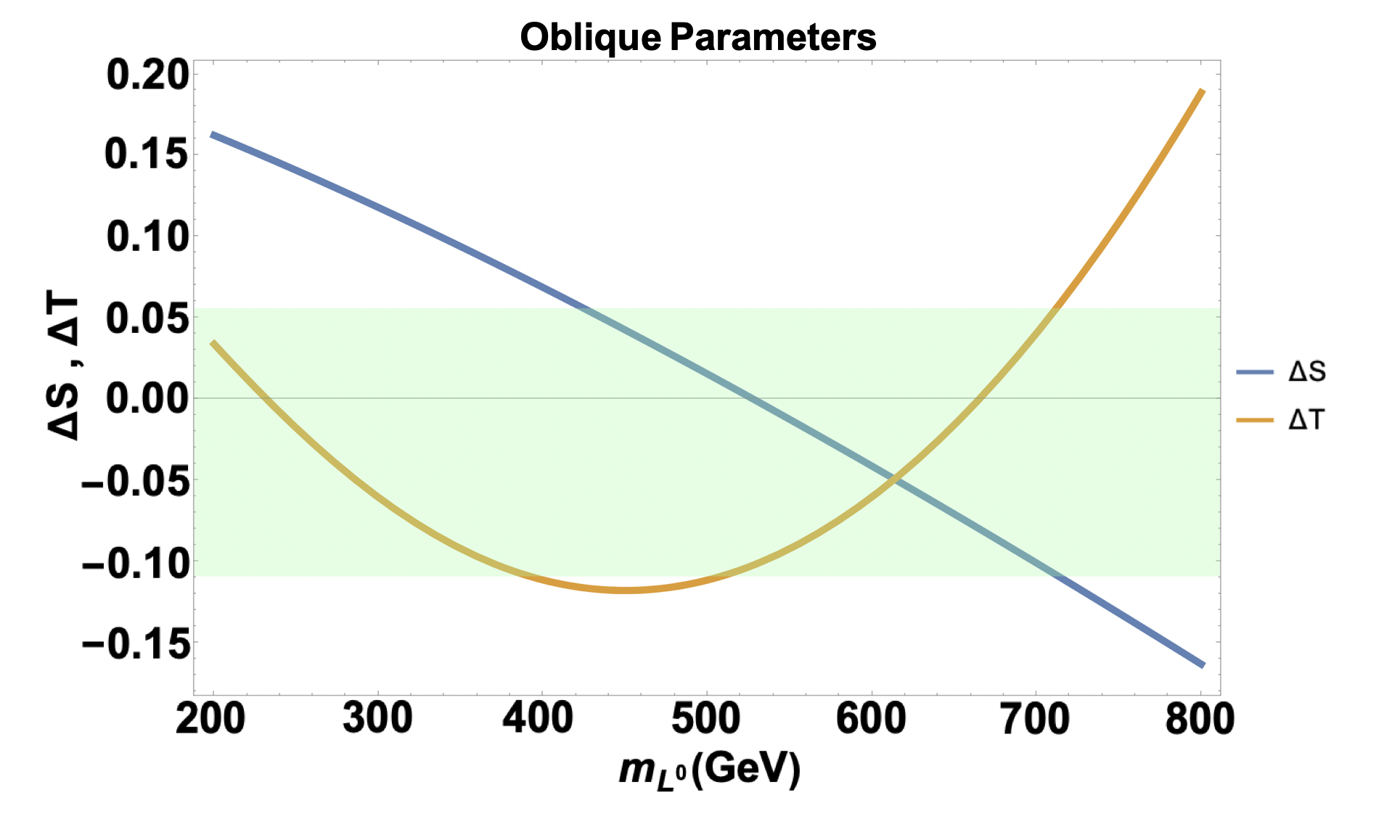}
		\caption{$\mathcal{D}_{1}$}
	\end{subfigure}\\
\begin{subfigure}{.5\textwidth}\hspace{-1.5cm}
		\includegraphics[height=2.0in]{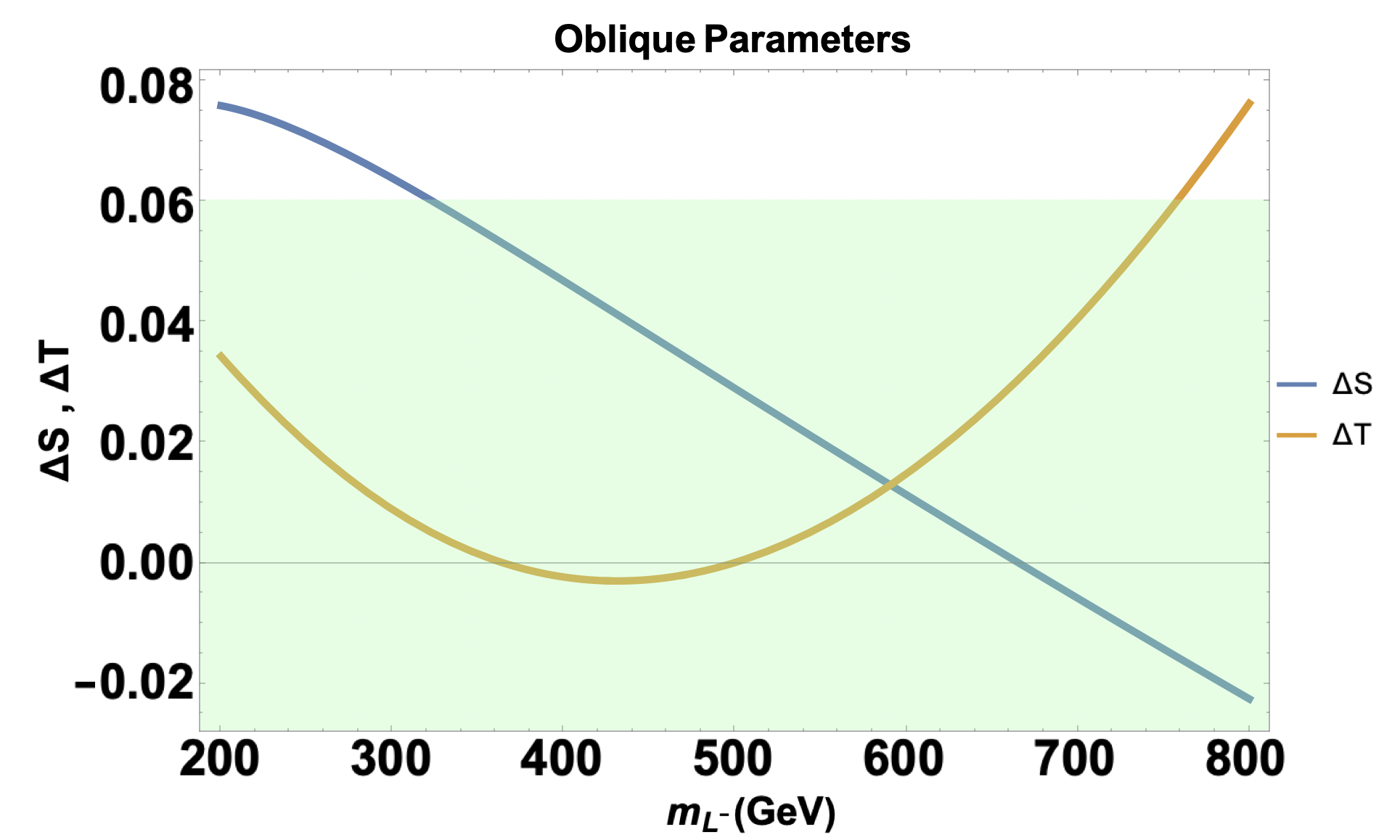}
		\caption{$\mathcal{D}_{2}$}
	\end{subfigure}\hspace{-1.1cm}
	\begin{subfigure}{.5\textwidth}
		\includegraphics[height=2.0in]{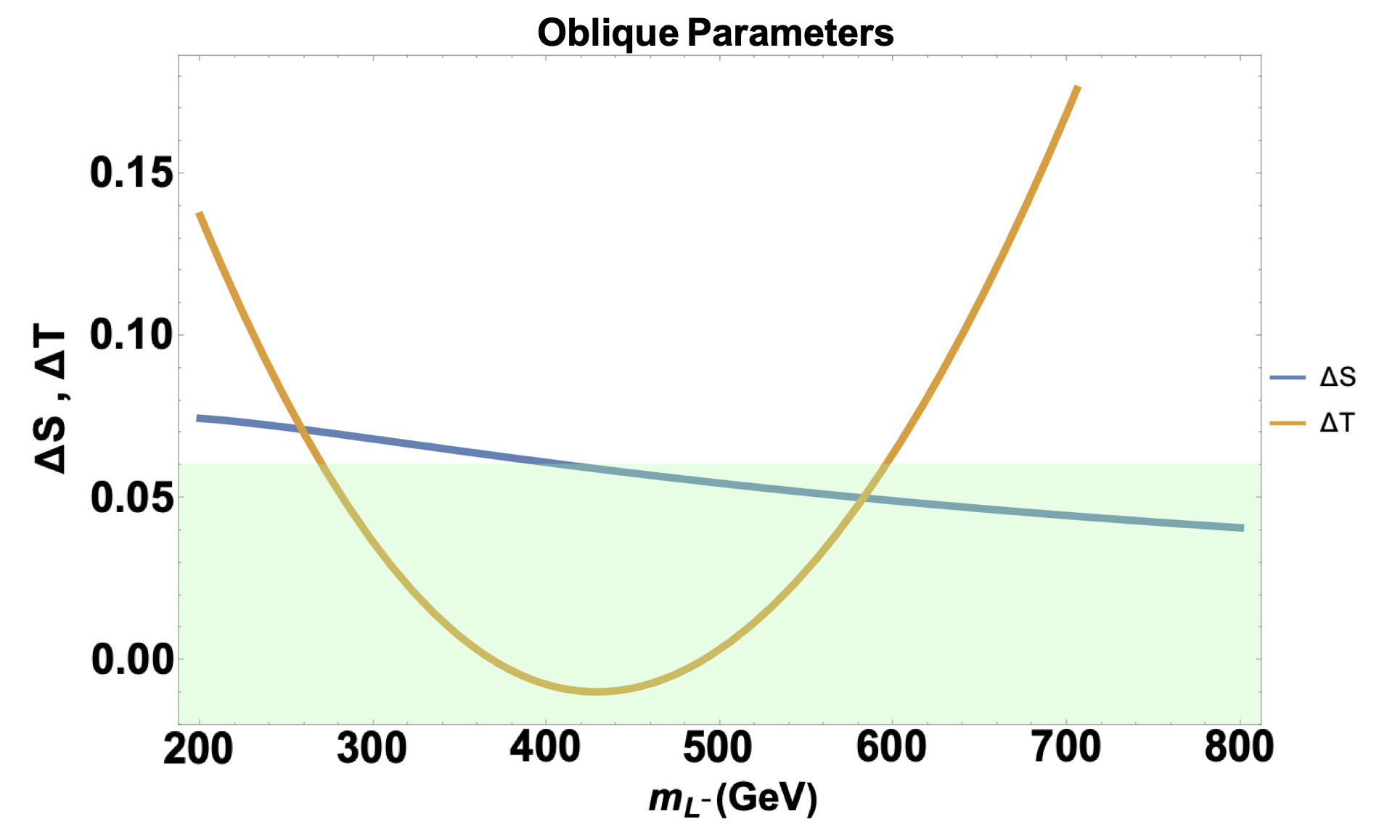}
		\caption{$\mathcal{D}_{2}$}
	\end{subfigure}	
  \caption{New physics contributions to the oblique parameters: $\mathbb{T}$ (orange) and $\mathbb{S}$ (blue) from doublet vector-like lepton representations for different VLL-SM lepton mixing $\sin \theta_L=0.05$ (left) and $\sin \theta_L=0.1$ (right).The green shaded region is the allowed space from the $\mathbb{S}$ and the $\mathbb{T}$ parameters in $2\sigma$ level. }
  \label{fig:VLLdoubletoblique}
\end{figure}

As shown in Fig. \ref{fig:VLLdoubletoblique}, the parameter space of $\mathcal{D}_{2}$ receives almost no constraint in the vicinity of the decoupling limit. The entire spectrum of $m_{L^-}$ scanned is allowed by both oblique parameters in the $2\sigma$ region for $\sin\theta=0.05$. However, the $\mathbb{S}$ parameter excludes  $m_{L^0}>730$ GeV in $\mathcal{D}_{1}$ model. The distinction between doublet models from the $\mathbb{S}$ parameter  is sharper than that for singlet models. With six different weak hypercharge choices, the $\mathbb{S}$ parameter constraints are significant, due to the extended number of lepton-modified gauge boson propagators. Increasing the lepton mixing to $\sin\theta=0.1$, the $\mathbb{T}$ parameter becomes restrictive for both doublet models. The mass of neutral VLL in $\mathcal{D}_{1}$ falls off the global fit of the oblique parameters for $m_{L^0}>760$ GeV, whereas the charged VLL mass of $\mathcal{D}_{2}$ model must be $m_{L^-}<670$ GeV to remain in $2\sigma$ region for the $\mathbb{T}$ parameter. In contrast, the $\mathbb{S}$ parameter imposes a minimum bound on $\mathcal{D}_{2}$ of $m_{L^-}> 400$ GeV.  
\begin{figure}[htbp]
	\centering
	\begin{subfigure}{.5\textwidth}\hspace{-1.5cm}
		\includegraphics[height=2.0in]{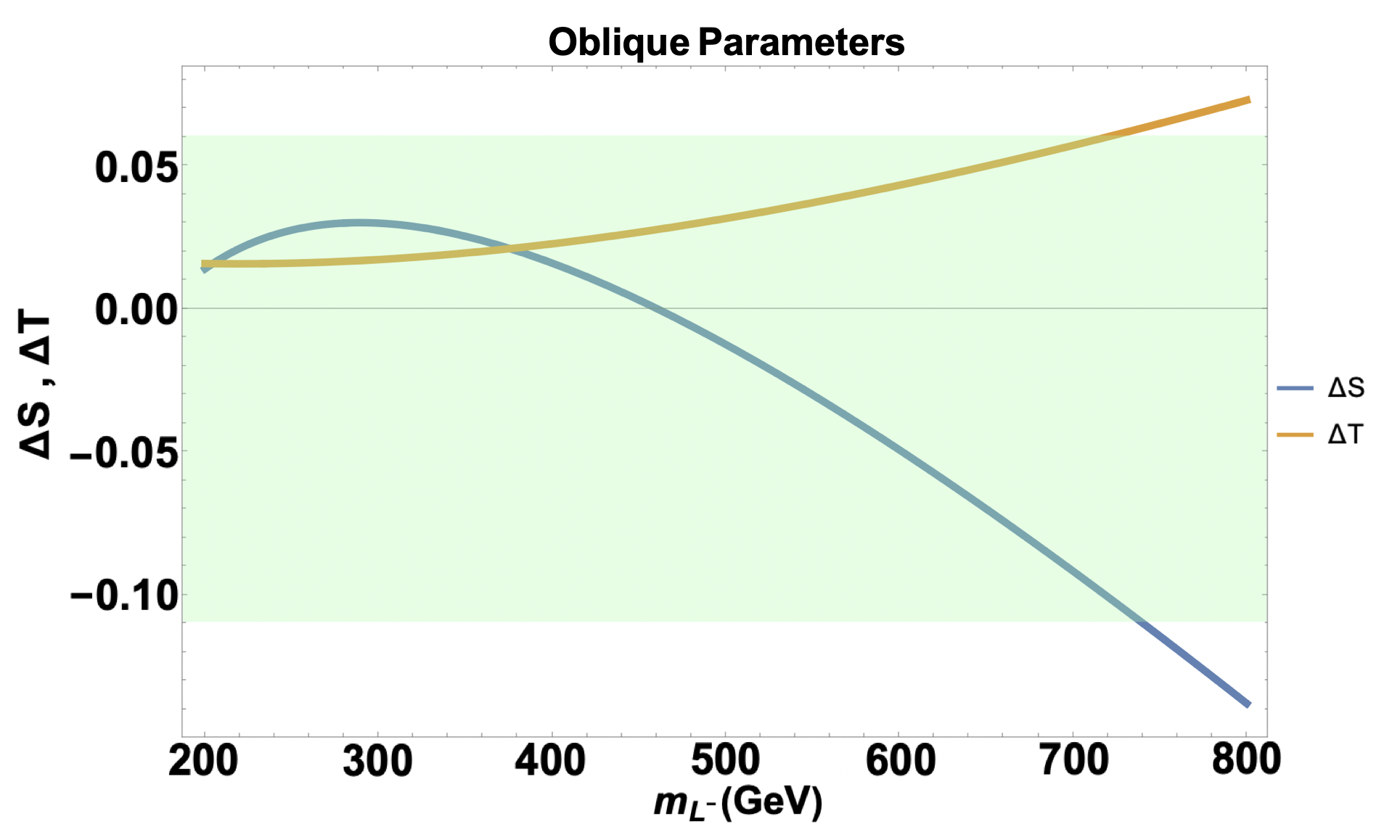}
		\caption{$\mathcal{T}_{1}$}
	\end{subfigure}\hspace{-1.2cm}
	\begin{subfigure}{.5\textwidth}
		\includegraphics[height=2.0in]{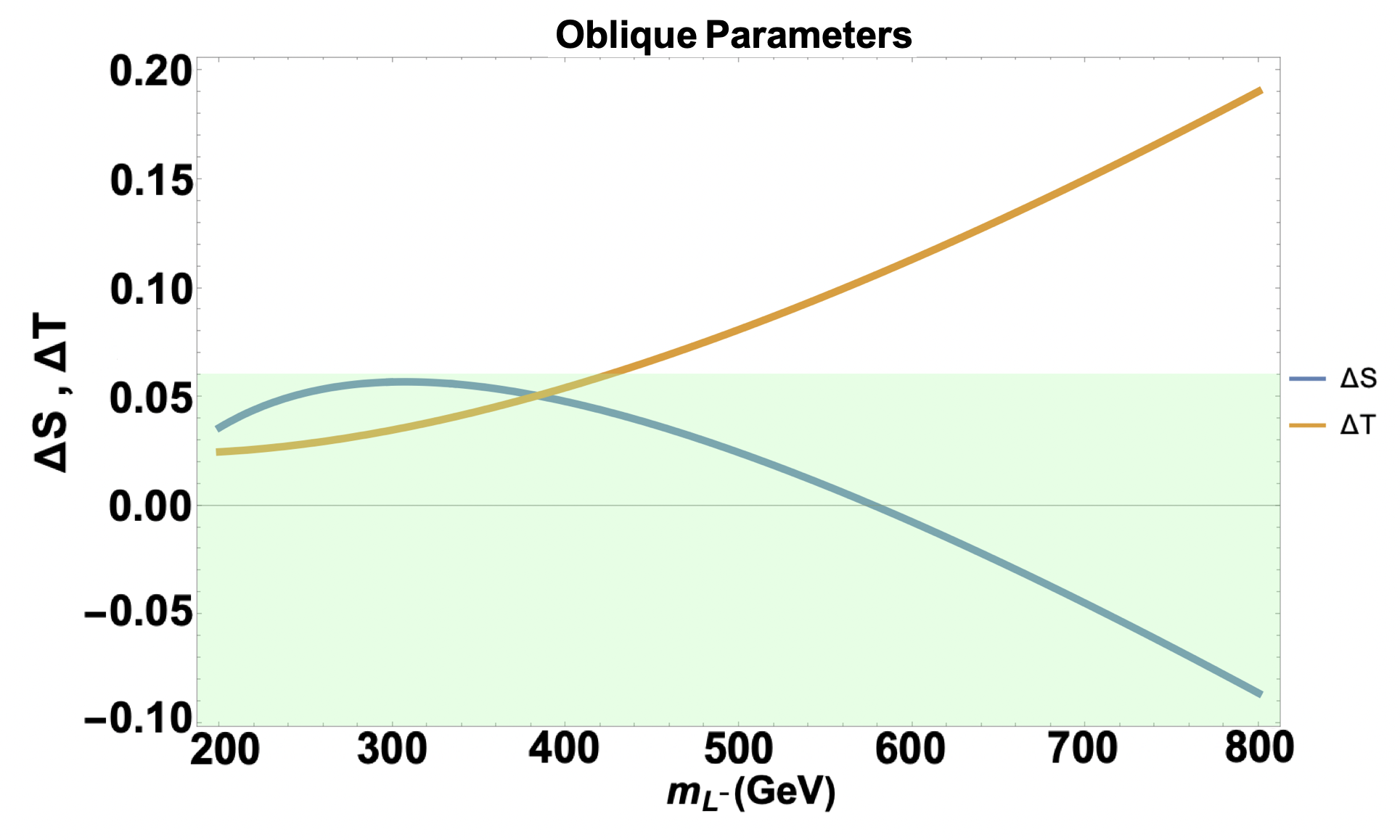}
		\caption{$\mathcal{T}_{1}$}
	\end{subfigure}\\
\begin{subfigure}{.5\textwidth}\hspace{-1.5cm}
		\includegraphics[height=2.0in]{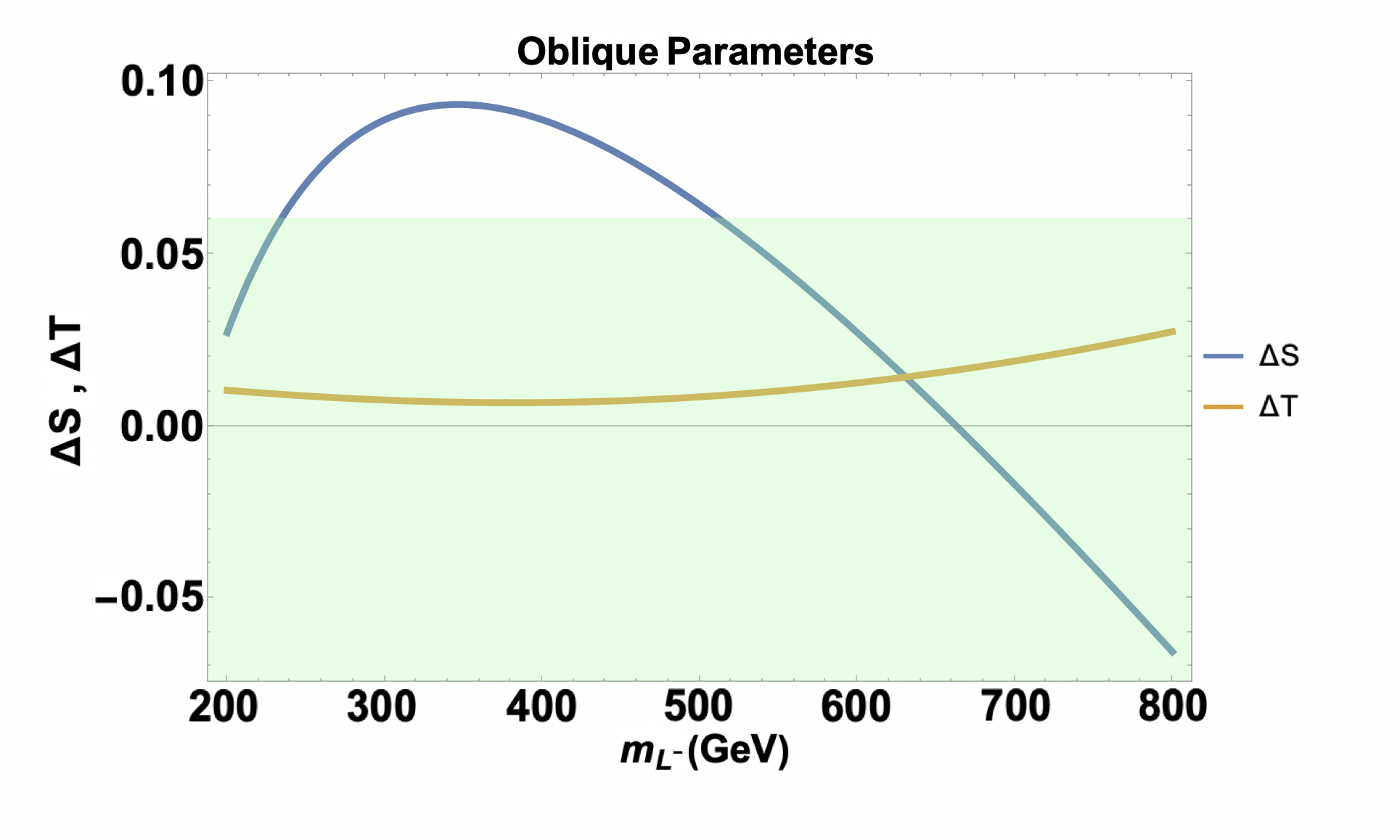}
		\caption{$\mathcal{T}_{2}$}
	\end{subfigure}\hspace{-1.2cm}
	\begin{subfigure}{.5\textwidth}
		\includegraphics[height=2.0in]{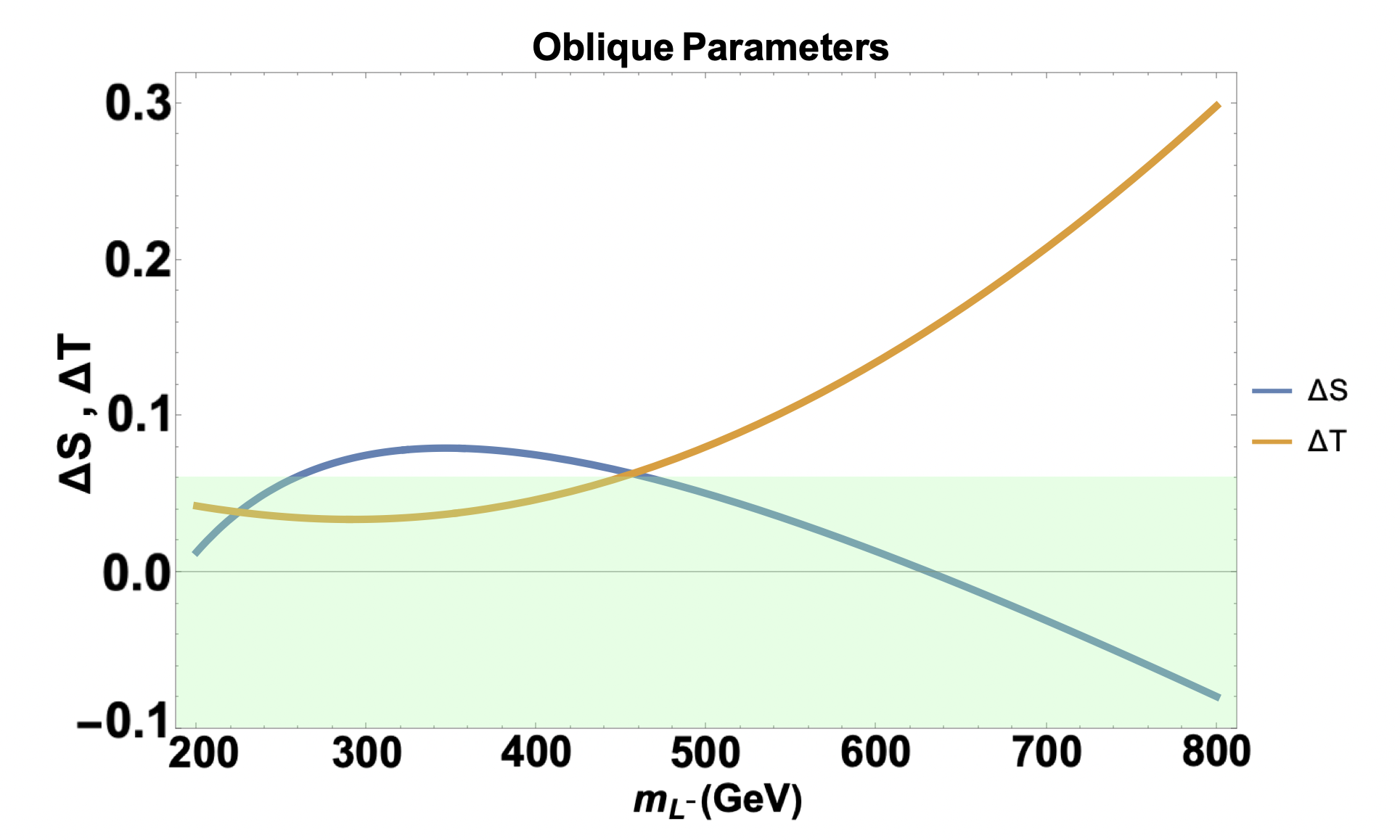}
		\caption{$\mathcal{T}_{2}$}
	\end{subfigure}	
  \caption{New physics contributions to the oblique parameters: $\mathbb{T}$ (orange) and $\mathbb{S}$ (blue) from triplet vector-like lepton representations for different VLL-SM lepton mixing $\sin \theta_L=0.05$ (left) and $\sin \theta_L=0.1$ (right).The green shaded region is the allowed space from the $\mathbb{S}$ and the $\mathbb{T}$ parameters in $2\sigma$ level. }
  \label{fig:VLLtripletoblique}
\end{figure}

Finally, by comparing triplet VLL models at $\sin\theta=0.05$, the allowed space of $\mathcal{T}_{1}$ is independent of the $\mathbb{S}$ and $\mathbb{T}$ parameters throughout the entire spectrum, as shown in Fig \ref{fig:VLLtripletoblique}. While the $\mathbb{T}$ parameter shows almost VLL mass-independent behavior for $\mathcal{T}_{2}$ model, the $\mathbb{S}$ parameter excludes $m_{L^-}=[250,510]$ GeV from constraints at the $3\sigma$ level. $\mathcal{T}_{1}$ has a similar $\mathbb{S}$ characteristic for larger mixing $\sin\theta=0.1$, thus $m_{L^-}$ is not limited. However, the upper bound from the $\mathbb{T}$ parameter occurs around $m_{L^-}\sim 650$ GeV. The constraints are more stringent on $\mathcal{T}_{2}$ representation as leptonic mixing increases due to its direct dependence on mass splitting between mass eigenstates. The upper bound from the $\mathbb{T}$ parameter occurs for $m_{L^-}\sim590$ GeV. 

As expected, a smaller mixing regime generates more relaxed constraints from the oblique parameters. We note that none of the representations is limited by the $\mathbb{T}$ parameter at $\sin\theta=0.05$, though this freedom is limited as the mixing gets larger. Thus, the oblique parameter $\mathbb{T}$ becomes more restrictive as $m_{VLL}$ gets larger. Furthermore, given different weak hypercharge choices, the $\mathbb{S}$ parameter also shows varying constraints but may not be as constraining as the $\mathbb{T}$ parameter unless the hypercharge choices lead to significant changes in the gauge boson propagators.
%%%%%%%%%%%%%%%%%
\newpage
%%%%%%%%%%%%%%%%%%%%%%%%%%%%%%%%%%%%%%%%%%%%%%%%%%%%%%%%%%%
\section{RGE Allowed Parameter Space of Vector-like Leptons}
\label{sec:rgef}
%%%%%%%%%%%%%%%%%%%%%%%%%%%%%%%%%%%%%%%%%%%%%%%%%%%%%%%%%%%%%
Theories with additional fermions customarily enhance  the instability of the Higgs self-coupling, driving it faster toward negative values at higher energy scales. This fundamentally signals the occurrence of an unbounded potential from below, thereby undermining vacuum stability. Such an outcome is already evident in the SM due to the top quark, which drives the Higgs quartic coupling negative around $\mu=10^{10}$ GeV at one loop \cite{Buttazzo_2013}. While supplementary scalar bosons can uphold the positivity of the Higgs self-coupling against the diminishing influences of the renormalization flow at higher energy scales, the introduction of vector-like fermions (VLFs) offers an intriguing alternative. These VLFs, through various gauge portals, have the potential to stabilize the electroweak vacuum. Therefore, the effects of VLFs have been studied in numerous extensions involving extensive scalar models. However, models with additional scalars are already promising when mixed with the Higgs fields, opening up a large parameter space due to considerable effects at the RGE level. The question remains, could one achieve vacuum stability {\it without} additional scalar(s)?

In this context, the inclusion of vector-like leptons (VLLs) exerts a strong influence on electroweak vacuum stability, predominantly through their effects on the Higgs quartic coupling via the weak hypercharge and isospin portals. In fact, additional charged fermions alone do not destabilize the Higgs potential.  Their gauge interactions stabilize it, while 
their Yukawa couplings to the SM Higgs introduce new instabilities.
 Unlike  the impact of vector-like quarks, VLLs engage differently with the gauge fields, leading to distinctive contributions to the RGEs of the Yukawa couplings $\Delta g_{1,2}$, due to the absence of the largest $\Delta g_3$ correction.  Scenarios that allow more than one generation  VLL  that do not exhibit such
couplings exist, and could be viable to stabilize the Higgs potential. The incorporation of VLLs introduces novel Yukawa interactions that generally serve to lower the Higgs quartic coupling. However, the gauge couplings $g_1$ and $g_2$, which are positively influenced by VLLs, are pivotal in counteracting this effect \cite{Tang:2013bz}. Furthermore, if VLL Yukawa couplings contribute sufficiently to balance the Higgs quartic coupling, compensating quartic effects $\lvert \lambda_{H }y^2_{L}\rvert > \lvert y^4_{L}\rvert$ along with gauge corrections,  this can generate a viable parameter space that keeps $\lambda_{H}>0$ for $\mu \leq M_{\rm Planck}$, preventing its descent into negativity at elevated scales \cite{Hiller_2022}. The intricate interplay between the Yukawa portal and the gauge couplings induced by the VLLs induces a non-trivial impact on the RGE flow, potentially unveiling regions of parameter space wherein the electroweak vacuum retains stability. This delicate equilibrium among the diverse contributions is paramount in determining the overall stability of the electroweak vacuum in the presence of VLLs.  Allowed by the experimental constraints, masses $m \sim \mathcal{O}$ (TeV) survive from stability constraints for VLQ   \cite{PhysRevD.107.036018,adhikary2024theoreticalconstraintsmodelsvectorlike}. However, we assume lighter scales $<\mathcal{O}$ (TeV) for VLL masses to obtain viable solutions that survive from the RGE flow. Hence, the examination of VLLs within the SM framework underscores a promising pathway for addressing vacuum stability without necessitating an extension of the scalar sector. To this end, we summarize our methodology as:

\begin{itemize}
\item We impose the minimum mass bounds on the neutral and charged sectors of VLLs and run RGEs over various models in Sec. \ref{sec:apprge} to generate the running of the Higgs and Yukawa couplings without encountering any Landau pole.
\item We also provide the allowed space for SM-VLL mixing versus $m_{VLL}$ by randomly generating $n_{VLL }+1$ parameter points as solutions to RGEs while enforcing stability and perturbative unitarity conditions on the couplings up to the Planck scale $\mu=M_{\rm Planck}$.
\item The initial conditions for the couplings appearing in the VLL representations are set at the energy scale $\mu_{0}=m_{t}$. 
\item Additionally, new physics corrections are also manifested through gauge boson loops in self-energy diagrams, namely the oblique parameters $\mathbb{S}$ and $\mathbb{T}$ from VLLs. We check the region of electroweak observables (EWPO) and discuss the scale favoring our findings from the RGE analyses. 
\end{itemize}
We now proceed to analyze the representations in Table \ref{tab:VLrepresentations}. 
%%%%%%%%%%%%%%%%%%%%%%%%%%%%%%%
\subsection{Singlet VLL: ${\cal S}_1$ and ${\cal S}_2$  }
\label{sec:rgeSresult}
\begin{figure}[htbp]
	\centering
	\begin{subfigure}{.5\textwidth}\hspace{-1.5cm}
		\includegraphics[height=2.2in]{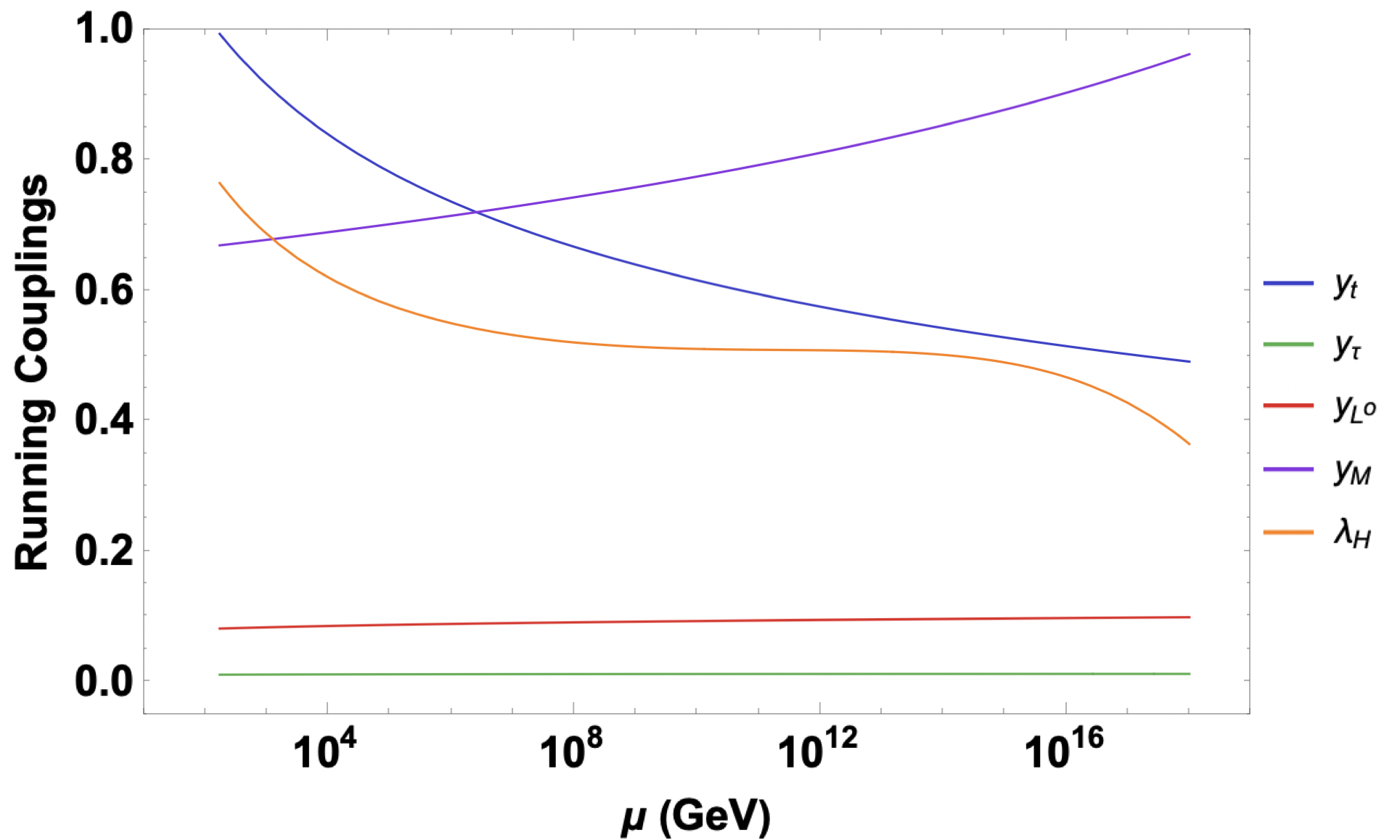}
		\caption{}
	\end{subfigure}\hspace{-0.3cm}
	\begin{subfigure}{.5\textwidth}
		\includegraphics[height=2.2in]{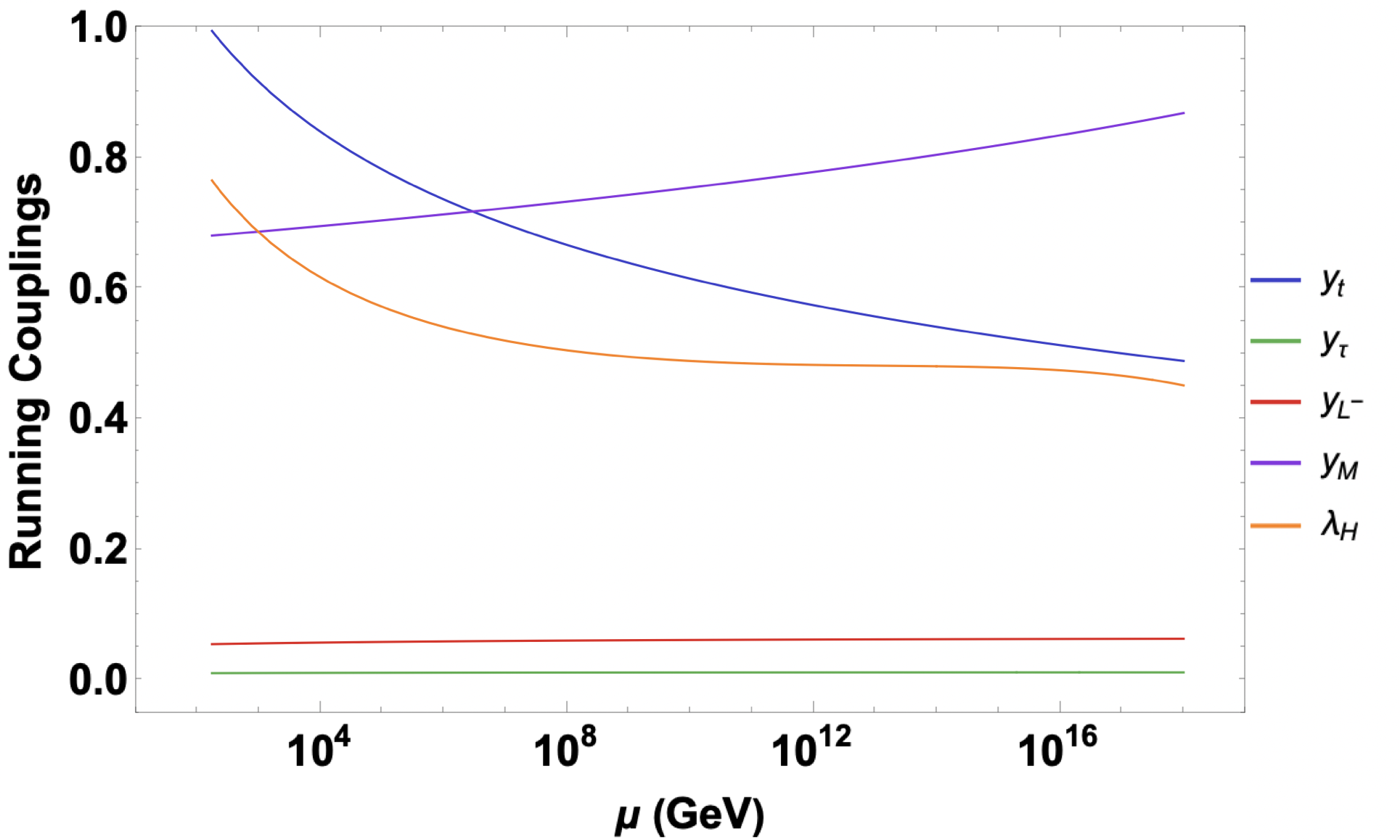}
		\caption{}
	\end{subfigure}
  \caption{The RGE running of the Yukawa and the Higgs coupling for models with singlet vector-like lepton representations. We show singlet vector-like representation, ${\cal S}_1$ (a), and ${\cal S}_2$ (b). For singlet models, we have set $m_{L^0}=120$ GeV, $m_{L^-}=125$ GeV  $\mu_0=m_t$, and $\sin \theta_L=0.1$.}
  \label{fig:rgeflowsinglets}
\end{figure}

\begin{figure}[htbp]
	\centering
	\begin{subfigure}{.5\textwidth}\hspace{-1.5cm}
		\includegraphics[height=2.4in]{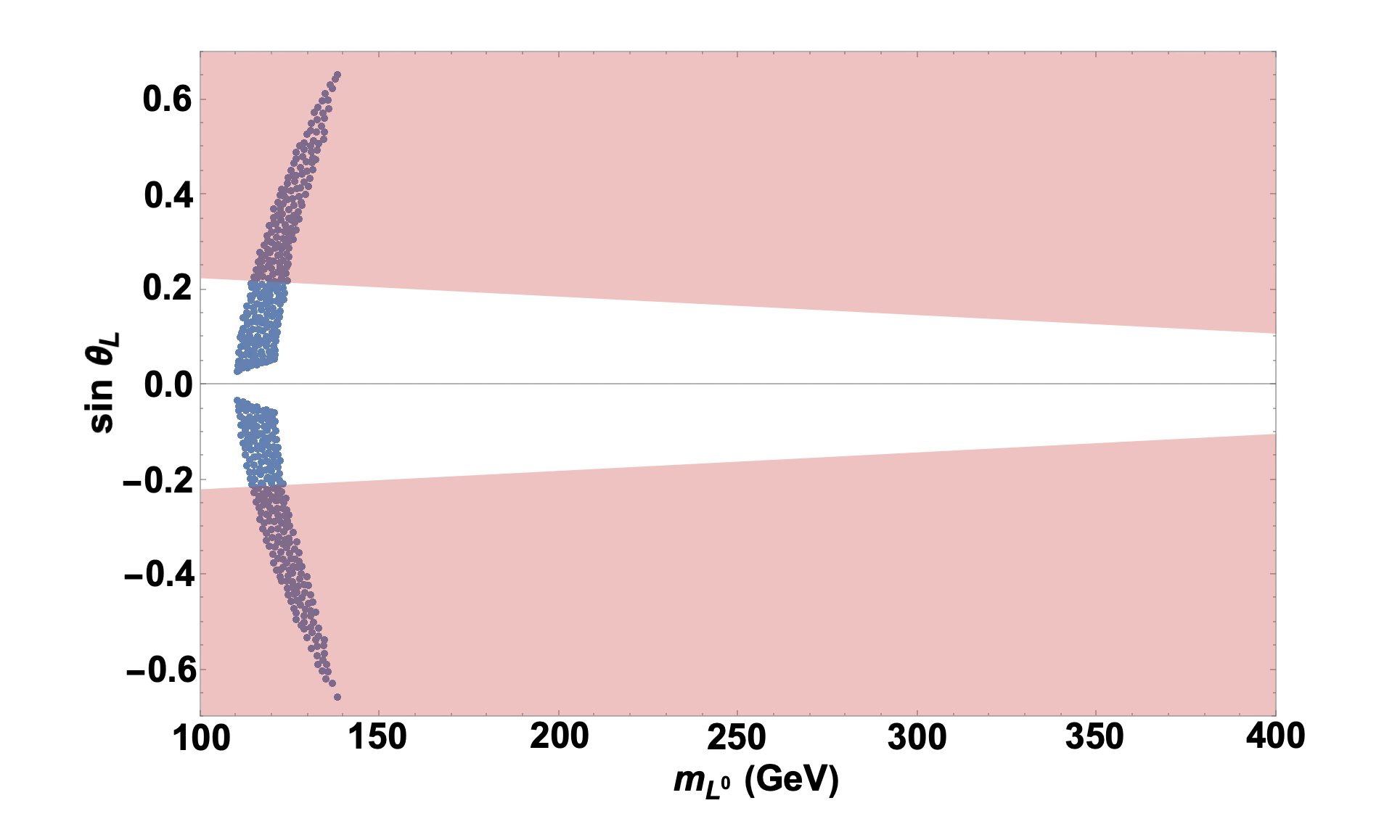}
		\caption{}
	\end{subfigure}\hspace{-0.3cm}
	\begin{subfigure}{.5\textwidth}
		\includegraphics[height=2.4in]{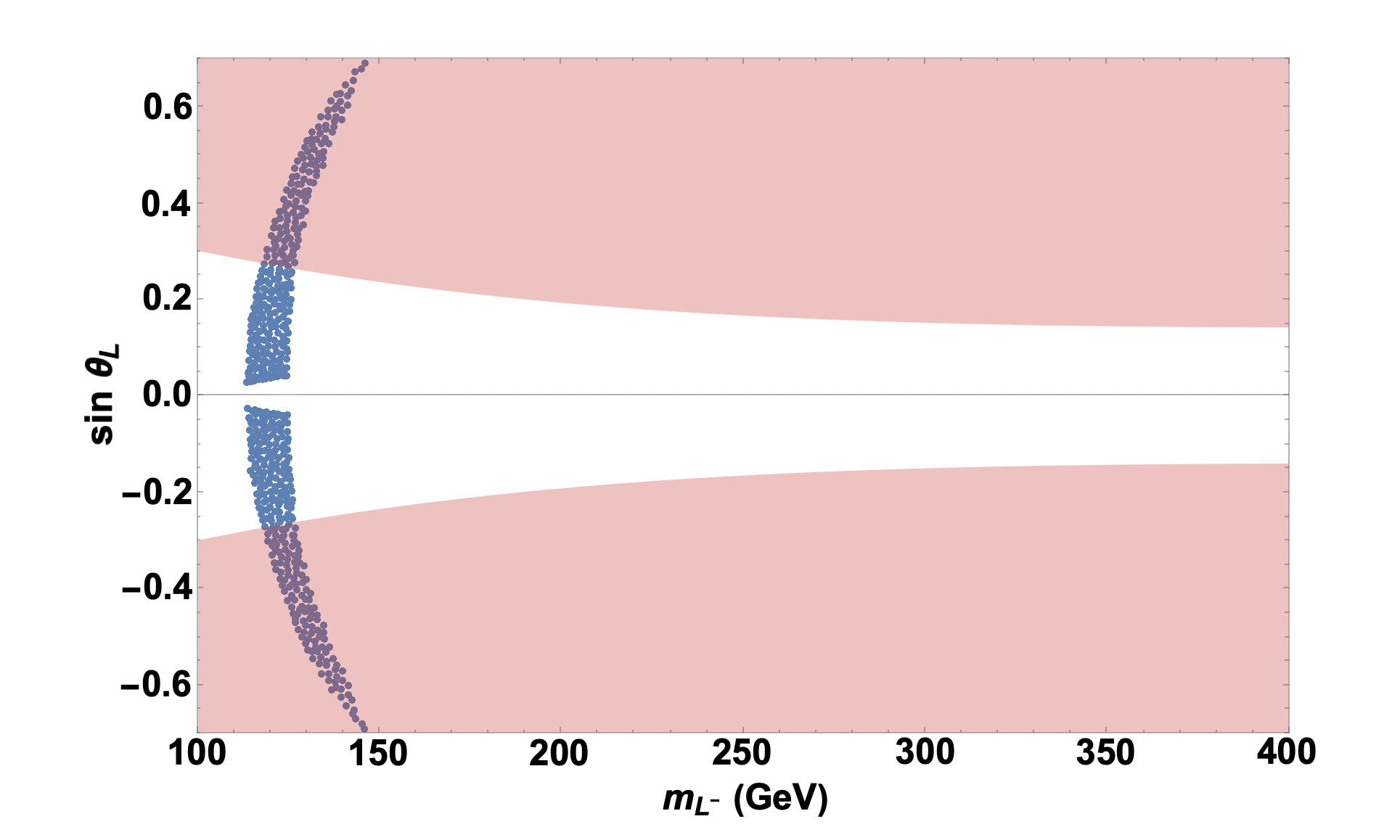}
		\caption{}
	\end{subfigure}
  \caption{The allowed parameter space extracted from theoretical constraints for the mass of vector-like leptons and its dependence on mixing angle to the third generation SM leptons for singlet vector-like lepton representations. We show singlet ${\cal S}_1$ vector-like model (a), and singlet vector-like ${\cal S}_2$ model (b). The region shaded in pink is disallowed by constraints coming from the electroweak precision observables as in Sec. \ref{sec:electroweakprecision}. }
  \label{fig:rgespacesinglets}
\end{figure}

%%%%%%%%%%%%%%%%%%%%%%%%%%%%
Singlet VLL extensions ${\cal S}_1$ and ${\cal S}_2$ of the SM generate the safest scenario for the Higgs quartic coupling among all the representations studied herein. Fixing the masses of neutral and charged vector fermions throughout our work to compare each model clearly showed that $\lambda$ is more prone to stay away from the vacuum instability scale in singlet VLL models, as seen in Fig. \ref{fig:rgeflowsinglets}.  The ${\cal S}_2$ model generates a relatively larger parameter space up to $m_{L^-}\sim 150$ GeV, as seen in Fig. \ref{fig:rgespacesinglets}, surviving all theoretical constraints. The distinction between ${\cal S}_1$ and ${\cal S}_2$ is fully attributed to weak hypercharge difference, where $U(1)_{Y}$ gauge portal $g_1$ receives no correction from the $L^0$ field alone, hence it narrows the allowed space. This can also be shown in the RG running of $y_M$ that defines the Higgs coupling to only VLLs. It strays dangerously close to the non-perturbative region in ${\cal S}_1$ model, for which the RGE controlling the $y_M$ coupling is completely dictated by the Yukawa terms in Eq. \ref{eq:rgeS1singletfermion}. The upward shift of the neutral Yukawa coupling in ${\cal S}_1$ compared to the charged Yukawa in ${\cal S}_2$ occurs due the fact that the mass splitting between neutrino and $L^0$ is larger than that between the $\tau$ and $L^-$, thus generating a larger initial condition as given in Eq. \ref{eq:IC}. Having the least number of free parameters used for RGE solutions, the ${\cal S}_1$ and ${\cal S}_2$ models are highly dependent on the reciprocal relation between the mass of the field and its mixing with the SM lepton. Although lighter mass scales are excluded by the experimental data \cite{10.1093/ptep/ptac097}, we found that $m_{L^{0,-}}\gtrsim110$ GeV; otherwise, the perturbativity of the Higgs coupling breaks down. Additionally, Fig. \ref{fig:rgespacesinglets} shows that the mixing between VLLs and SM leptons remains non-zero, serving as the most important condition for stabilizing the electroweak vacuum in the presence of new fields. The absence of the color charge is prominent in Fig. \ref{fig:rgeflowsinglets}, producing a unique feature of VLL Yukawa couplings that differ significantly from those of vector-like quarks \cite{PhysRevD.109.036016}. Furthermore, we checked that, except for highly exotic weak hypercharge choices $Y>\lvert3/2\rvert$ \cite{Altmannshofer_2014}, VLL Yukawa couplings tend to increase over energy scales, as the remain insensitive to the largest gauge portal correction $g_3$\footnote{The relative strength of $\tau$ Yukawa is too small and the running coupling appears almost flat compared to other couplings in the model.}.  In this graph, we indicate the region shaded in pink which is disallowed by constraints coming from the electroweak precision observables as in Sec. \ref{sec:electroweakprecision}.
%%%%%%%%%%%%%%%%%%%%%%%%%%%%%%%%%%%%%%
\subsection{Doublet VLL: ${\cal D}_1$ and ${\cal D}_2$  }
\label{sec:rgeDresult}
\begin{figure}[htbp]
	\centering
	\begin{subfigure}{.5\textwidth}\hspace{-1.5cm}
		\includegraphics[height=2.2in]{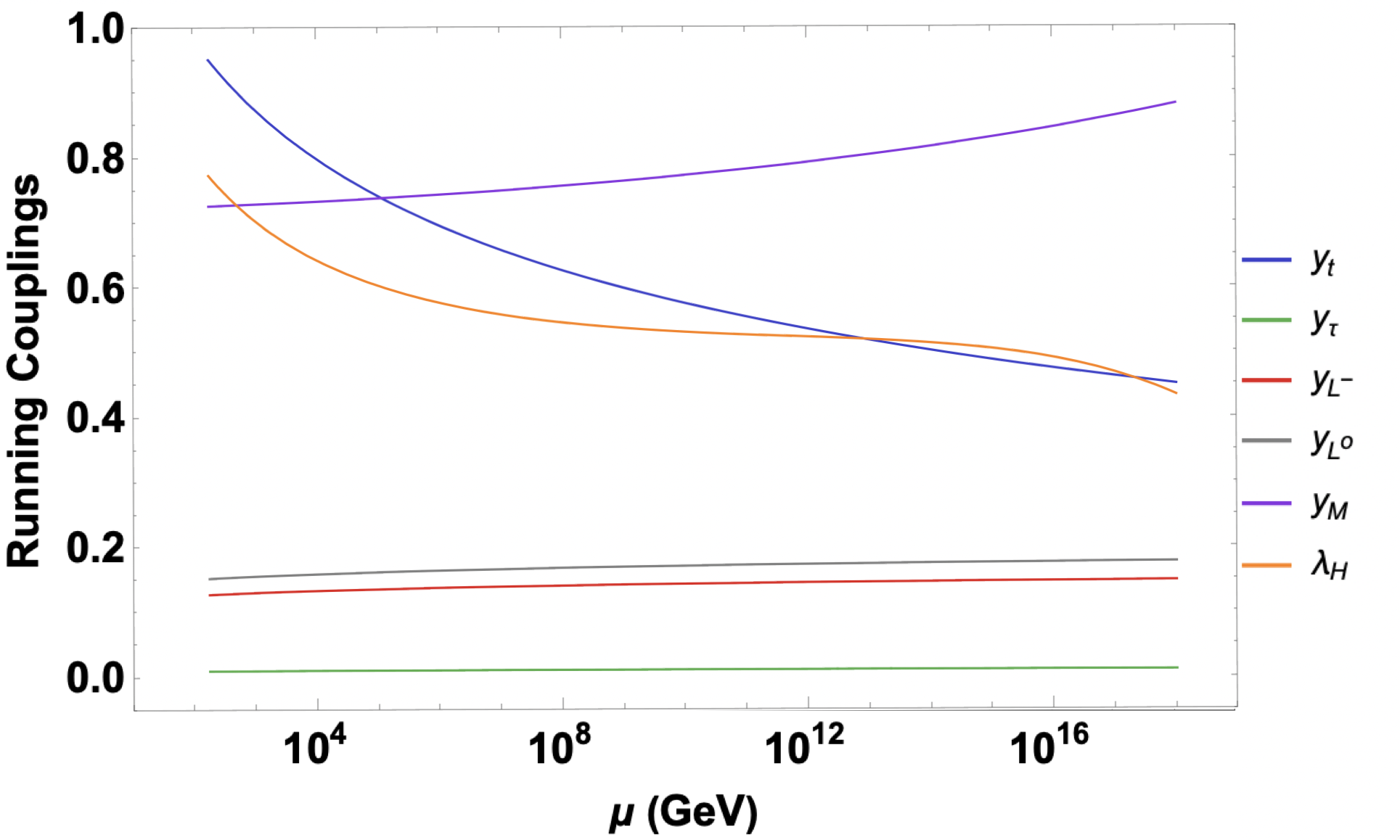}
		\caption{}
	\end{subfigure}\hspace{-0.3cm}
	\begin{subfigure}{.5\textwidth}
		\includegraphics[height=2.2in]{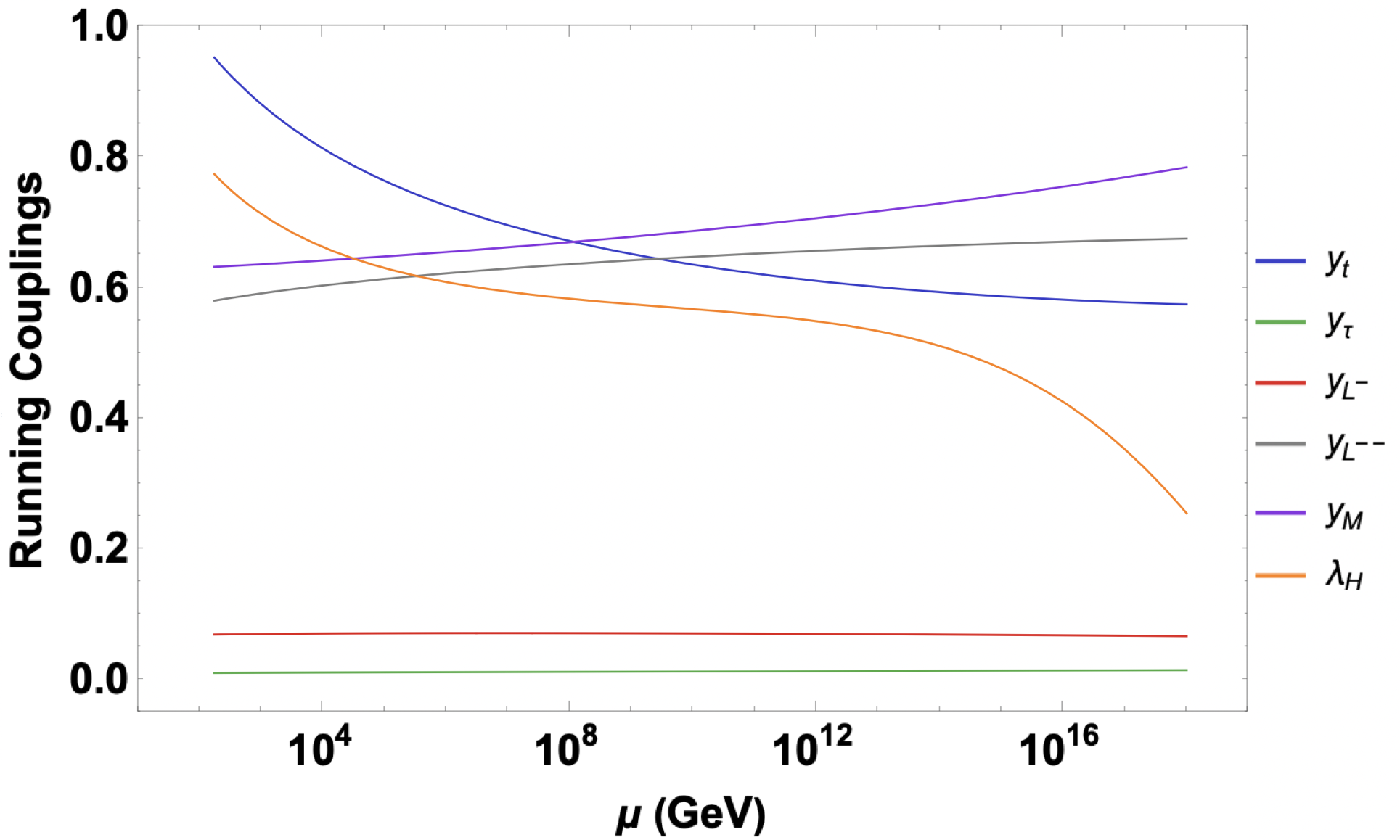}
		\caption{}
	\end{subfigure}
  \caption{The RGE running of the Yukawa and the Higgs coupling for models with doublet vector-like leptons. We show doublet vector-like representation, ${\cal D}_1$ (a), and ${\cal D}_2$ (b). For doublet models, we have set $m_{L^0}=150$ GeV, $m_{L^-}=130$ GeV, $m_{L^{--}}=160$ GeV,  $\mu_0=m_t$, and $\sin \theta_L=0.1$.}
  \label{fig:rgeflowdoublets}
\end{figure}

\begin{figure}[htbp]
	\centering
	\begin{subfigure}{.5\textwidth}\hspace{-1.5cm}
		\includegraphics[height=2.4in]{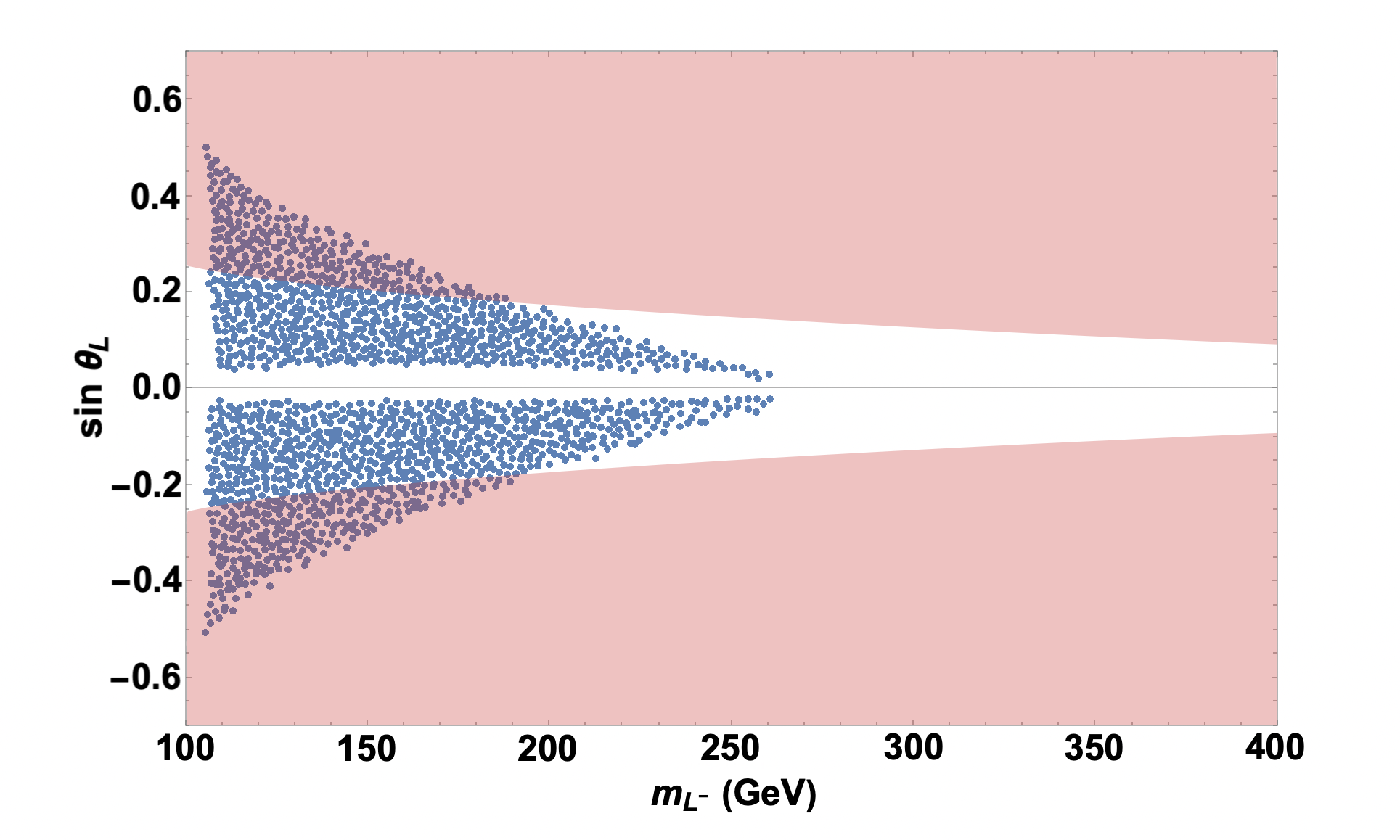}
		\caption{}
	\end{subfigure}\hspace{-0.3cm}
	\begin{subfigure}{.5\textwidth}
		\includegraphics[height=2.4in]{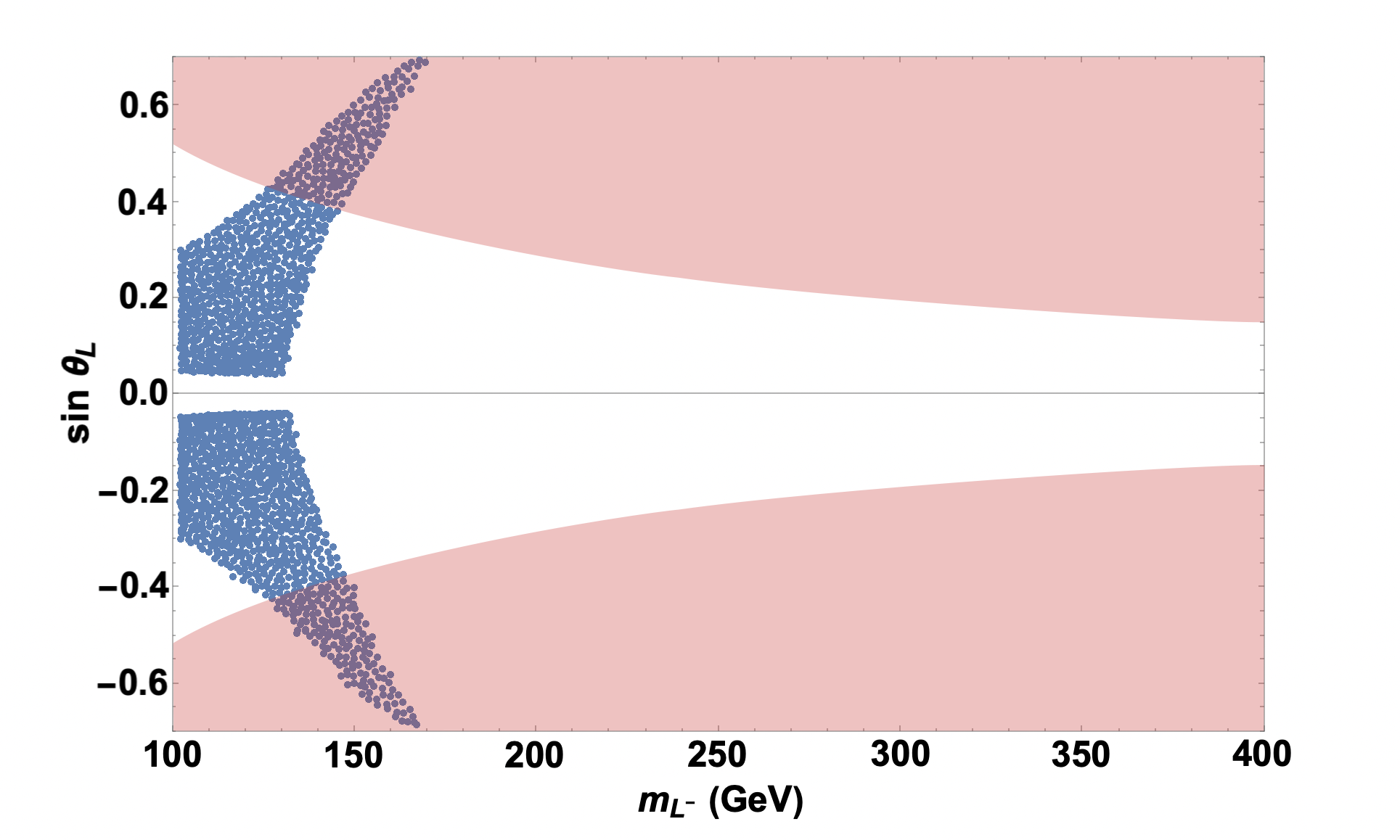}
		\caption{}
	\end{subfigure}
  \caption{The allowed parameter space extracted from theoretical constraints for the mass of vector-like leptons and its dependence on mixing angle to the third generation SM leptons for doublet vector-like representations. We show doublet ${\cal D}_1$ vector-like model (a), and doublet vector-like ${\cal D}_2$ model (b).  The region shaded in pink is disallowed by constraints coming from the electroweak precision observables as in Sec. \ref{sec:electroweakprecision}.}
  \label{fig:rgespacedoublets}
\end{figure}

In contrast to singlet models, RGE solutions of doublet models ${\cal D}_1$ and ${\cal D}_2$ exhibit a  more sensitive behaviour with respect to the Higgs coupling, especially in the presence of non-SM-like charges. The uncoupled nature of doubly charged VLLs drastically adjusts the starting value of the running coupling $y_{L^{--}}$ as illustrated in Fig. \ref{fig:rgeflowdoublets}. Consequently, this adjustment affects the Higgs RGE more significantly than for fields that mix with SM leptons across all multiplets. However, this phenomenon also imposes a soft upper bound on the mass of exotic leptons, constrained by perturbativity to $m_{L^{--}}< m_t$. As mentioned earlier, larger hypercharge values for VLLs can cause Yukawa couplings to decrease with increasing energy, similar to quark couplings in renormalization theory. As shown in the right panel of Fig. \ref{fig:rgeflowdoublets}, $y_L$ begins to decrease around $\mu\gtrsim10^{13}$ GeV as might be expected from the effect of the largest hypercharge-carrying field ${\cal D}_2$. The vacuum stability condition requires smaller mixing angles to counterbalance initial conditions due to the mass increment; however, the Yukawa coupling $y_M$ increases as the VLL-SM mixing scale approaches the decoupling region. Therefore, representations that exclude both neutral and charged VLLs simultaneously are more sensitive to the value of $y_M$ due to the indirect effects of uncoupled leptons via RGEs. This sensitivity results in distinct parameter spaces for ${\cal S}_1$, ${\cal S}_2$ and ${\cal D}_2$ compared to other models. On the other hand,  the model ${\cal D}_1$, including both $L^0$ and $L^-$,  provides more space as both up and down sector mixings vary between extreme ends while maintaining $\lambda$ in the vacuum stability regime. The extension of the RGE parameter space is also related to the additional number of positive quadratic and negative quartic Yukawa terms. The limits on doublet models are more relaxed compared to singlet VLLs with $m_{L^-}$ upper bound reaching  approximately $170$ GeV for ${\cal D}_2$, and the mass of the charged lepton rising up to $\sim$ 260 GeV in the ${\cal D}_1$ model, allowed by mixing angle $\sin\theta<0.05$. In fact, we  verified numerically that a scale $m_{VLL}>260$ GeV breaks the perturbativity of Yukawa couplings before the Higgs quartic coupling becomes negative. Thus Fig. \ref{fig:rgespacedoublets} shows that the upper bound on the mass of charged VLLs for ${\cal D}_2$ model corresponds to a critical value where $\lambda$ starts to run to negative values before Yukawa couplings become non-perturbative. Finally, non-SM-like multiplets can generate heavier mass values that meet theoretical requirements; however, the Higgs constraints from VLLs limit the mixing domain, which is in particular constrained by the Higgs diphoton decay rate \cite{Barducci_2023}. As before, in this graph, we indicate the region shaded in pink which is disallowed by constraints coming from the electroweak precision observables as in Sec. \ref{sec:electroweakprecision}.
%%%%%%%%%%%%%%%%%%%%%%%%%%%%%%
\subsection{Triplet VLL: ${\cal T}_1$ and ${\cal T}_2$  }
\label{sec:rgeTresult}
\begin{figure}[htbp]
	\centering
	\begin{subfigure}{.5\textwidth}\hspace{-1.5cm}
		\includegraphics[height=2.2in]{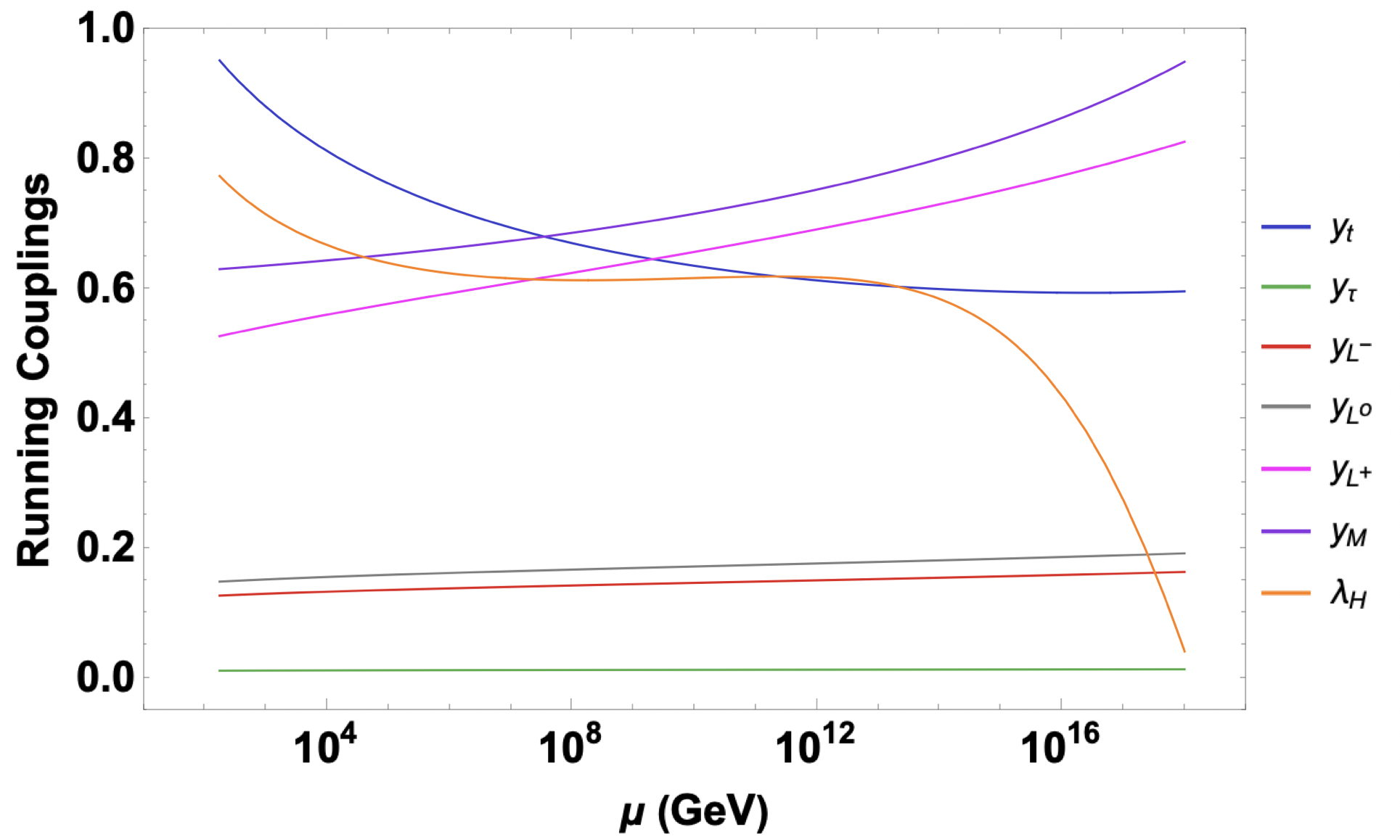}
		\caption{}
	\end{subfigure}\hspace{-0.3cm}
	\begin{subfigure}{.5\textwidth}
		\includegraphics[height=2.2in]{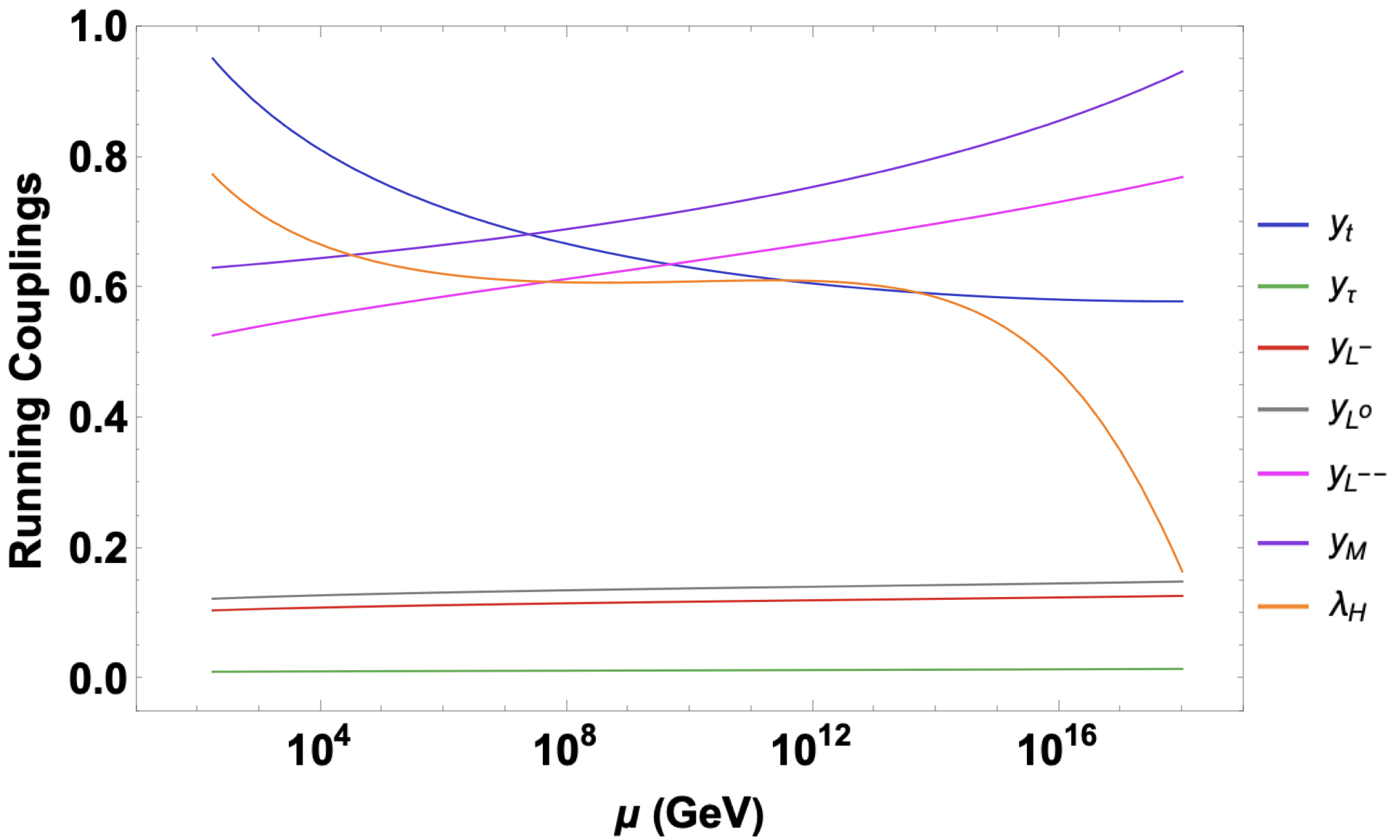}
		\caption{}
	\end{subfigure}
  \caption{The RGE running of the Yukawa and the Higgs coupling for models with triplet vector-like leptons. We show triplet vector-like representation, ${\cal T}_1$ (a), and ${\cal T}_2$ (b). For triplet models, we have set $m_{L^0}=150$ GeV, $m_{L^-}=200$ GeV, $m_{L^+}=170$ GeV, $m_{L^{--}}=170$ GeV,  $\mu_0=m_t$, and $\sin \theta_L=0.1$.}
  \label{fig:rgeflowtriplets}
\end{figure}

\begin{figure}[htbp]
	\centering
	\begin{subfigure}{.5\textwidth}\hspace{-1.5cm}
		\includegraphics[height=2.4in]{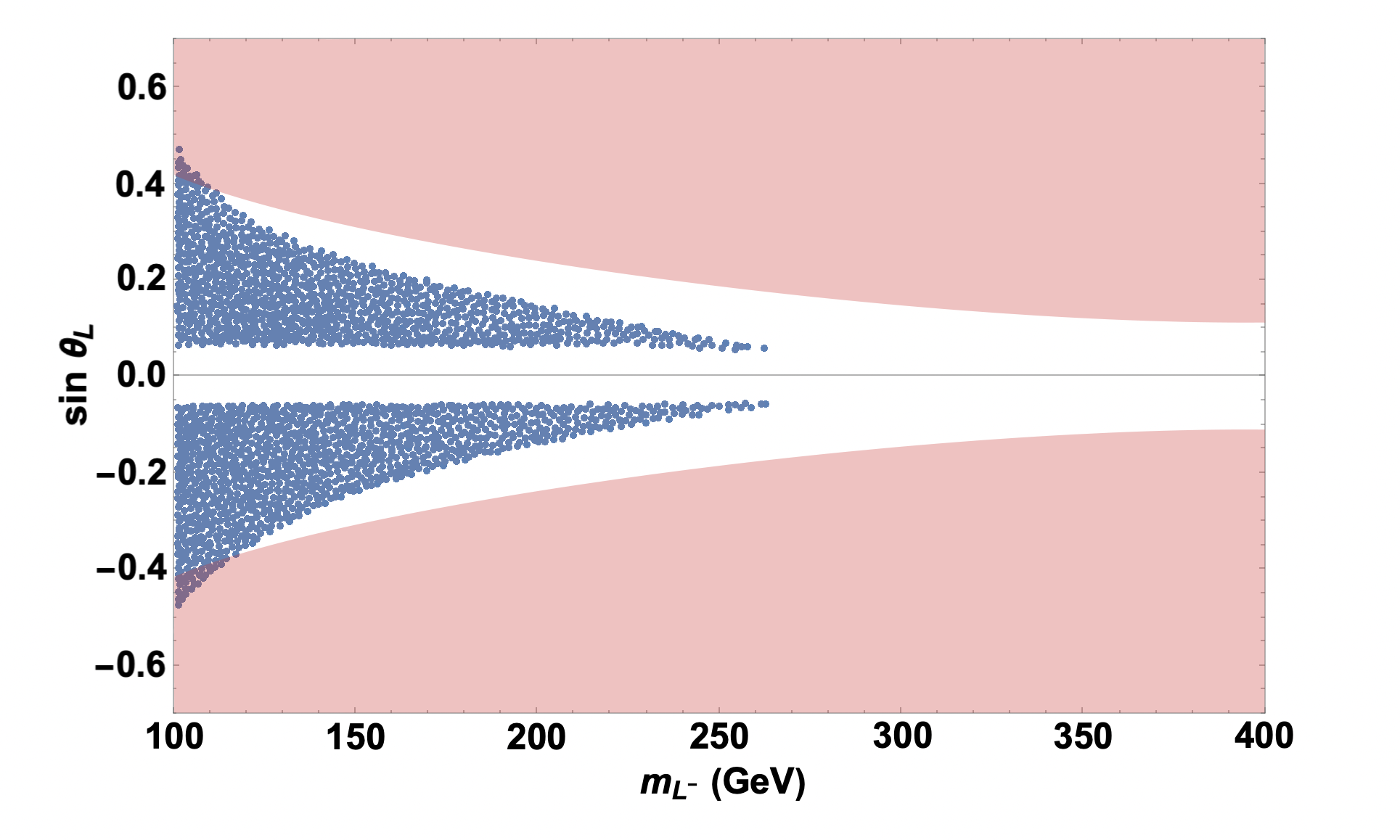}
		\caption{}
	\end{subfigure}\hspace{-0.3cm}
	\begin{subfigure}{.5\textwidth}
		\includegraphics[height=2.4in]{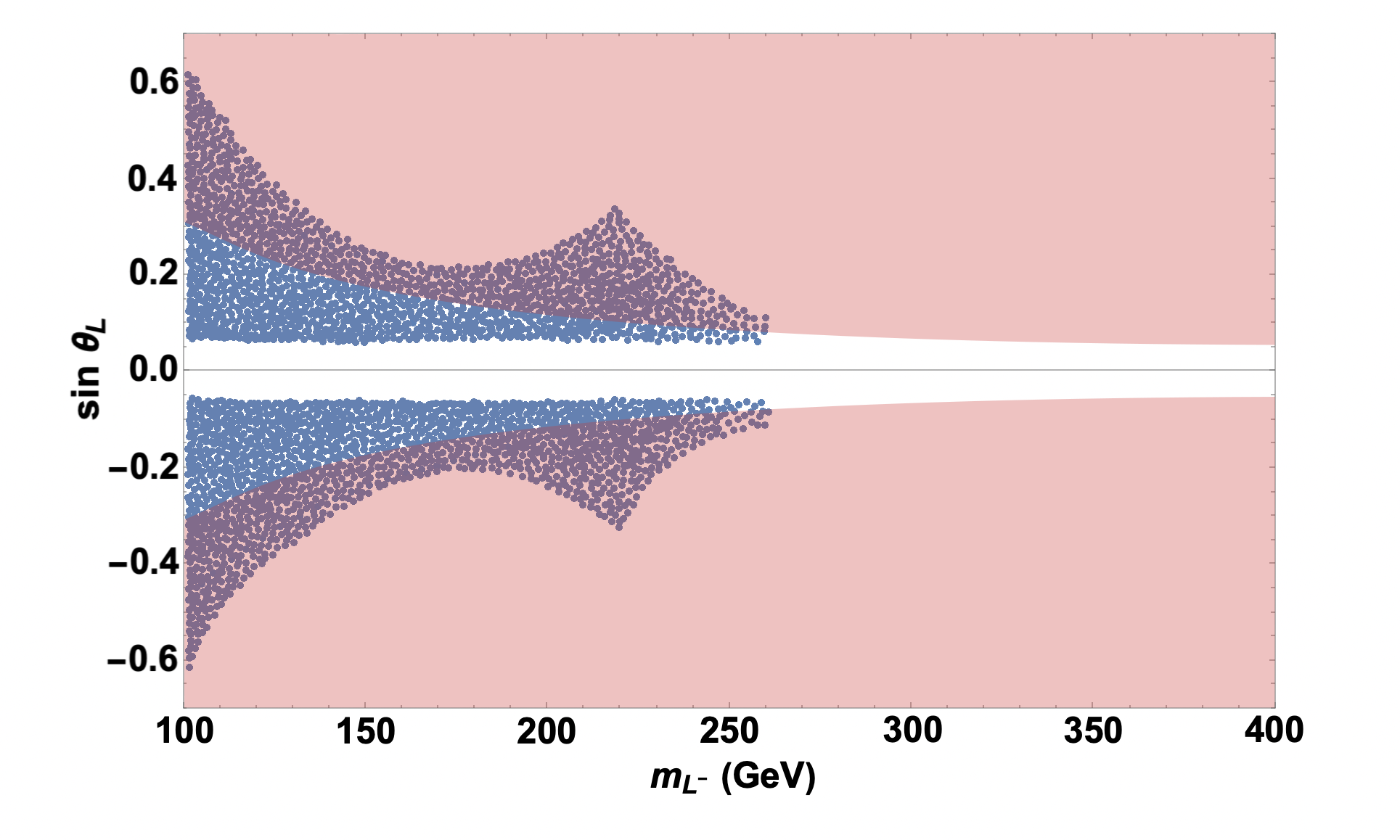}
		\caption{}
	\end{subfigure}
  \caption{The allowed parameter space extracted from theoretical constraints for the mass of  vector-like leptons and its dependence on mixing angle to the third generation SM leptons for triplet vector-like representations. We show triplet ${\cal T}_1$ vector-like model (a) and triplet vector-like ${\cal T}_2$ model (b).  The region shaded in pink is disallowed by constraints coming from the electroweak precision observables as in Sec. \ref{sec:electroweakprecision}.}
  \label{fig:rgespacetriplets}
\end{figure}

The case of triplets vector-like representations is affected by both the exotic $L^+$, $L^{--}$, and SM-like vector partners.  In Fig. \ref{fig:rgespacetriplets} the mixing is allowed to be either small or large in the low mass region, whereas larger masses generally require smaller mixing for both triplet models. The mass spectrum reaches up to $270$ GeV, though the theoretical minimum would be allowed to be lower in our work, but is excluded by experimental constraints. Similar to singlet models, the distinction between parameter spaces arises from weak hypercharge. Nevertheless, RGE constraints on the triplet model ${\cal T}_1$ model are less relaxed due to the absence of $g_1$ correction, which imposes a smaller mixing regime compared to the ${\cal T}_2$ model. The minimum value of the mixing angle required to ensure vacuum stability is slightly larger than for all other representations. This feature is analytically motivated by the fact that triplets rely simultaneously on both neutral and charged Yukawa RGEs, thus requiring a relatively larger minimum mixing across the entire mass spectrum. Moreover,  $\lambda$ in Fig. \ref{fig:rgeflowtriplets}  approaches to zero in ${\cal T}_1$ while $y_M$ is the largest correction among the models, directly correlating the allowed space to the inverse relationship between VLL mass and mixing angle shown in Fig. \ref{fig:rgespacetriplets}.  Finally, we can conclude that VLL triplets are more promising for stabilizing the vacuum, as they scan over a larger spectrum while satisfying both stability and perturbative unitarity constraints.  As for the case of singlets and doublets,  we shade in pink the region which is disallowed by constraints coming from the electroweak precision observables as in Sec. \ref{sec:electroweakprecision}.
\newpage
\subsection{Effects of Two Loop Corrections to the RGE }
\label{sec:2looprgeresult}
Here we analyse the effect of using the RGEs to two loop accuracy, and whether the next-to-next-to-leading order (NNLO) couplings evolve in such a way as to extend the VLL parameter space. The transition from one loop to two loop RGEs introduces nuanced changes to the running of all couplings, due to  higher-order interactions and mixed terms that appear in two loop corrections.  Effectively this allows all terms in the RGE to be influenced by the presence of different sectors, generating fully coupled equations. At the one loop level gauge couplings generally increase with energy due to contributions from additional fermions which enhances the gauge portal for vacuum stability concerns. When moving to two loop corrections, gauge couplings receive additional positive contributions from themselves, further enhancing their growth. However the Yukawa couplings mitigate the gauge coupling runnings to  order  -$g_i^4y_{F}^2$. While the interplay between couplings becomes more complex at the two loop level, making deterministic interpretation difficult, it is expected that the gauge portal weakens as the multiplets include more fermions.

\begin{figure}[htbp]
	\centering
	\begin{subfigure}{.5\textwidth}\hspace{-1.5cm}
		\includegraphics[height=2.2in]{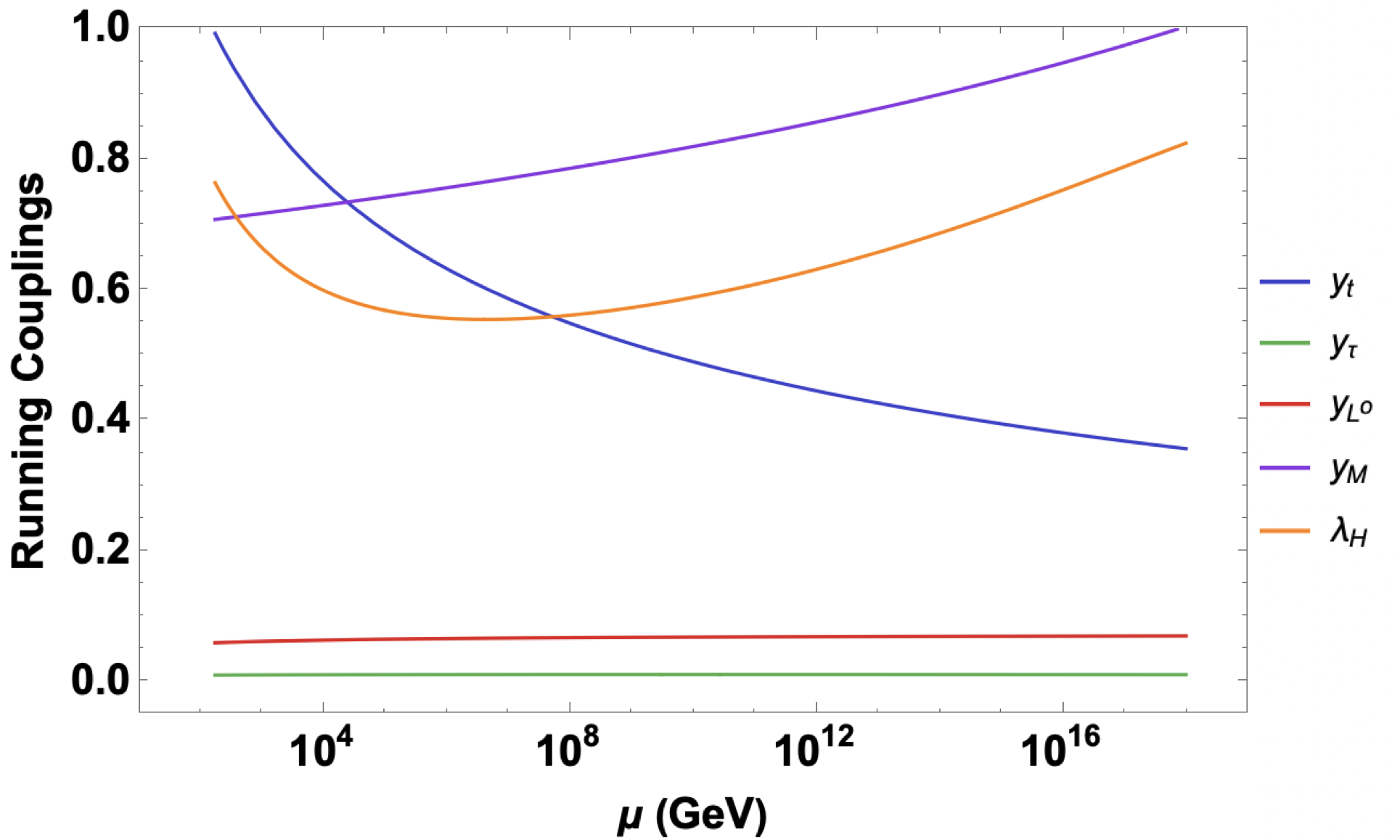}
		\caption{}
	\end{subfigure}\hspace{-0.3cm}
	\begin{subfigure}{.5\textwidth}
		\includegraphics[height=2.2in]{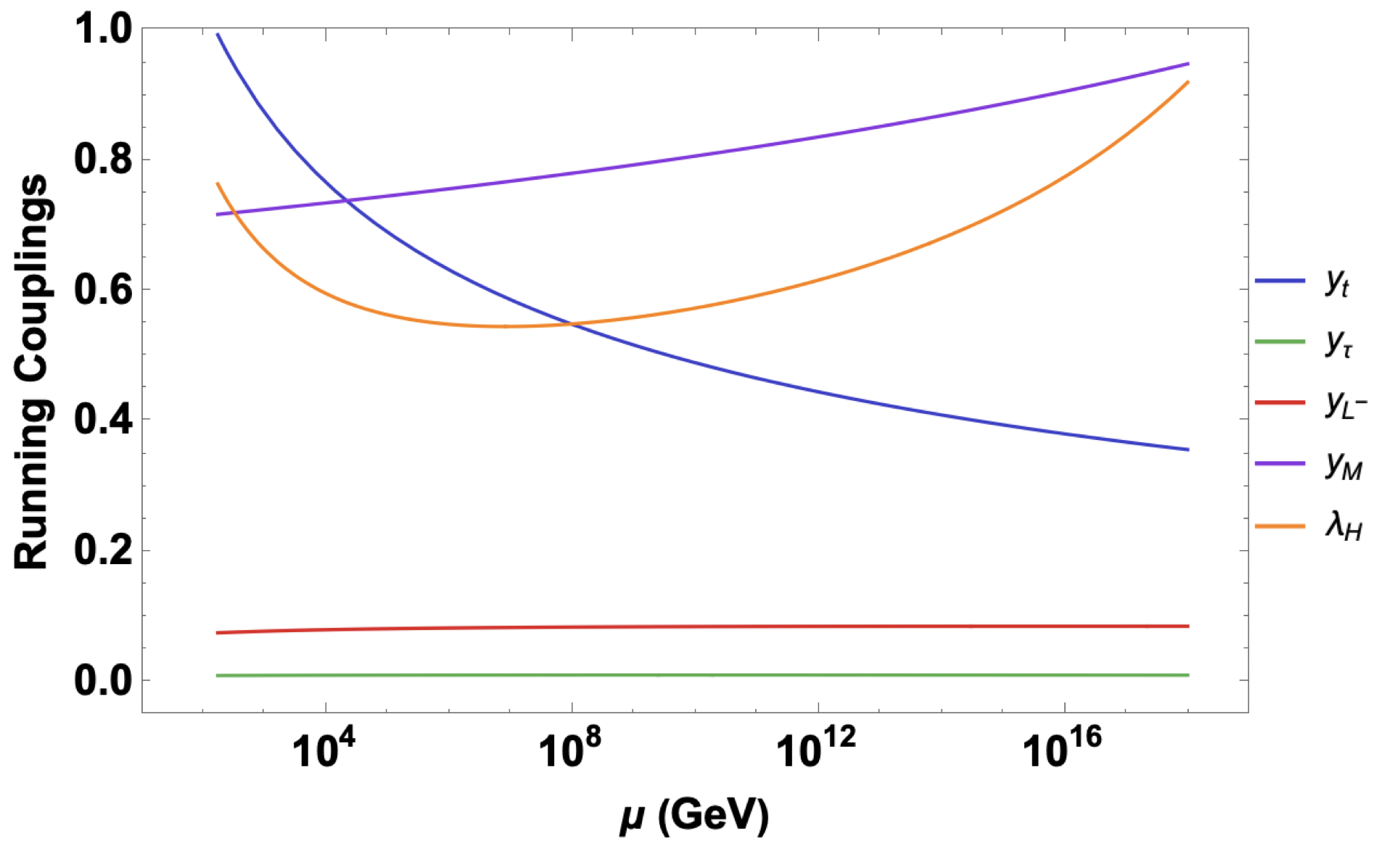}
		\caption{}
	\end{subfigure}
  \caption{The two loop RGE running of the Yukawa and the Higgs coupling for models with singlet vector-like leptons. We show singlet vector-like representation, (a) ${\cal S}_1$, and (b) ${\cal S}_2$. For singlet models, we have set $m_{L^0}=150$ GeV, $m_{L^-}=160$ GeV  $\mu_0=m_t$, and $\sin \theta_L=0.1$.}
  \label{fig:2looprgesinglets}
\end{figure}

\begin{figure}[htbp]
	\centering
	\begin{subfigure}{.5\textwidth}\hspace{-1.5cm}
		\includegraphics[height=2.2in]{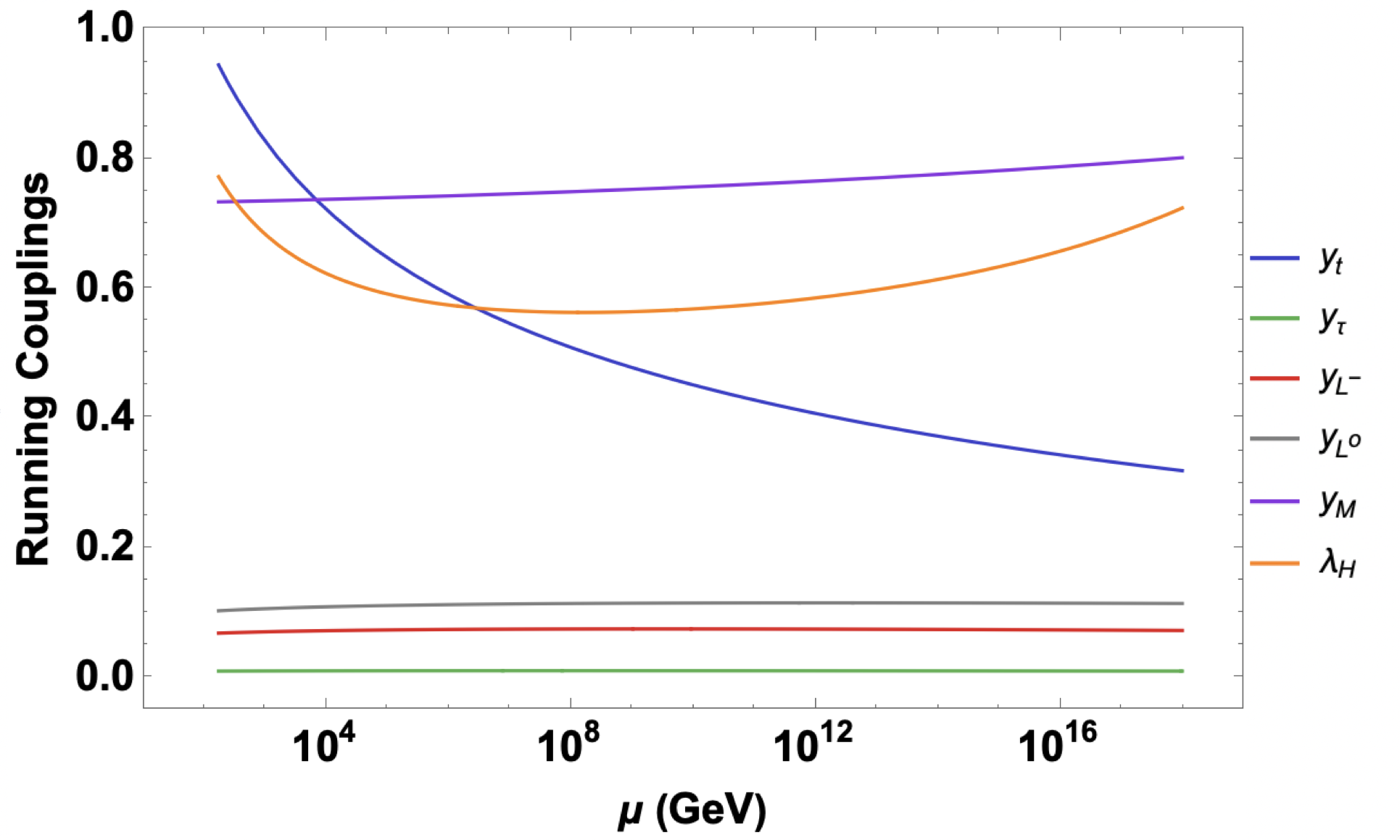}
		\caption{}
	\end{subfigure}\hspace{-0.3cm}
	\begin{subfigure}{.5\textwidth}
		\includegraphics[height=2.2in]{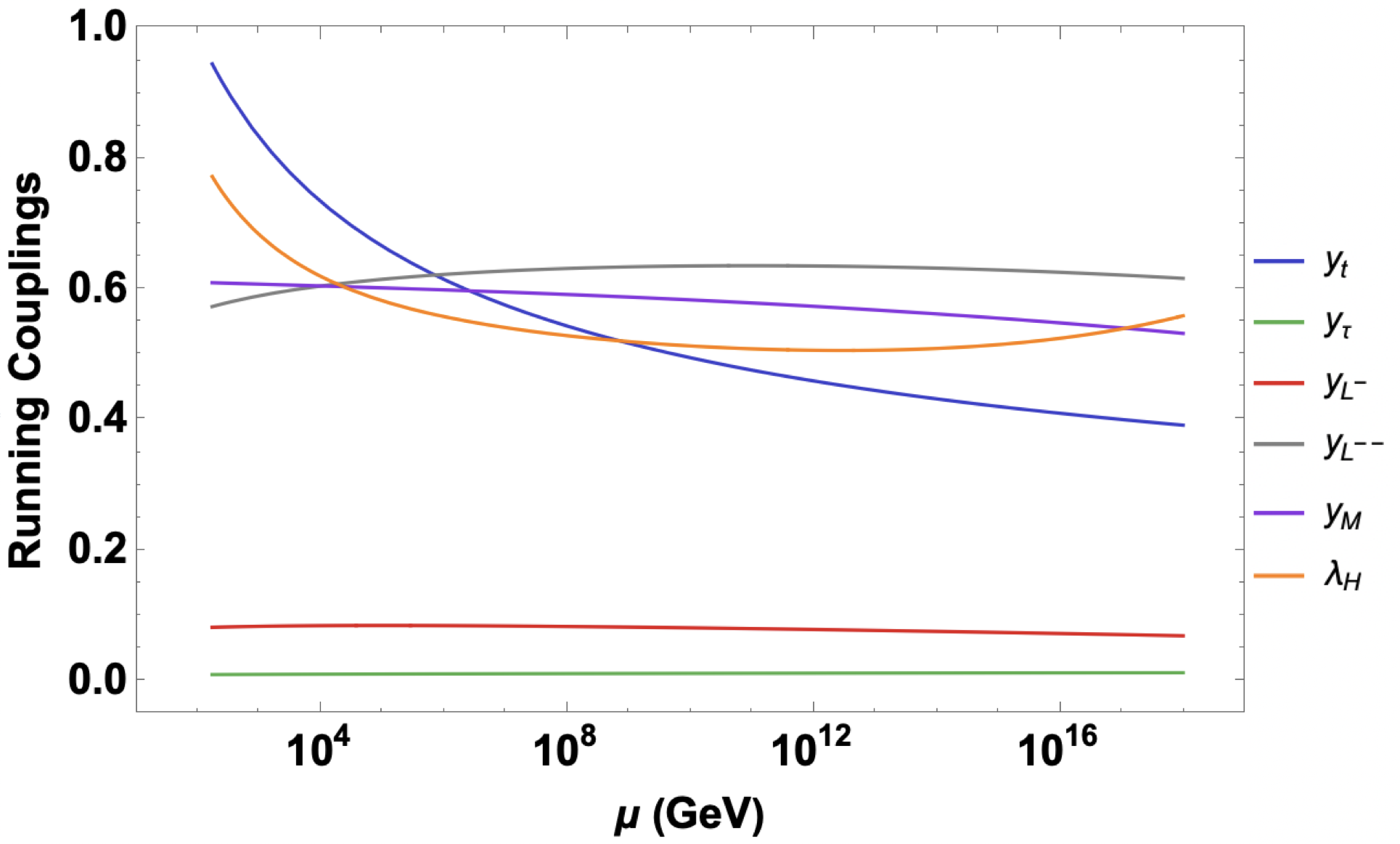}
		\caption{}
	\end{subfigure}
  \caption{The two loop RGE running of the Yukawa and the Higgs coupling for models with doublet vector-like leptons. We show doublet vector-like representation, (a) ${\cal D}_1$, and (b) ${\cal D}_2$. For doublet models, we have set $m_{L^0}=200$ GeV, $m_{L^-}=220$ GeV, $m_{L^{--}}=170$ GeV,  $\mu_0=m_t$, and $\sin \theta_L=0.1$.}
  \label{fig:2looprgedoublets}
\end{figure}

\begin{figure}[htbp]
	\centering
	\begin{subfigure}{.5\textwidth}\hspace{-1.5cm}
		\includegraphics[height=2.2in]{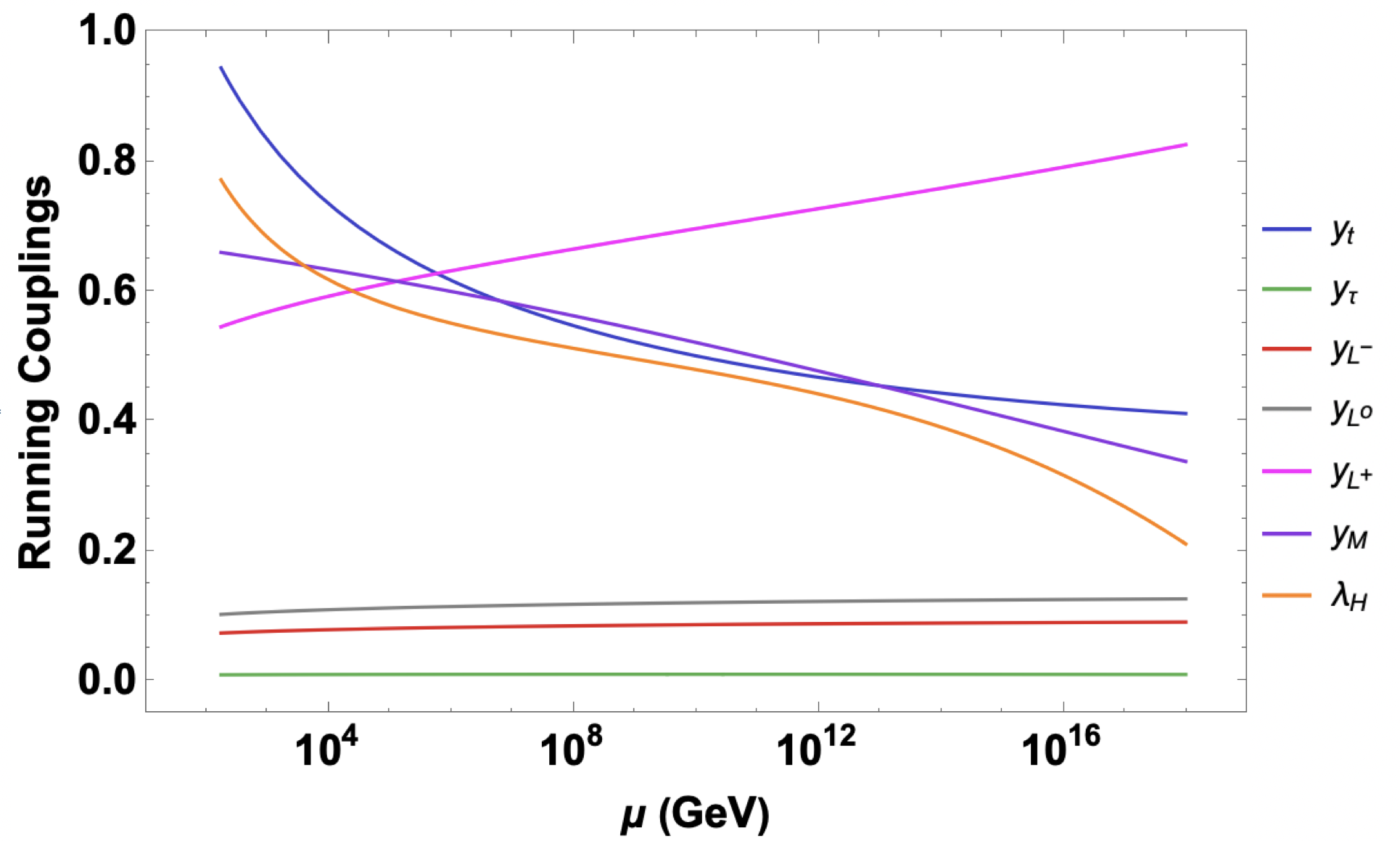}
		\caption{}
	\end{subfigure}\hspace{-0.3cm}
	\begin{subfigure}{.5\textwidth}
		\includegraphics[height=2.2in]{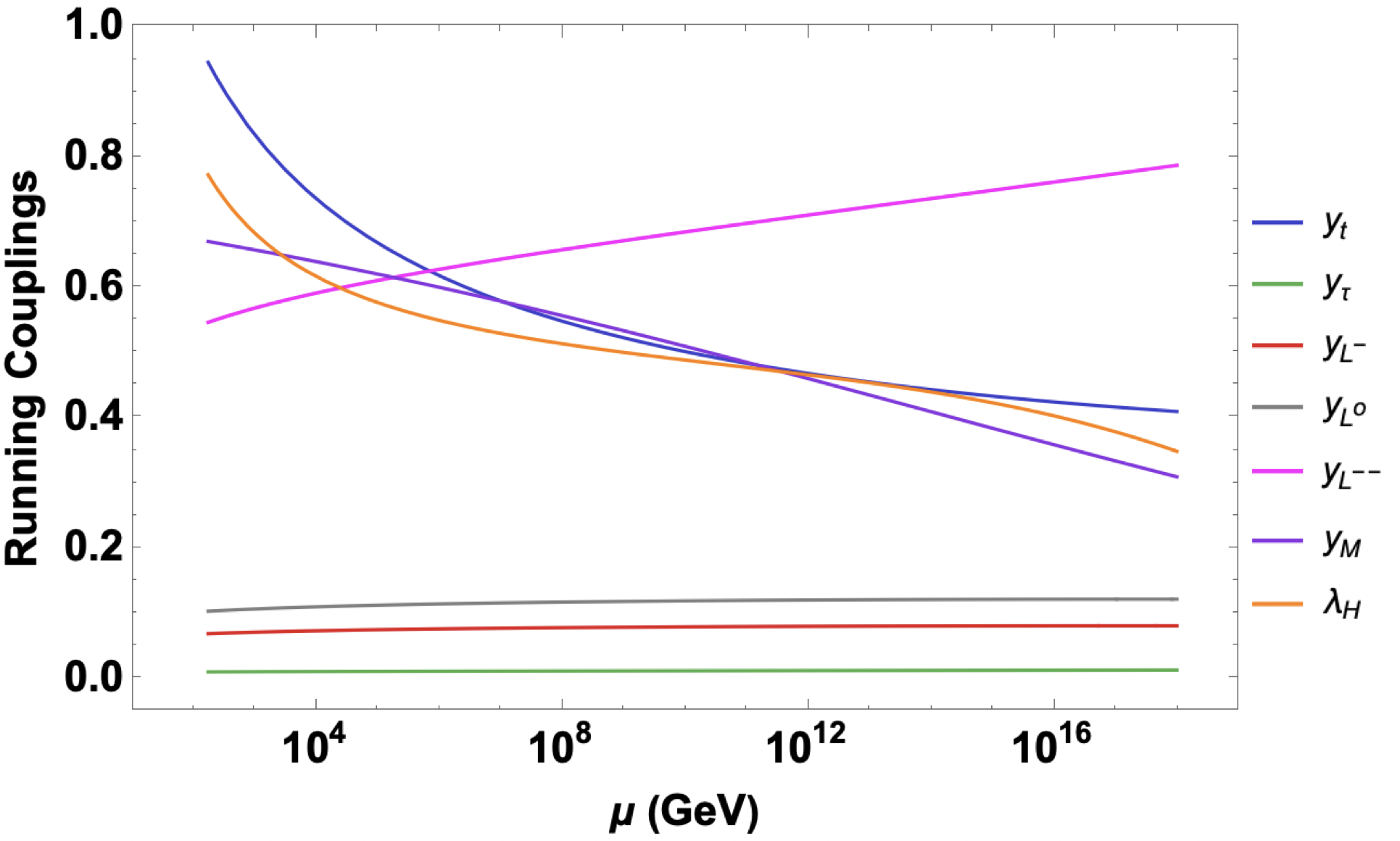}
		\caption{}
	\end{subfigure}
  \caption{The two loop RGE running of the Yukawa and the Higgs coupling for models with triplet vector-like leptons. We show triplet vector-like representation,  (a) ${\cal T}_1$, and  (b) ${\cal T}_2$. For triplet models, we have set $m_{L^0}=200$ GeV, $m_{L^-}=220$ GeV, $m_{L^+}=170$ GeV, $m_{L^{--}}=170$ GeV,  $\mu_0=m_t$, and $\sin \theta_L=0.1$.}
  \label{fig:2looprgetriplets}
\end{figure}

Even in the absence of VLL, the top Yukawa becomes smaller due to $-y_t^2g_3^4$ correction. Additionally all VLLs contribute negatively  to the overall running of the top Yukawa as seen in Figs. \ref{fig:2looprgesinglets}-\ref{fig:2looprgetriplets}. Hence, the overall decline of the top Yukawa coupling  directly affects the Higgs quartic coupling. For the two loop analysis, we found that all the VLL Yukawa couplings except $y_M$ have a basically insignificant effect on RGEs from any type of cross term $y_i^2y_j^4$,  due to their relatively small initial values. Thus we only emphasize the effects of the terms that  affect the two loop level and also strengthen the Higgs quartic coupling up to the Planck scale.  

For singlet models, the $SU(2)_W$ portal from VLLs does not contribute to the top Yukawa coupling, so the only gauge portal correction from two loop is $g_1^4y_{L^-}^2$, hence the overall difference of the top Yukawa running between two singlet models is almost insignificant, as seen in Fig. \ref{fig:2looprgesinglets}. We increased the masses of singlet VLL to $m_{L^0}=150$ GeV and $m_{L^-}=160$ GeV for ${\cal S}_1$ and ${\cal S}_2$ models as the greatest effect to $\lambda^{(2)}_{\text{RGE}}$ comes from $y_M^2y_L^2\lambda$, and it is additive for the number of VLLs in each representation. This is also seen from the two loop behaviour of $y_M$ throughout all the VLL multiplets. In the next leading order, the Yukawa couplings do not decrease, an effect expected for $y_{L^-}$ in ${\cal D}_2$ model because negative effects are fully associated with gauge couplings $g_1$ and $g_2$. On the other hand, the  coupling of Yukawa terms is negative at two loop order, thus forcing $y_M$ to decrease with respect to the energy scale. However, this feature is not manifested for the singlet models, as the dominance of one loop effects do not allow $y_M$ to run downwards because the self coupling effects of this Yukawa term become larger with respect to the number of VLLs in a model.  In an agreement with the one loop RGE results, ${\cal S}_2$ model allows larger VLL masses compared to ${\cal S}_1$ model, while  the Higgs quartic coupling is safer away from instability region, mainly due to the effect of $\lambda g_1^4$. Although the results from two loop RGE extended the parameter space for VLL masses, we found that $\lambda$ changes direction around $\mu\sim 10^{14}$ GeV when $m_{VLL}>170$ GeV and eventually hits the instability region.

The top Yukawa coupling becomes even smaller in the ${\cal D}_1$ model, leading to a greater boost for the parameter space surviving  the stability constraint by ameliorating the effect on the Higgs quartic coupling. This is mainly due to the fact that $y_t^2g_1^4$ effect is nine times stronger for ${\cal D}_2$ model. Thus, the relatively stronger top Yukawa coupling running in ${\cal D}_2$ exerts more pressure on $\lambda$ up to the Planck scale. The most striking feature of two loop corrections for the  Yukawa couplings is present in the models with non-SM like charges. The quartic and hexic Yukawa terms couple negatively  in two loop and the coefficients rapidly multiply in RGE level due to large number of VLL members. As the VLLs with exotic charges do not couple to SM leptons, their effect on the initial values to $y_{L^{+}}$ and $y_{L^{--}}$ is more effective in $y_{M}$ on both quadratic and quartic scale. Combined with the one loop $-Y^2g_1^2$ effect, $y_M$ runs downward throughout the entire spectrum in ${\cal D}_2$ as seen from Fig. \ref{fig:2looprgedoublets}. The consequence of this is to indirectly decrease the effect of $y_M^2y_L^2\lambda$ on $\lambda^{(2)}$, which can be counted as an additional reason why ${\cal D}_1$ allows larger parameter space to survive the vacuum stability condition. The two loop corrections to all doublet VLL RGEs extend the upper bound to $m_{VLL}\leq 290$ GeV. However, higher VLL masses either break the perturbativity of the Yukawa couplings, or destabilize the vacuum, depending on the leptonic mixing scale.

Finally, in triplet models, the absence of hypercharge terms in the ${\cal T}_1$ model uniquely determines the difference in the parameter space and in the running of couplings, whereas the Yukawa effects at the two loop level are similar, due to the particle content. The intricate interplay between coupled terms at the two loop RGE level results in the largest mass bounds for triplets. However, with fixed VLL inputs, the Higgs quartic coupling runs to its smallest values in triplet models as shown in Fig. \ref{fig:2looprgetriplets}. The two loop corrections open up more space, up to $m_{L^{0}}<290$ GeV and $m_{L^{-}}<310$ GeV compared to the one loop results in Fig. \ref{fig:rgespacetriplets}.
Thus the inclusion of two loop renormalization group equations in the analysis has proven to  improve the predictive accuracy of the model's coupling evolution. As demonstrated, the two loop RGE study not only refines the running of the couplings but also extends the available parameter space by up to 20$\%$ for all six $SU(2)$ representations. This extension is primarily due to the introduction of more complex couplings inherent in the two loop structure, which effectively capture higher-order effects missing in the one loop approximation.

\newpage
%%%%%%%%%%%%%%%%%%%%%%%%%%%%%%%%%%
\newpage
\section{Conclusions}
\label{sec:conclusion}
%%%%%%%%%%%%%%%%%%%%%%%%%%%%%%%%%%%%%%%%%%%%%%%%%
We have studied SM extensions with six different vector-like lepton representations. Our main focus has been to study the effects of new vector-like lepton fields on electroweak vacuum stability, while also satisfying perturbative unitarity conditions for all the couplings appearing in various representations. We concentrated on the answering the question of whether, unlike the case where vector-like quarks are introduced, and stability requires introduction of an additional scalar field, one can achieve stability with VLL only, and without introducing any additional fields. If this is possible, this would be a novel feature of the SM with VLLs.

Our analysis shows that, while VLLs can stabilize the Higgs quartic coupling up to the Planck scale under certain conditions, significant constraints on Yukawa couplings and VLL masses exist. Specifically, with a particular choice of Yukawa couplings to the SM Higgs field, and given that $\lambda y_M^2$ surpasses the large quartic terms at the RGE level, vector-like leptons have an allowed but limited parameter space that prevents the Higgs quartic coupling from diverging up to the Planck scale. The absence of the colour charge and the large number of VLLs lead to unconventional behaviour, causing Yukawa couplings to increase with the energy scale if the hypercharge  is not sufficiently large. If the scale of new physics is very high and the number of flavours $n_{F}$ is too small, the RG evolutions enter the non-perturbative region prematurely before rising again. Hence we assumed here VLLs masses of $\ll\mathcal{O}$(TeV) scale. Allowing all lepton generations from the SM to mix with VLLs could strengthen the gauge portal $\beta_{\Delta g_i}$. However, third-generation leptons are less constrained by flavour physics experiments compared to the first and second generations, where flavour-changing neutral currents (FCNCs) and lepton flavour violation (LFV) processes tightly constrain any mixings for the lighter generations. In addition, even if allowed, these mixings will be negligibly small due to the smallness of the first and second generation lepton masses.

The allowed strength of mixing between the SM and VLLs, according to RGE solutions, is determined by the presence of both neutral and charged VLL partners simultaneously, specifically the coexistence of mass splitting initial conditions. We found that large mixing is required if a model excludes both $L^0$ and $L^-$. Consequently, the relative weight on $\lambda(\mu)$ with respect to the largest Yukawa $y_M$ becomes smaller, while its initial condition starts at a lower value. Even though singlet VLLs have a similar RGE structure, the difference in hypercharge eventually leads to a wider parameter space allowed for the charged sector $m_{L^-}$. Our findings from RGE analysis are consistent with the data, considering the minimum bound $m_{VLL}\gtrsim110$ GeV. Doublet and triplet VLL models open up more allowed parameter space, as expected from additional terms at the RGE level, reaching around $m_{VLL}\sim 270$ GeV, while the leptonic mixing has a minimum bound $\sin\theta\gtrsim0.05$.  We further extended the RGE analysis to two loop corrections to check the amount of improvement to the running couplings in all models. The interplay between the number of higher order interactions generated an extra space for VLL masses, enlarging the maximum bound up to $20\%$, now reaching 310 GeV for charged VLL. The running of the Yukawa term $y_M$ that connects left- and right-handed part of a VLL played a major role in the two loop evolution, along with the gauge couplings $g_1,\, g_2$ with the latter two becoming  coupled at two loop level. Since new leptonic fields with exotic charges do not couple to the SM leptons through the Higgs, the upper bounds on their masses depend on perturbative unitarity conditions. Therefore, we assume $m_{L^{+,--}}<m_t$ in order to keep RG evolutions manageable. Although our renormalization analysis uses only two sets of free parameters, $m_{VLL}$ and $\sin\theta^{u,d}_{L,R}$, further studies can modify unconventional initial conditions for exotic $VLL_{L,R}$ fields in order to extend the allowed space from RGE solutions. 

We also scanned the oblique parameters, ensuring that mixing constraints from the Higgs channel are considered. Generally, the $\mathbb{T}$ does not constrain VLL masses at $\sin\theta=0.05$; however, the $\mathbb{T}$ parameter becomes restrictive with larger mixing $\sin\theta=0.1$, rendering $m_{VLL}>600$ GeV into the $3\sigma$ region. Moreover, the overall constraint from the $\mathbb{S}$ parameter relies more on hypercharge effects  than on the mixing variance within the same multiplet,  showing less severe differences as $\sin\theta$ increases. Limits from EWPO at a fixed mixing scale are more relaxed compared to vacuum stability bounds. Additionally, the parameter space obtained from RGE solutions does not exclude our results from the oblique parameters at $\sin\theta\leq0.1$ as there is always a solution in $m_{VLL}=[100,300]$ GeV throughout RGE level. Nevertheless, the oblique parameters rapidly deviate from the global fit for large mixings due to their direct dependence on mass splitting within multiplets. 

Future investigations may build on this study by incorporating analyses of non-perturbative effects and higher-order corrections and by examining more intricate VLL representations. Furthermore, exploring the implications of varied initial conditions and potential new physics beyond VLLs could yield additional insights. This study enhances our understanding of the influence of novel fermion fields on electroweak vacuum stability and offers valuable guidance for the search for new particles and interactions in forthcoming experimental endeavours.

Finally, we note that adding a scalar singlet field undoubtedly increases the allowed mass range for  VLL masses that satisfy RGE constraints (stability and perturbative unitarity) because the extra parameters from scalar sector allow to extend the range of VLL parameters.   This approach was explored before, and, in our opinion, does not have much different or newer features  than the model with vector-like quarks and an extra scalar, because apart from overall factors in RGE level, all that is different in Yukawa RGE sector are $g_3$ corrections terms, and these do not do not appear in scalar RGE terms. 

%%%%%%%%%%%%%%%%%%%%%%%%%%%%%%%%%%%%%%%%%%%%%%%%%%%%%
%%%%%%%%%%%%%%%%%%%%%%%%%%%%%%%%%%%%%%%%%%%%%%%%%%%%%
\begin{acknowledgments}
%%%%%%%%%%%%%%%%%%%%%%%%%%%%%%%%%%%%%%%%%%%%%%%%%%%%%%
 This work is funded in part by NSERC  under grant number SAP105354.
\end{acknowledgments}

 %%%%%%%%%%%%%%%%%%%%%%%%%%%%%%%%%%%%%%%%%%%%%%%%%%%%%
\newpage
 \section{Appendix}
 \label{sec:appendix}
 %%%%%%%%%%%%%%%%%%%%%%%%%%%%%%%%%%%%%%%%%%%%%%%%
 %%%%%%%%%%%%%%%%%%%%%%%%%%%%%%%%%%%%%%%%%%%%%%%%%%%%%%
In the subsequent appendices, we provide the renormalization group equations pertinent to the vector-like lepton (VLL) representations analyzed in this study. Additionally, we present electroweak couplings under VLL modifications to be used in the $\mathbb{S}$ and $\mathbb{T}$ parameters, along with the relevant Passarino-Veltman integrals employed in the calculations for thoroughness.
%%%%%%%%%%%%%%%%%%%%%%%%%%%%%%%%%%%%%
\subsection{RGEs for Vector-like Leptons}
\label{sec:apprge}
%%%%%%%%%%%%%%%%%%%%%%%%%%%%%%%%%%%%%%%%%%%%%%%%%%%%%%
\subsubsection{Singlet ${\cal S}_1~(L^0)$, $Y=0$}
%%%%%%%%%%%%%%%%%%%%%%%%%%%%%%%%%%%%%%%%%%%%%%%%%%%%%%
The relevant RGE for the Yukawa couplings are
\begin{eqnarray}
\label{eq:rgeS1singletfermion}
\frac{dy_t^2}{d \ln \mu^2}&=& \frac{y_t^2}{16 \pi^2}\left (\frac{9y_t^2}{2}+y_\tau^2+2y_{L^0}^2-\frac{17g_1^2}{20}-\frac{9g_2^2}{4}- 8g_3^2 \right)\, ,\nonumber\\
\frac{dy_\tau^2}{d \ln \mu^2}&=& \frac{y_\tau^2}{16 \pi^2}\left (\frac{5y_\tau^2}{2}+3y_t^2+2y_{L^0}^2-\frac{9g_1^2}{4}-\frac{9g_2^2}{4} \right)\, ,\nonumber\\
\frac{dy_{L^0}^2}{d \ln \mu^2}&=& \frac{y_{L^0}^2}{16 \pi^2}\left (3y_t^2+ \frac{5y_{L^0}^2}{2}+\frac{y_\tau^2}{2}+\frac{y_M^2}{4}-\frac{9g_1^2}{20}-\frac{9g_2^2}{4} \right)\, ,\nonumber\\
\frac{dy_M^2}{d \ln \mu^2}&=& \frac{y_M^2}{16 \pi^2}\left (y_{L^0}^2+\frac{5y_M^2}{2}\right).
\end{eqnarray}
The Higgs sector RGE, describing the interactions between the scalar boson and all fermions:
\begin{eqnarray}
\label{eq:rgeS1singletscalar}
\frac{d \lambda}{d \ln \mu^2}&=& \frac{1}{16 \pi^2} \left[ \lambda \left (-\frac{9g_{1}^{2}}{10} - \frac{9g_{2}^{2}}{2}+6y_t^2+4y_{L^0}^2+4y_{M}^2+2y_{\tau}^2\right) +12\lambda_{1}^{2}+\frac{27g_1^4}{400}+\frac{9g_2^4}{16}+\frac{9g_1^2g_2^2}{40}
 \right. \nonumber \\
 &-& \left. 6y_t^4-4y_\tau^4-4y_{L^0}^4-4y_M^4 -4y_M^2y_{L^0}^2\right ]
 \end{eqnarray}
%%%%%%%%%%%%%%%%%%%%%%%%%%%%%%%%%%%%%%%%%%%%%%%%%%%%%%
\subsubsection{Singlet ${\cal S}_2~(L^-)$, $Y=-1$}
%%%%%%%%%%%%%%%%%%%%%%%%%%%%%%%%%%%%%%%%%%%%%%%%%%%%%%
The relevant RGE for the Yukawa couplings are
\begin{eqnarray}
\label{eq:rgeS2singletfermion}
\frac{dy_t^2}{d \ln \mu^2}&=& \frac{y_t^2}{16 \pi^2}\left (\frac{9y_t^2}{2}+y_\tau^2+2y_{L^-}^2-\frac{17g_1^2}{20}-\frac{9g_2^2}{4}- 8g_3^2 \right)\, ,\nonumber\\
\frac{dy_\tau^2}{d \ln \mu^2}&=& \frac{y_\tau^2}{16 \pi^2}\left (\frac{5y_\tau^2}{2}+3y_t^2+2y_{L^-}^2-\frac{9g_1^2}{4}-\frac{9g_2^2}{4} \right)\, ,\nonumber\\
\frac{dy_{L^-}^2}{d \ln \mu^2}&=& \frac{y_{L^-}^2}{16 \pi^2}\left (3y_t^2+ \frac{5y_{L^-}^2}{2}+\frac{y_\tau^2}{2}+\frac{y_M^2}{4}-\frac{9g_1^2}{4}-\frac{9g_2^2}{4} \right)\, ,\nonumber\\
\frac{dy_M^2}{d \ln \mu^2}&=& \frac{y_M^2}{16 \pi^2}\left (y_{L^-}^2+\frac{5y_M^2}{2}-\frac{18}{5}g_1^2\right).
\end{eqnarray}
The Higgs sector RGE, describing the interactions between the scalar boson and all fermions:
\begin{eqnarray}
\label{eq:rgeS2singletscalar}
\frac{d \lambda}{d \ln \mu^2}&=& \frac{1}{16 \pi^2} \left[ \lambda \left (-\frac{9g_{1}^{2}}{10} - \frac{9g_{2}^{2}}{2}+6y_t^2+4y_{L^-}^2+4y_{M}^2+2y_{\tau}^2\right) +12\lambda_{1}^{2}+\frac{27g_1^4}{400}+\frac{9g_2^4}{16}+\frac{9g_1^2g_2^2}{40}
 \right. \nonumber \\
 &-& \left. 6y_t^4-4y_\tau^4-4y_{L^-}^4-4y_M^4-4y_\tau^2y_{L^-}^2 -4y_M^2y_{L^-}^2\right ]
 \end{eqnarray}
 Finally the coupling constants gain additional terms due to the new fermion, for both models ${\cal S}_1,~{\cal S}_2$ with singlet fermions as follows:
\begin{eqnarray}
\frac{dg_1^2}{d \ln \mu^2}&=&\frac{g_1^4}{16 \pi^2}\left(\frac{41}{10}+\frac{4}{5}\right )\, ,\qquad (\Delta g^{S_1}_1=0) \nonumber \\
\frac{dg_2^2}{d \ln \mu^2}&=&\frac{g_2^4}{16 \pi^2}\left( -\frac{19}{6} \right ) \, , \nonumber \\
\frac{dg_3^2}{d \ln \mu^2}&=&\frac{g_3^4}{16 \pi^2}\left(-7\right )\, .
\end{eqnarray}
%%%%%%%%%%%%%%%%%%%%%%%%%%%%%%%%%%%%%%%%%%%%%%%%%%%%%%
\subsubsection{Doublet ${\cal D}_1$ $(L^0,L^-), \,Y=-1/2$}
%%%%%%%%%%%%%%%%%%%%%%%%%%%%%%%%%%%%%%%%%%%%%%%%%%%%%%
The relevant RGE for the Yukawa couplings are
\begin{eqnarray}
\label{eq:rgeD1doubletfermion}
\frac{dy_t^2}{d \ln \mu^2}&=& \frac{y_t^2}{16 \pi^2}\left (\frac{9y_t^2}{2}+y_\tau^2+2y_{L^-}^2+2y_{L^0}^2-\frac{17g_1^2}{20}-\frac{9g_2^2}{4}- 8g_3^2 \right)\, ,\nonumber\\
\frac{dy_\tau^2}{d \ln \mu^2}&=& \frac{y_\tau^2}{16 \pi^2}\left (\frac{5y_\tau^2}{2}+\frac{5y_{L^-}^2}{2}+3y_t^2+\frac{y_{L^0}^2}{2}-\frac{9g_1^2}{4}-\frac{9g_2^2}{4} \right)\, ,\nonumber\\
\frac{dy_{L^0}^2}{d \ln \mu^2}&=& \frac{y_{L^0}^2}{16 \pi^2}\left (3y_t^2+ \frac{5y_{L^0}^2}{2}+\frac{y_{L^-}^2}{2}+y_\tau^2+\frac{y_M^2}{2}-\frac{9g_1^2}{4}-\frac{9g_2^2}{4} \right)\, ,\nonumber\\
\frac{dy_{L^-}^2}{d \ln \mu^2}&=& \frac{y_{L^-}^2}{16 \pi^2}\left (3y_t^2+ \frac{5y_{L^-}^2}{2}+\frac{y_{L^0}^2}{2}+y_\tau^2+\frac{y_M^2}{2}-\frac{15g_1^2}{4}-\frac{9g_2^2}{4} \right)\, ,\nonumber\\
\frac{dy_M^2}{d \ln \mu^2}&=& \frac{y_M^2}{16 \pi^2}\left (y_{L^-}^2+y_{L^0}^2+\frac{7y_M^2}{2}-\frac{9}{10}g_1^2-\frac{9}{2}g_2^2\right).
\end{eqnarray}
The Higgs sector RGE, describing the interactions between the scalar boson and all fermions:
\begin{eqnarray}
\label{eq:rgeD1doubletscalar}
\frac{d \lambda}{d \ln \mu^2}&=& \frac{1}{16 \pi^2} \left[ \lambda \left (-\frac{9g_{1}^{2}}{5} -9g_2^2+6y_t^2+4y_{L^-}^2+4y_{L^0}^2+4y_{M}^2+2y_{\tau}^2\right) +12\lambda_{1}^{2}+\frac{27g_1^4}{200}+\frac{9g_2^4}{8}+\frac{9g_1^2g_2^2}{20}
 \right. \nonumber \\
 &-& \left. 6y_t^4-4y_\tau^4-4y_{L^-}^4-4y_{L^0}^4-4y_M^4-4y_\tau^2y_{L^-}^2 -4y_M^2y_{L^-}^2-4y_M^2y_{L^0}^2\right ]
 \end{eqnarray}

%%%%%%%%%%%%%%%%%%%%%%%%%%%%%%%%%%%%%%%%%%%%%%%%%%%%%%
%%%%%%%%%%%%%%%%%%%%%%%%%%%%%%%%%%%%%%%%%%%%%%%%%%%%%%
\subsubsection{Doublet ${\cal D}_2$ $(L^-,L^{--}), \,Y=-3/2$}
%%%%%%%%%%%%%%%%%%%%%%%%%%%%%%%%%%%%%%%%%%%%%%%%%%%%%%
The relevant RGE for the Yukawa couplings are
\begin{eqnarray}
\label{eq:rgeD2doubletfermion}
\frac{dy_t^2}{d \ln \mu^2}&=& \frac{y_t^2}{16 \pi^2}\left (\frac{9y_t^2}{2}+y_\tau^2+2y_{L^-}^2+2y_{L^{--}}^2-\frac{17g_1^2}{20}-\frac{9g_2^2}{4}- 8g_3^2 \right)\, ,\nonumber\\
\frac{dy_\tau^2}{d \ln \mu^2}&=& \frac{y_\tau^2}{16 \pi^2}\left (\frac{5y_\tau^2}{2}+\frac{5y_{L^{--}}^2}{2}+3y_t^2+\frac{y_{L^-}^2}{2}-\frac{9g_1^2}{4}-\frac{9g_2^2}{4} \right)\, ,\nonumber\\
\frac{dy_{L^-}^2}{d \ln \mu^2}&=& \frac{y_{L^-}^2}{16 \pi^2}\left (3y_t^2+ \frac{5y_{L^-}^2}{2}+\frac{y_{L^{--}}^2}{2}+y_\tau^2+\frac{y_M^2}{2}-\frac{27g_1^2}{2}-\frac{9g_2^2}{4} \right)\, ,\nonumber\\
\frac{dy_{L^{--}}^2}{d \ln \mu^2}&=& \frac{y_{L^{--}}^2}{16 \pi^2}\left (3y_t^2+ \frac{5y_{L^{--}}^2}{2}+\frac{y_{L^-}^2}{2}+y_\tau^2+\frac{y_M^2}{2}-\frac{45g_1^2}{4}-\frac{9g_2^2}{4} \right)\, ,\nonumber\\
\frac{dy_M^2}{d \ln \mu^2}&=& \frac{y_M^2}{16 \pi^2}\left (y_{L^-}^2+y_{L^{--}}^2+\frac{7y_M^2}{2}-\frac{27}{2}g_1^2-\frac{9}{4}g_2^2\right).
\end{eqnarray}
The Higgs sector RGE, describing the interactions between the scalar boson and all fermions:
\begin{eqnarray}
\label{eq:rgeD2doubletscalar}
\frac{d \lambda}{d \ln \mu^2}&=& \frac{1}{16 \pi^2} \left[ \lambda \left (-\frac{9g_{1}^{2}}{5} -9g_2^2+6y_t^2+4y_{L^-}^2+4y_{L^{--}}^2+4y_{M}^2+2y_{\tau}^2\right) +12\lambda_{1}^{2}+\frac{27g_1^4}{200}+\frac{9g_2^4}{8}+\frac{9g_1^9g_2^2}{20}
 \right. \nonumber \\
 &-& \left. 6y_t^4-4y_\tau^4-4y_{L^-}^4-4y_{L^{--}}^4-4y_M^4-4y_\tau^2y_{L^-}^2 -4y_M^2y_{L^-}^2-4y_M^2y_{L^{--}}^2\right ]
 \end{eqnarray}
The coupling constants gain additional terms due to the new fermion in all doublet models as follows:
\begin{eqnarray}
\frac{dg_1^2}{d \ln \mu^2}&=&\frac{g_1^4}{16 \pi^2}\left(\frac{41}{10}+\frac{18}{5}\right )\, ,\qquad (\Delta g^{D_2}_1=27/5)  \nonumber \\
\frac{dg_2^2}{d \ln \mu^2}&=&\frac{g_2^4}{16 \pi^2}\left( -\frac{19}{6}+\frac{7}{3} \right ) \, , \nonumber \\
\frac{dg_3^2}{d \ln \mu^2}&=&\frac{g_3^4}{16 \pi^2}\left(-7\right )\, .
\end{eqnarray}
%%%%%%%%%%%%%%%%%%%%%%%%%%%%%%%%%%%%%%%%%%%%%%%%%%%%%%

\subsubsection{Triplet ${\cal T}_1$ $(L^+, L^0, L^-), \,Y=0$}
%%%%%%%%%%%%%%%%%%%%%%%%%%%%%%%%%%%%%%%%%%%%%%%%%%%%%%

The relevant RGE for the Yukawa couplings are
\begin{eqnarray}
\label{eq:rgeX1tripletfermion}
\frac{dy_t^2}{d \ln \mu^2}&=& \frac{y_t^2}{16 \pi^2}\left (\frac{9y_t^2}{2}+y_\tau^2+2y_{L^+}^2+2y_{L^-}^2+2y_{L^0}^2-\frac{17g_1^2}{20}-\frac{9g_2^2}{4}- 8g_3^2 \right)\, ,\nonumber\\
\frac{dy_\tau^2}{d \ln \mu^2}&=& \frac{y_\tau^2}{16 \pi^2}\left (\frac{5y_\tau^2}{2}+\frac{5y_{L^0}^2}{2}+\frac{5y_{L^-}^2}{2}+3y_t^2+\frac{y_{L^+}^2}{2}-\frac{9g_1^2}{4}-\frac{9g_2^2}{4} \right)\, ,\nonumber\\
\frac{dy_{L^+}^2}{d \ln \mu^2}&=& \frac{y_{L^+}^2}{16 \pi^2}\left (3y_t^2+ \frac{5y_{L^+}^2}{2}+ \frac{y_{L^0}^2}{2}+\frac{y_{L^-}^2}{2}+y_\tau^2+\frac{y_M^2}{4}-\frac{9g_1^2}{20}-\frac{9g_2^2}{4} \right)\, ,\nonumber\\
\frac{dy_{L^0}^2}{d \ln \mu^2}&=& \frac{y_{L^0}^2}{16 \pi^2}\left (3y_t^2+ \frac{5y_{L^0}^2}{2}+\frac{y_{L^+}^2}{2}+\frac{y_{L^-}^2}{2}+y_\tau^2+\frac{y_M^2}{4}-\frac{9g_1^2}{20}-\frac{9g_2^2}{4} \right)\, ,\nonumber\\
\frac{dy_{L^-}^2}{d \ln \mu^2}&=& \frac{y_{L^-}^2}{16 \pi^2}\left (3y_t^2+ \frac{5y_{L^-}^2}{2}+\frac{y_{L^0}^2}{2}+\frac{y_{L^+}^2}{2}+y_\tau^2+\frac{y_M^2}{4}-\frac{9g_1^2}{20}-\frac{9g_2^2}{4} \right)\, ,\nonumber\\
\frac{dy_M^2}{d \ln \mu^2}&=& \frac{y_M^2}{16 \pi^2}\left (y_{L^+}^2+y_{L^-}^2+y_{L^0}^2+\frac{9y_M^2}{2}-12g_2^2\right).
\end{eqnarray}
The Higgs sector RGE, describing the interactions between the scalar boson and all fermions:
\begin{eqnarray}
\label{eq:rgeX1tripletscalar}
\frac{d \lambda}{d \ln \mu^2}&=& \frac{1}{16 \pi^2} \left[ \lambda \left (-\frac{9g_{1}^{2}}{10} -\frac{9g_2^2}{2}+6y_t^2+4y_{L^-}^2+4y_{L^+}^2+4y_{L^0}^2+4y_{M}^2+2y_{\tau}^2\right) +12\lambda_{1}^{2}+\frac{27g_1^4}{400}+\frac{9g_2^4}{16}+\frac{9g_1^2g_2^2}{20}
 \right. \nonumber \\
 &-& \left. 6y_t^4-4y_\tau^4-4y_{L^+}^4-4y_{L^-}^4-4y_{L^0}^4-4y_M^4-4y_\tau^2y_{L^-}^2-4y_M^2y_{L^+}^2 -4y_M^2y_{L^-}^2-4y_M^2y_{L^0}^2\right ]
 \end{eqnarray}

%%%%%%%%%%%%%%%%%%%%%%%%%%%%%%%%%%%%%%%%%%%%%%%%%%%%%%
\subsubsection{Triplet ${\cal T}_2$ $( T, B, Y), \, Y=-1$}
%%%%%%%%%%%%%%%%%%%%%%%%%%%%%%%%%%%%%%%%%%%%%%%%%%%%%%

\begin{eqnarray}
\label{eq:rgeX2tripletfermion}
\frac{dy_t^2}{d \ln \mu^2}&=& \frac{y_t^2}{16 \pi^2}\left (\frac{9y_t^2}{2}+y_\tau^2+2y_{L^{--}}^2+2y_{L^-}^2+2y_{L^0}^2-\frac{17g_1^2}{20}-\frac{9g_2^2}{4}- 8g_3^2 \right)\, ,\nonumber\\
\frac{dy_\tau^2}{d \ln \mu^2}&=& \frac{y_\tau^2}{16 \pi^2}\left (\frac{5y_\tau^2}{2}+\frac{5y_{L^-}^2}{2}+\frac{5y_{L^{--}}^2}{2}+3y_t^2+\frac{y_{L^0}^2}{2}-\frac{9g_1^2}{4}-\frac{9g_2^2}{4} \right)\, ,\nonumber\\
\frac{dy_{L^0}^2}{d \ln \mu^2}&=& \frac{y_{L^0}^2}{16 \pi^2}\left (3y_t^2+ \frac{5y_{L^0}^2}{2}+ \frac{y_{L^-}^2}{2}+\frac{y_{L^{--}}^2}{2}+y_\tau^2+\frac{y_M^2}{4}-\frac{9g_1^2}{4}-\frac{9g_2^2}{4} \right)\, ,\nonumber\\
\frac{dy_{L^-}^2}{d \ln \mu^2}&=& \frac{y_{L^-}^2}{16 \pi^2}\left (3y_t^2+ \frac{5y_{L^-}^2}{2}+\frac{y_{L^0}^2}{2}+\frac{y_{L^{--}}^2}{2}+y_\tau^2+\frac{y_M^2}{4}-\frac{9g_1^2}{4}-\frac{9g_2^2}{4} \right)\, ,\nonumber\\
\frac{dy_{L^{--}}^2}{d \ln \mu^2}&=& \frac{y_{L^{--}}^2}{16 \pi^2}\left (3y_t^2+ \frac{5y_{L^{--}}^2}{2}+\frac{y_{L^0}^2}{2}+\frac{y_{L^-}^2}{2}+y_\tau^2+\frac{y_M^2}{4}-\frac{9g_1^2}{4}-\frac{9g_2^2}{4} \right)\, ,\nonumber\\
\frac{dy_M^2}{d \ln \mu^2}&=& \frac{y_M^2}{16 \pi^2}\left (y_{L^0}^2+y_{L^-}^2+y_{L^{--}}^2+\frac{9y_M^2}{2}-\frac{81}{20}g_1^2-12g_2^2\right).
\end{eqnarray}
The Higgs sector RGE, describing the interactions between the scalar boson and all fermions:
\begin{eqnarray}
\label{eq:rgeX2tripletscalar}
\frac{d \lambda}{d \ln \mu^2}&=& \frac{1}{16 \pi^2} \left[ \lambda \left (-\frac{9g_{1}^{2}}{5} -9g_2^2+6y_t^2+4y_{L^-}^2+4y_{L^{--}}^2+4y_{L^0}^2+4y_{M}^2+2y_{\tau}^2\right) +12\lambda_{1}^{2}+\frac{27g_1^4}{400}+\frac{9g_2^4}{16}+\frac{9g_1^2g_2^2}{20}
 \right. \nonumber \\
 &-& \left. 6y_t^4-4y_\tau^4-4y_{L^{--}}^4-4y_{L^-}^4-4y_{L^0}^4-4y_M^4-4y_\tau^2y_{L^-}^2-4y_M^2y_{L^{--}}^2 -4y_M^2y_{L^-}^2-4y_M^2y_{L^0}^2\right ]
 \end{eqnarray}

The coupling constants gain additional terms due to the new fermion in all doublet models as follows:
\begin{eqnarray}
\frac{dg_1^2}{d \ln \mu^2}&=&\frac{g_1^4}{16 \pi^2}\left(\frac{41}{10}+\frac{12}{5}\right )\, , \qquad (\Delta g^{{\cal T}_1}_1=0) \nonumber \\
\frac{dg_2^2}{d \ln \mu^2}&=&\frac{g_2^4}{16 \pi^2}\left( -\frac{19}{6}+\frac{16}{3} \right ) \, , \nonumber \\
\frac{dg_3^2}{d \ln \mu^2}&=&\frac{g_3^4}{16 \pi^2}\left(-7\right )\, .
\end{eqnarray}

\subsection{Electroweak couplings of VLL and the SM leptonss}
\label{sec:EWcouplings}
Couplings of the SM gauge bosons to fermions are uniquely modified with new mass eigenstates of vector-like leptons in terms of mass splitting expressions. We give the complete list of electroweak couplings used in calculation of  Peskin-Takeuchi parameters.
\subsubsection{Singlet ${\cal S}_1~(L^0)$, $Y=0$}
\label{eq:EWCsingletS1}
\begin{align}
\Omega^L_{W\nu\tau}&=\frac{ec_L^u}{\sqrt{2}s_W}           &  \Omega^L_{Z\nu\nu} &=\frac{ec_L^{u^2}}{2s_Wc_W}        \nonumber   \\
\Omega^R_{W\nu\tau}&=0          &  \Omega^R_{Z\nu\nu}&=0  \nonumber  \\
\Omega^L_{W\tau L^0}&=\frac{es_L^u}{\sqrt{2}s_W}  &  \Omega^L_{Z\tau\tau}&=\frac{e}{2s_Wc_W}(-1+2s_W^2)    \nonumber \\
\Omega^R_{W\tau L^0}&=0 &      \Omega^R_{Z\tau\tau}&=\frac{es_W}{c_W}  \nonumber  \\
&& \Omega^L_{Z\nu L^0} &=\frac{es_L^uc_L^u}{2s_Wc_W} \nonumber \\
&& \Omega^R_{Z\nu L^0} &=0 \nonumber \\
&& \Omega^L_{ZL^0L^0} &=\frac{es_L^{u^2}}{2s_Wc_W} \nonumber \\
&& \Omega^R_{ZL^0L^0} &=0
\end{align}

\subsubsection{Singlet ${\cal S}_2~(L^-)$, $Y=-1$}
\label{eq:EWCsingletS2}
\begin{align}
\Omega^L_{W\nu\tau}&=\frac{ec_L^d}{\sqrt{2}s_W}           &  \Omega^L_{Z\nu\nu} &=\frac{e}{2s_Wc_W}          \nonumber   \\
\Omega^R_{W\nu\tau}&=0          &  \Omega^R_{Z\nu\nu}&=0  \nonumber  \\
\Omega^L_{W\nu L^-}&=\frac{es_L^d}{\sqrt{2}s_W}  &  \Omega^L_{Z\tau\tau}&=\frac{e}{2s_Wc_W}(-c_L^{d^2}+2s_W^2)    \nonumber \\
\Omega^R_{W\nu L^-}&=0 &      \Omega^R_{Z\tau\tau}&=\frac{es_W}{c_W}  \nonumber  \\
&& \Omega^L_{Z\tau L^-} &=-\frac{es_L^dc_L^d}{2s_Wc_W} \nonumber \\
&& \Omega^R_{Z\tau L^-} &=0 \nonumber \\
&& \Omega^L_{ZL^-L^-} &=\frac{e}{2s_Wc_W}(-s_L^{d^2}+2s_W^2) \nonumber \\
&& \Omega^R_{ZL^-L^-} &=\frac{es_W}{c_W}
\end{align}

\subsubsection{Doublet ${\cal D}_1$ $(L^0,L^-), \,Y=-1/2$}
\label{eq:EWCdoubletD1}
\begin{align}
\Omega^L_{W\nu\tau}&=\frac{e}{\sqrt{2}s_W}(c_L^uc_L^d+s_L^us_L^d)           &  \Omega^L_{Z\nu\nu} &=\frac{e}{2s_Wc_W}         \nonumber   \\
\Omega^R_{W\nu\tau}&=\frac{es_R^us_R^d}{\sqrt{2}s_W}        &  \Omega^R_{Z\nu\nu}&=\frac{es_R^{u^2}}{2s_Wc_W}  \nonumber  \\
\Omega^L_{W\tau L^0}&=\frac{e}{\sqrt{2}s_W}(c_L^ds_L^u-c_L^us_L^d)  &  \Omega^L_{Z\tau\tau}&=\frac{e}{2s_Wc_W}(-1+2s_W^2)    \nonumber \\
\Omega^R_{W\tau L^0}&=-\frac{ec_R^us_R^d}{\sqrt{2}s_W} &      \Omega^R_{Z\tau\tau}&=\frac{e}{2s_Wc_W}(-s_R^{d^2}+2s_W^2)   \nonumber  \\
\Omega^L_{W\nu L^-}&=\frac{e}{\sqrt{2}s_W}(c_L^us_L^d-c_L^ds_L^u)& \Omega^L_{Z\nu L^0} &=0\nonumber \\
\Omega^R_{W\nu L^-}&=-\frac{ec_R^ds_R^u}{\sqrt{2}s_W}& \Omega^R_{Z\nu L^0} &=-\frac{ec_R^us_R^u}{2s_Wc_W} \nonumber \\
\Omega^L_{WL^0L^-}&=\frac{e}{\sqrt{2}s_W}(c_L^uc_L^d+s_L^us_L^d) & \Omega^L_{Z\tau L^-} &=0 \nonumber \\
\Omega^R_{WL^0L^-}&=\frac{ec_R^uc_R^d}{\sqrt{2}s_W} & \Omega^R_{Z\tau L^-} &=\frac{ec_R^ds_R^d}{2s_Wc_W} \nonumber \\
&& \Omega^L_{ZL^0L^0} &=\frac{e}{2s_Wc_W}          \nonumber   \\
&& \Omega^R_{ZL^0L^0}&=\frac{ec_R^{u^2}}{2s_Wc_W}  \nonumber  \\
&& \Omega^L_{ZL^-L^-}&=\frac{e}{2s_Wc_W}(-1+2s_W^2)    \nonumber \\
&&\Omega^R_{ZL^-L^-}&=\frac{e}{2s_Wc_W}(-c_R^{d^2}+2s_W^2)
\end{align}

\subsubsection{Doublet ${\cal D}_2$ $(L^-,L^{--}), \,Y=-3/2$}
\label{eq:EWCdoubletD2}
\begin{align}
\Omega^L_{W\nu\tau}&=\frac{ec_L^d}{\sqrt{2}s_W}           &  \Omega^L_{Z\nu\nu} &=\frac{e}{2s_Wc_W}         \nonumber   \\
\Omega^R_{W\nu\tau}&=0        &  \Omega^R_{Z\nu\nu}&=0 \nonumber  \\
\Omega^L_{W\nu L^-}&=\frac{es_L^d}{\sqrt{2}s_W}  &  \Omega^L_{Z\tau\tau}&=\frac{e}{2s_Wc_W}(2s_L^{d^2}+2s_W^2-1)    \nonumber \\
\Omega^R_{W\nu L^-}&=0 &      \Omega^R_{Z\tau\tau}&=\frac{e}{2s_Wc_W}(s_R^{d^2}+2s_W^2)   \nonumber  \\
\Omega^L_{W\tau L^{--}}&=-\frac{es_L^d}{\sqrt{2}s_W}& \Omega^L_{Z\tau L^-} &=-\frac{ec_L^ds_L^d}{s_Wc_W}\nonumber \\
\Omega^R_{W\tau L^{--}}&=-\frac{es_R^d}{\sqrt{2}s_W}& \Omega^R_{Z\tau L^-} &=-\frac{ec_R^ds_R^d}{2s_Wc_W} \nonumber \\
\Omega^L_{WL^-L^{--}}&=\frac{ec_L^d}{\sqrt{2}s_W} & \Omega^L_{ZL^- L^-} &=\frac{e}{2s_Wc_W}(2c_L^{d^2}+2s_W^2-1) \nonumber \\
\Omega^R_{WL^-L^{--}}&=\frac{ec_R^d}{\sqrt{2}s_W} & \Omega^R_{ZL^- L^-} &=\frac{e}{2s_Wc_W}(c_R^{d^2}+2s_W^2) \nonumber \\
&& \Omega^L_{ZL^{--}L^{--}}&=\frac{e}{2s_Wc_W}(-1+4s_W^2)    \nonumber \\
&&\Omega^R_{ZL^{--}L^{--}}&=\frac{e}{2s_Wc_W}(-1+4s_W^2)
\end{align}

\subsubsection{Triplet ${\cal T}_1$ $(L^+, L^0, L^-), \,Y=0$}
\label{eq:EWCtripletX1}
\begin{align}
\Omega^L_{W\nu\tau}&=\frac{e}{\sqrt{2}s_W}(c_L^uc_L^d+\sqrt{2}s_L^us_L^d)        &  \Omega^L_{Z\nu\nu} &=\frac{ec_L^{u^2}}{2s_Wc_W}          \nonumber   \\
\Omega^R_{W\nu\tau}&=\frac{es_R^us_R^d}{s_W}        &  \Omega^R_{Z\nu\nu}&=0  \nonumber  \\
\Omega^L_{W\nu L^+}&=-\frac{es_L^u}{s_W}        &  \Omega^L_{Z\tau\tau} &=-\frac{e}{2s_Wc_W}(s_L^{d^2}-2s_W^2+1)  \nonumber \\
\Omega^R_{W\nu L^+}&=-\frac{es_R^u}{s_W}      &  \Omega^R_{Z\tau\tau}&=-\frac{e}{2s_Wc_W}(2s_R^{d^2}-2s_W^2)    \nonumber  \\
\Omega^L_{W\nu L^-}&=\frac{e}{\sqrt{2}s_W}(c_L^us_L^d-\sqrt{2}c_L^ds_L^u) &  \Omega^L_{Z\nu L^0} &=\frac{ec_L^us_L^u}{2s_Wc_W}\nonumber \\
\Omega^R_{W\nu L^-}&=-\frac{ec_R^ds_R^u}{s_W} &  \Omega^R_{Z\nu L^0} &=0\nonumber \\
\Omega^L_{W\tau L^0}&=\frac{e}{\sqrt{2}s_W}(c_L^ds_L^u-\sqrt{2}c_L^us_L^d)  & \Omega^L_{Z\tau L^-} &=\frac{ec_L^ds_L^d}{2s_Wc_W}\nonumber \\
\Omega^R_{W\tau L^0}&=-\frac{ec_R^us_R^d}{s_W} & \Omega^R_{Z\tau L^-} &=\frac{ec_R^ds_R^d}{s_Wc_W}\nonumber \\
\Omega^L_{WL^+L^0}&=\frac{ec_L^u}{s_W} & \Omega^L_{ZL^+L^+} &=\frac{e}{s_Wc_W}(1-s_W^2) \nonumber \\
\Omega^R_{WL^+L^0}&=\frac{ec_R^u}{s_W} &  \Omega^R_{ZL^+L^+}&=\frac{e}{s_Wc_W}(1-s_W^2)  \nonumber  \\
\Omega^L_{WL^0L^-}&=\frac{e}{\sqrt{2}s_W}(s_L^us_L^d+\sqrt{2}c_L^uc_L^d) &  \Omega^L_{ZL^0L^0} &=\frac{es_L^{u^2}}{2s_Wc_W} \nonumber \\ 
\Omega^R_{WL^0L^-}&=\frac{ec_R^uc_R^d}{s_W} &  \Omega^R_{ZL^0L^0} &=0 \nonumber \\ 
&&\Omega^L_{ZL^-L^-}&=-\frac{e}{2s_Wc_W}(c_L^{d^2}-2s_W^2+1)    \nonumber \\
&&\Omega^R_{ZL^-L^-}&=-\frac{e}{2s_Wc_W}(2c_R^{d^2}-2s_W^2)
\end{align}

\subsubsection{Triplet ${\cal T}_2$ $(L^0, L^-,L^{--} ), \, Y=-1$}
\label{eq:EWCtripletX2}
\begin{align}
\Omega^L_{W\nu\tau}&=\frac{e}{\sqrt{2}s_W}(c_L^uc_L^d+\sqrt{2}s_L^us_L^d)        &  \Omega^L_{Z\nu\nu} &=\frac{e}{2s_Wc_W}(1+s_L^{u^2})          \nonumber   \\
\Omega^R_{W\nu\tau}&=\frac{es_R^us_R^d}{s_W}        &  \Omega^R_{Z\nu\nu}&=\frac{es_R^{u^2}}{s_Wc_W}   \nonumber  \\
\Omega^L_{W\nu L^-}&=\frac{e}{\sqrt{2}s_W}(c_L^us_L^d-\sqrt{2}s_L^uc_L^d)        &  \Omega^L_{Z\tau\tau} &=\frac{e}{2s_Wc_W}(-c_L^{d^2}+2s_W^2)  \nonumber \\
\Omega^R_{W\nu L^-}&=-\frac{es_R^uc_R^d}{s_W}      &  \Omega^R_{Z\tau\tau}&=\frac{es_W}{c_W}   \nonumber  \\
\Omega^L_{W\tau L^0}&=\frac{e}{\sqrt{2}s_W}(c_L^ds_L^u-\sqrt{2}c_L^us_L^d) &  \Omega^L_{Z\nu L^0} &=-\frac{es_L^uc_L^u}{2s_Wc_W}\nonumber \\
\Omega^R_{W\tau L^0}&=-\frac{es_R^dc_R^u}{s_W} &  \Omega^R_{Z\nu L^0} &=-\frac{es_R^uc_R^u}{s_Wc_W}\nonumber \\
\Omega^L_{W\tau L^{--}}&=-\frac{es_L^d}{s_W}  & \Omega^L_{Z\tau L^-} &=-\frac{ec_L^ds_L^d}{2s_Wc_W}\nonumber \\
\Omega^R_{W\tau L^{--}}&=-\frac{es_R^d}{s_W} & \Omega^R_{Z\tau L^-} &=0\nonumber \\
\Omega^L_{WL^0L^-}&=\frac{e}{\sqrt{2}s_W}(s_L^us_L^d+\sqrt{2}c_L^uc_L^d) & \Omega^L_{ZL^0L^0} &=\frac{e}{2s_Wc_W}(1+c_L^{u^2}) \nonumber \\
\Omega^R_{WL^0L^-}&=\frac{ec_R^uc_R^d}{s_W} &  \Omega^R_{ZL^0L^0}&=\frac{ec_R^{u^2}}{s_Wc_W}  \nonumber  \\
\Omega^L_{WL^-L^{--}}&=\frac{ec_L^d}{s_W} &  \Omega^L_{ZL^-L^-} &=\frac{e}{2s_Wc_W}(-s_L^{d^2}+2s_W^2) \nonumber \\ 
\Omega^R_{WL^-L^{--}}&=\frac{ec_R^d}{s_W} &  \Omega^R_{ZL^-L^-} &=\frac{es_W}{c_W} \nonumber \\ 
&&\Omega^L_{ZL^-L^{--}}&=\frac{e}{s_Wc_W}(-1+2s_W^2)    \nonumber \\
&&\Omega^R_{ZL^-L^{--}}&=\frac{e}{s_Wc_W}(-1+2s_W^2)
\end{align}

\newpage
\subsection{Passarino-Veltman Integrals}
\label{sec:PVinteg}
The analytical expressions of frequently used PV functions are defined as
\begin{eqnarray}
A_0(m^2)&=&m^2\left(1-\ln\frac{m^2}{\mu^2}\right), \nonumber \\ 
B_0(0,m_1^2,m_2^2)&=&\frac{A_0(m_1^2)-A_0(m_2^2)}{m_1^2-m_2^2},\nonumber \\
B_0(0,m_1^2,m_1^2)&=&\frac{A_0(m_1^2)}{m_1^2}-1, \nonumber \\
B_0(m_1^2,0,m_1^2)&=&\frac{A_0(m_1^2)}{m_1^2}+1, \nonumber \\
B_1(0,m_1^2,m_2^2)&=&\frac{2y^2\ln y_2-4y_2\ln y_2-y_2^2+4y_2-3}{4(y_2-1)^2}+\frac{1}{2}\ln \frac{m_1^2}{\mu^2},\nonumber \\
B_{00}(m_1^2,m_2^2,m_3^2)&=&\frac{(m_1-m_2-m_3)(m_1+m_2-m_3)(m_1-m_2+m_3)(m_1+m_2+m_3)B_0(m_1^2,m_2^2,m_3^2)}{4(1-D)m_1^2}\, \nonumber\\
&+&\frac{A_0(m_2^2)(m_1^2+m_2^2-m_3^2)}{4(1-D)m_1^2}-\frac{A_0(m_3^2)(m_1^2-m_2^2+m_3^2)}{4(1-D)m_1^2},\nonumber \\
B_{00}(0,m_2^2,m_3^2)&=&-\frac{A_0(m_3^2)}{2(1-D)}-\frac{m_2^2B_0(0,m_2^2,m_3^2)}{1-D}-\frac{(m_2^2-m_3^2)B_1(0,m_2^2,m_3^2)}{2(1-D)}\nonumber,\\
B_{00}(0,m^2,m^2)&=&-\frac{A_0(m^2)}{2(1-D)}-\frac{m^2B_0(0,m^2,m^2)}{1-D}, \nonumber \\
B_{00}(m_1^2,m_2^2,m_2^2)&=&\frac{(m_1^2-4m_2^2)B_0(m_1^2,m_2^2,m_2^2)}{4(1-D)}-\frac{A_0(m_2^2)}{2(1-D)},\nonumber \\ 
C_0(m_1^2,m_2^2,m_3^2)&=&\frac{1}{m_1^2}\frac{y_2\ln y_2-y_3\ln y_3-y_2y_3\ln y_2+y_2y_3\ln y_3}{(y_2-1)(y_3-1)(y_2-y_3)},\nonumber \\
C_0(m_1^2,m_2^2,m_2^2)&=&\frac{1}{m_1^2}\frac{\ln y_2-y_2+1}{(y_2-1)^2},\nonumber \\ 
C_0(m^2,m^2,m^2)&=&-\frac{1}{2m^2},
\end{eqnarray}

and the mass ratio parameter 
\begin{equation*}
y_i=\frac{m_i^2}{m_1^2}.
\end{equation*}
It is useful to isolate the divergent part of the Passarino-Veltman integrals:
\begin{eqnarray}
\mathrm{Div} \, \left[ A_0(m^2) \right] &=& \Delta_\epsilon \, m^2, \nonumber \\
\mathrm{Div} \, \left[ B_0(p_{21}^2, m_1^2, m_2^2) \right] &=& \Delta_\epsilon,\nonumber \\
\mathrm{Div} \, \left[ B_1(p_{21}^2, m_1^2, m_2^2) \right] &=& -\frac{1}{2} \, \Delta_\epsilon,\nonumber \\
\mathrm{Div} \, \left[ B_{00}(p_{21}^2, m_1^2, m_2^2) \right] &=& \frac{1}{12} \, \Delta_\epsilon \, \left( 3 m_1^2 + 3 m_2^2 - p_{21}^2 \right),\nonumber \\
\mathrm{Div} \, \left[ B_{11}(p_{21}^2, m_1^2, m_2^2) \right] &=& \frac{1}{3} \, \Delta_\epsilon,\nonumber \\
\mathrm{Div} \, \left[ B_{00}(m_1^2, m_2^2, m_3^2) \right] &=& \frac{1}{4} \, \Delta_\epsilon,
\end{eqnarray}
where the divergent term in MS scheme is given by 
\begin{equation}
\Delta_{\epsilon}=\frac{2}{\epsilon}-\gamma_E+\ln 4\pi\,.
\end{equation}

Finally, the complementary relations to the definitions above can be summarized with the following four scalar functions:
\begin{eqnarray}
B_2(p^2,m_1^2,m_2^2)&=&B_{21}(p^2,m_1^2,m_2^2),\nonumber \\
B_3(p^2,m_1^2,m_2^2)&=&-B_1(p^2,m_1^2,m_2^2)-B_{21}(p^2,m_1^2,m_2^2),\nonumber \\
B_4(p^2,m_1^2,m_2^2)&=&-m_1^2B_1(p^2,m_2^2,m_1^2)-m_2^2B_1(p^2,m_1^2,m_2^2),\nonumber \\
B_5(p^2,m_1^2,m_2^2)&=&A_0(m_1^2)+A_0(m_2^2)-4B_{22}(p^2,m_1^2,m_2^2).
\end{eqnarray}

%%%%%%%%%%%%%%%%%%%%%%%%%%%%%%%%%%%%%%%%%%%%%%%%%%%%%%
\newpage

\bibliography{VLLSM}

\begin{thebibliography}{10}

\bibitem{CMS:2022dwd}
A.~Tumasyan et~al.,
\newblock Nature {\bf 607}, 60 (2022),
\newblock [Erratum: Nature 623, (2023)].

\bibitem{ATLAS:2022vkf}
G.~Aad et~al.,
\newblock Nature {\bf 607}, 52 (2022),
\newblock [Erratum: Nature 612, E24 (2022)].

\bibitem{Sher:1988mj}
M.~Sher,
\newblock Phys. Rept. {\bf 179}, 273 (1989).

\bibitem{Sher:1993mf}
M.~Sher,
\newblock Phys. Lett. B {\bf 317}, 159 (1993),
\newblock [Addendum: Phys.Lett.B 331, 448--448 (1994)].

\bibitem{Elias-Miro:2011sqh}
J.~Elias-Miro et~al.,
\newblock Phys. Lett. B {\bf 709}, 222 (2012).

\bibitem{Degrassi:2012ry}
G.~Degrassi et~al.,
\newblock JHEP {\bf 08}, 098 (2012).

\bibitem{Anchordoqui:2012fq}
L.~A. Anchordoqui et~al.,
\newblock JHEP {\bf 02}, 074 (2013).

\bibitem{Lebedev:2012zw}
O.~Lebedev,
\newblock Eur. Phys. J. C {\bf 72}, 2058 (2012).

\bibitem{Bednyakov:2015sca}
A.~V. Bednyakov, B.~A. Kniehl, A.~F. Pikelner, and O.~L. Veretin,
\newblock Phys. Rev. Lett. {\bf 115}, 201802 (2015).

\bibitem{Alekhin_2012}
S.~Alekhin, A.~Djouadi, and S.~Moch,
\newblock Physics Letters B {\bf 716}, 214–219 (2012).

\bibitem{Gonderinger:2012rd}
M.~Gonderinger, H.~Lim, and M.~J. Ramsey-Musolf,
\newblock Phys. Rev. D {\bf 86}, 043511 (2012).

\bibitem{Falkowski:2015iwa}
A.~Falkowski, C.~Gross, and O.~Lebedev,
\newblock JHEP {\bf 05}, 057 (2015).

\bibitem{Khan:2014kba}
N.~Khan and S.~Rakshit,
\newblock Phys. Rev. D {\bf 90}, 113008 (2014).

\bibitem{Han:2015hda}
H.~Han and S.~Zheng,
\newblock JHEP {\bf 12}, 044 (2015).

\bibitem{Garg:2017iva}
I.~Garg, S.~Goswami, K.~N. Vishnudath, and N.~Khan,
\newblock Phys. Rev. D {\bf 96}, 055020 (2017).

\bibitem{PhysRevD.107.036018}
A.~Arsenault, K.~Y. Cingiloglu, and M.~Frank,
\newblock Phys. Rev. D {\bf 107}, 036018 (2023).

\bibitem{Borah:2020nsz}
D.~Borah, R.~Roshan, and A.~Sil,
\newblock Phys. Rev. D {\bf 102}, 075034 (2020).

\bibitem{Toma:2013bka}
T.~Toma,
\newblock Phys. Rev. Lett. {\bf 111}, 091301 (2013).

\bibitem{Giacchino:2013bta}
F.~Giacchino, L.~Lopez-Honorez, and M.~H.~G. Tytgat,
\newblock JCAP {\bf 10}, 025 (2013).

\bibitem{Giacchino:2014moa}
F.~Giacchino, L.~Lopez-Honorez, and M.~H.~G. Tytgat,
\newblock JCAP {\bf 08}, 046 (2014).

\bibitem{Ibarra:2014qma}
A.~Ibarra, T.~Toma, M.~Totzauer, and S.~Wild,
\newblock Phys. Rev. D {\bf 90}, 043526 (2014).

\bibitem{Barman:2019tuo}
B.~Barman, S.~Bhattacharya, P.~Ghosh, S.~Kadam, and N.~Sahu,
\newblock Phys. Rev. D {\bf 100}, 015027 (2019).

\bibitem{Barman:2019aku}
B.~Barman, D.~Borah, P.~Ghosh, and A.~K. Saha,
\newblock JHEP {\bf 10}, 275 (2019).

\bibitem{Barman:2019oda}
B.~Barman, A.~Dutta~Banik, and A.~Paul,
\newblock Phys. Rev. D {\bf 101}, 055028 (2020).

\bibitem{Giacchino:2015hvk}
F.~Giacchino, A.~Ibarra, L.~Lopez~Honorez, M.~H.~G. Tytgat, and S.~Wild,
\newblock JCAP {\bf 02}, 002 (2016).

\bibitem{Baek:2016lnv}
S.~Baek, P.~Ko, and P.~Wu,
\newblock JHEP {\bf 10}, 117 (2016).

\bibitem{Baek:2017ykw}
S.~Baek, P.~Ko, and P.~Wu,
\newblock JCAP {\bf 07}, 008 (2018).

\bibitem{Colucci:2018vxz}
S.~Colucci et~al.,
\newblock Phys. Rev. D {\bf 98}, 035002 (2018).

\bibitem{Biondini:2019int}
S.~Biondini and S.~Vogl,
\newblock JHEP {\bf 11}, 147 (2019).

\bibitem{Martin:2009bg}
S.~P. Martin,
\newblock Phys. Rev. D {\bf 81}, 035004 (2010).

\bibitem{Zheng:2019kqu}
S.~Zheng,
\newblock Eur. Phys. J. C {\bf 80}, 273 (2020).

\bibitem{Graham:2009gy}
P.~W. Graham, A.~Ismail, S.~Rajendran, and P.~Saraswat,
\newblock Phys. Rev. D {\bf 81}, 055016 (2010).

\bibitem{Endo:2011mc}
M.~Endo, K.~Hamaguchi, S.~Iwamoto, and N.~Yokozaki,
\newblock Phys. Rev. D {\bf 84}, 075017 (2011).

\bibitem{Araz:2018uyi}
J.~Y. Araz, S.~Banerjee, M.~Frank, B.~Fuks, and A.~Goudelis,
\newblock Phys. Rev. D {\bf 98}, 115009 (2018).

\bibitem{Kong:2010qd}
K.~Kong, S.~C. Park, and T.~G. Rizzo,
\newblock JHEP {\bf 07}, 059 (2010).

\bibitem{Huang:2012kz}
G.-Y. Huang, K.~Kong, and S.~C. Park,
\newblock JHEP {\bf 06}, 099 (2012).

\bibitem{Schwaller:2013hqa}
P.~Schwaller, T.~M.~P. Tait, and R.~Vega-Morales,
\newblock Phys. Rev. D {\bf 88}, 035001 (2013).

\bibitem{Halverson:2014nwa}
J.~Halverson, N.~Orlofsky, and A.~Pierce,
\newblock Phys. Rev. D {\bf 90}, 015002 (2014).

\bibitem{Bahrami:2016has}
S.~Bahrami, M.~Frank, D.~K. Ghosh, N.~Ghosh, and I.~Saha,
\newblock Phys. Rev. D {\bf 95}, 095024 (2017).

\bibitem{Bhattacharya:2018fus}
S.~Bhattacharya, P.~Ghosh, N.~Sahoo, and N.~Sahu,
\newblock Front. in Phys. {\bf 7}, 80 (2019).

\bibitem{Agashe:2008fe}
K.~Agashe, T.~Okui, and R.~Sundrum,
\newblock Phys. Rev. Lett. {\bf 102}, 101801 (2009).

\bibitem{Redi:2013pga}
M.~Redi,
\newblock JHEP {\bf 09}, 060 (2013).

\bibitem{Falkowski:2013jya}
A.~Falkowski, D.~M. Straub, and A.~Vicente,
\newblock JHEP {\bf 05}, 092 (2014).

\bibitem{Frank:2014aca}
M.~Frank, C.~Hamzaoui, N.~Pourtolami, and M.~Toharia,
\newblock Phys. Lett. B {\bf 742}, 178 (2015).

\bibitem{Xiao:2014kba}
M.-L. Xiao and J.-H. Yu,
\newblock Phys. Rev. D {\bf 90}, 014007 (2014),
\newblock [Addendum: Phys.Rev.D 90, 019901 (2014)].

\bibitem{Hiller:2022rla}
G.~Hiller, T.~H\"ohne, D.~F. Litim, and T.~Steudtner,
\newblock Phys. Rev. D {\bf 106}, 115004 (2022).

\bibitem{Egana-Ugrinovic:2017jib}
D.~Egana-Ugrinovic,
\newblock JHEP {\bf 12}, 064 (2017).

\bibitem{He:2022zjz}
S.-P. He,
\newblock Chin. Phys. C {\bf 47}, 043102 (2023).

\bibitem{Hiller:2019mou}
G.~Hiller, C.~Hormigos-Feliu, D.~F. Litim, and T.~Steudtner,
\newblock Phys. Rev. D {\bf 102}, 071901 (2020).

\bibitem{PhysRevD.106.055042}
J.~Cao, L.~Meng, L.~Shang, S.~Wang, and B.~Yang,
\newblock Phys. Rev. D {\bf 106}, 055042 (2022).

\bibitem{Dhuria:2015ufo}
M.~Dhuria and G.~Goswami,
\newblock Phys. Rev. D {\bf 94}, 055009 (2016).

\bibitem{Zhang:2015uuo}
J.~Zhang and S.~Zhou,
\newblock Chin. Phys. C {\bf 40}, 081001 (2016).

\bibitem{Gabrielli:2013hma}
E.~Gabrielli et~al.,
\newblock Phys. Rev. D {\bf 89}, 015017 (2014).

\bibitem{Bond:2017wut}
A.~D. Bond, G.~Hiller, K.~Kowalska, and D.~F. Litim,
\newblock JHEP {\bf 08}, 004 (2017).

\bibitem{Hiller:2023bdb}
G.~Hiller, T.~H\"ohne, D.~F. Litim, and T.~Steudtner,
\newblock {Vacuum Stability as a Guide for Model Bulding},
\newblock in {\em {57th Rencontres de Moriond on Electroweak Interactions and
  Unified Theories}}, 2023.

\bibitem{Hiller:2024zjp}
G.~Hiller, T.~H\"ohne, D.~F. Litim, and T.~Steudtner,
\newblock (2024).

\bibitem{Khoze:2014xha}
V.~V. Khoze, C.~McCabe, and G.~Ro,
\newblock JHEP {\bf 08}, 026 (2014).

\bibitem{L3:2001xsz}
P.~Achard et~al.,
\newblock Phys. Lett. B {\bf 517}, 75 (2001).

\bibitem{DELPHI:2003uqw}
J.~Abdallah et~al.,
\newblock Eur. Phys. J. C {\bf 31}, 421 (2003).

\bibitem{10.1093/ptep/ptac097}
P.~D. Group et~al.,
\newblock Progress of Theoretical and Experimental Physics {\bf 2022}, 083C01
  (2022).

\bibitem{Ellis:2014dza}
S.~A.~R. Ellis, R.~M. Godbole, S.~Gopalakrishna, and J.~D. Wells,
\newblock JHEP {\bf 09}, 130 (2014).

\bibitem{Mann:2017wzh}
R.~Mann et~al.,
\newblock Phys. Rev. Lett. {\bf 119}, 261802 (2017).

\bibitem{PhysRevD.7.1888}
S.~Coleman and E.~Weinberg,
\newblock Phys. Rev. D {\bf 7}, 1888 (1973).

\bibitem{PhysRevD.2.1541}
C.~G. Callan,
\newblock Phys. Rev. D {\bf 2}, 1541 (1970).

\bibitem{Chiappini_2023}
M.~Chiappini and on~behalf of~the MEG II~collaboration,
\newblock Journal of Instrumentation {\bf 18}, C10020 (2023).

\bibitem{Pezzullo_2017}
G.~Pezzullo,
\newblock Nuclear and Particle Physics Proceedings {\bf 285?286}, 3?7 (2017).

\bibitem{Blechman2010TheFP}
A.~E. Blechman, A.~A. Petrov, and G.~K. Yeghiyan,
\newblock Journal of High Energy Physics {\bf 2010}, 1 (2010).

\bibitem{Bizot:2015zaa}
N.~Bizot and M.~Frigerio,
\newblock JHEP {\bf 01}, 036 (2016).

\bibitem{CMS:2019hsm}
A.~M. Sirunyan et~al.,
\newblock Phys. Rev. D {\bf 100}, 052003 (2019).

\bibitem{CMS:2022nty}
A.~Tumasyan et~al.,
\newblock Phys. Rev. D {\bf 105}, 112007 (2022).

\bibitem{CMS:2022cpe}
A.~Tumasyan et~al.,
\newblock Phys. Lett. B {\bf 846}, 137713 (2023).

\bibitem{ATLAS:2023sbu}
G.~Aad et~al.,
\newblock JHEP {\bf 07}, 118 (2023).

\bibitem{Sultansoy:2019xiw}
S.~Sultansoy,
\newblock (2019).

\bibitem{CMS:2024bni}
A.~Hayrapetyan et~al.,
\newblock (2024).

\bibitem{Peskin:1991sw}
M.~E. Peskin and T.~Takeuchi,
\newblock Phys. Rev. D {\bf 46}, 381 (1992).

\bibitem{Cynolter:2008ea}
G.~Cynolter and E.~Lendvai,
\newblock Eur. Phys. J. C {\bf 58}, 463 (2008).

\bibitem{Lavoura:1992np}
L.~Lavoura and J.~P. Silva,
\newblock Phys. Rev. D {\bf 47}, 2046 (1993).

\bibitem{garg2013vectorlikeleptonsextended}
S.~K. Garg and C.~S. Kim,
\newblock Vector like leptons with extended higgs sector, 2013.

\bibitem{Hahn:1999mt}
T.~Hahn,
\newblock Acta Phys. Polon. B {\bf 30}, 3469 (1999).

\bibitem{Shtabovenko:2020gxv}
V.~Shtabovenko, R.~Mertig, and F.~Orellana,
\newblock Comput. Phys. Commun. {\bf 256}, 107478 (2020).

\bibitem{Buttazzo_2013}
D.~Buttazzo et~al.,
\newblock Journal of High Energy Physics {\bf 2013} (2013).

\bibitem{Tang:2013bz}
Y.~Tang,
\newblock Mod. Phys. Lett. A {\bf 28}, 1330002 (2013).

\bibitem{Hiller_2022}
G.~Hiller, T.~Höhne, D.~F. Litim, and T.~Steudtner,
\newblock Physical Review D {\bf 106} (2022).

\bibitem{adhikary2024theoreticalconstraintsmodelsvectorlike}
A.~Adhikary, M.~Olechowski, J.~Rosiek, and M.~Ryczkowski,
\newblock Theoretical constraints on models with vector-like fermions, 2024.

\bibitem{PhysRevD.109.036016}
K.~Y. Cingiloglu and M.~Frank,
\newblock Phys. Rev. D {\bf 109}, 036016 (2024).

\bibitem{Altmannshofer_2014}
W.~Altmannshofer, M.~Bauer, and M.~Carena,
\newblock Journal of High Energy Physics {\bf 2014} (2014).

\bibitem{Barducci_2023}
D.~Barducci, L.~Di~Luzio, M.~Nardecchia, and C.~Toni,
\newblock Journal of High Energy Physics {\bf 2023} (2023).

\end{thebibliography}

\end{document}